\documentclass{aa}  

\usepackage{graphicx}
\usepackage{txfonts}

\usepackage{hyperref} 
\newcommand{\lp}{ \;\left(}
\newcommand{\rp}{ \right)}
\newcommand{\lc}{ \left[}
\newcommand{\rc}{ \right]}

\begin{document} 

    \title{Properties of observable mixed inertial and gravito-inertial modes in $\gamma$~Doradus stars}


    \author{Marion Galoy
          \and
          François Lignières
          \and
          Jérôme Ballot
          }

    \institute{IRAP, Université de Toulouse, CNRS, CNES, UPS, 14 avenue Edouard Belin, 31400 Toulouse, France\\
              \email{marion.galoy@irap.omp.eu}
             }

    \date{Received 13 Mars 2024 / Accepted 3 July 2024}

 
    \abstract
   {The space missions \textit{Kepler} and TESS provided a large number of highly detailed time series 
   for 
   main-sequence stars, including $\gamma$ Doradus stars. Additionally, numerous $\gamma$ Doradus stars are to be observed in the near future thanks to the upcoming PLATO mission. 
   In $\gamma$ Doradus stars, gravito-inertial modes in the radiative zone and inertial modes in the convective core can interact resonantly, 
   which translates into the appearance of dip structures in the period spacing of modes. Those dips are information-rich, as they are related to the star core characteristics.}
   {Our aim is to characterise these dips according to stellar properties and thus to develop new
   seismic diagnostic tools to constrain the internal structure of $\gamma$~Doradus stars, especially their cores.}
   {We used the two-dimensional oscillation code TOP to compute sectoral prograde and axisymmetric dipolar modes in $\gamma$ Doradus stars at different rotation rates and evolutionary stages. We then characterised the dips we obtained by their width and location on the period spacing diagram.}
   {We found that the width and the location of the dips 
   depend quasi-linearly on the ratio of the rotation rate and the Brunt-Väisälä frequency at the core interface. This allowed us to determine empirical relations between the width and location of dips as well as the resonant inertial mode frequency in the core
   and the Brunt-Väisälä frequency at the interface between the convective core and the radiative zone. 
   We propose an approximate theoretical model to support and discuss these empirical relations.}
   {The empirical relations we established could be applied to dips observed in data, which would allow for the estimation of frequencies of resonant inertial modes in the core and of the Brunt-Väisälä jump at the interface between the core and the radiative zone. As those two parameters are both related to the evolutionary stage of the star, their determination could lead to more accurate estimations of stellar ages.}


\maketitle
\nolinenumbers 
%

\section{Introduction}
    

Asteroseismology has proven to be a crucial tool in unravelling the mysteries of stellar evolution and determining the ages and evolutionary stages of stars \citep[e.g.][and references therein]{ Lebreton2014b, Aerts2021}. Estimating the evolutionary stages of stars is particularly significant to addressing the complex question of angular momentum transport within stars \citep{Aerts2019}.\\
Among pulsating stars, $\gamma$ Doradus (hereafter $\gamma$ Dor) stars are main-sequence stars that typically have a mass of 1.4--2.0\:M$_\odot$ and a radius of 1.4--2.3\:R$_\odot$ \citep{1999Kaye, Uytterhoeven2011}. Their structure is characterised by a convective core, a large radiative zone, and a small sub-surface convective envelope. They exhibit non-radial gravity modes oscillating with a period of 0.3 to 3\:d \citep{1999Kaye}. Thanks to high-precision photometry provided by space-borne instruments, especially by the \emph{Kepler} mission \citep{Borucki2010} and the Transiting Exoplanet Survey Satellite \citep[TESS;][]{Ricker2014}, a large quantity of good quality data is now available, and it has enabled the study of numerous $\gamma$ Dor stars.\\
Gravito-inertial modes have been identified in 611 $\gamma$ Dor stars observed by \emph{Kepler} \citep{GangLiCatalogue} and in 106 $\gamma$ Dor stars observed by TESS \citep{Garcia2022}. These observed modes have enabled both the rotation rate near the bottom of the radiative zone and the buoyancy radius for a large number of $\gamma$ Dor stars to be determined \citep[see][]{VanReeth2015b,VanReeth2016,Ouazzani2017, VanReeth2018, Saio2018,Christophe2018, Mombarg2019,Ouazzani2019, Li2019b, GangLiCatalogue, Takata2020vrai}, which has led to unprecedented constraints on the models of angular momentum transport \citep{Ouazzani2019, GangLiCatalogue}.
These seismic analyses used an approximation of the rotation effects on gravity modes called the traditional approximation of rotation \citep[TAR; see][]{Unno1989,Lee1989,Lee1997,Townsend2003, Bouabid2013}. While generally convenient for modes confined in radiative zones \citep{Ballot2012, Ouazzani2017, Ouazzani2020}, the TAR cannot describe modes oscillating in the convective cores of $\gamma$ Dor stars.\\

Using an oscillation code that includes a full treatment of the Coriolis force, \citet{Ouazzani2020} showed the existence of gravito-inertial modes resonantly interacting with pure inertial modes confined in the convective core.
In the otherwise smooth curve representing the period spacing between modes of consecutive order $\Delta P$ as a function of the period $P$, this interaction produces observable dips localised near the period of the inertial mode. 
This discovery constitutes a major breakthrough since such a phenomenon can help probe the convective core of $\gamma$ Dor stars. \citet{Saio2021} analysed dips in 16 $\gamma$~Dor stars 
observed by \emph{Kepler} and estimated their core rotation rate. Moreover, \citet{TT22} proposed a theoretical model of the shape of resonant dips in the $\Delta P$ - $P$ curve. 

In this article, we numerically study the resonance between gravito-inertial and pure inertial modes of the convective core 
with the aim of extracting as much information as possible on the convective core from seismic data. 
For the different series of observed gravito-inertial modes \citep[e.g. ][]{GangLiCatalogue},
we studied the evolution of the dips with 
the rotation rate 
and the evolutionary stage of the star. This led us to develop seismic diagnostic tools that are usable on observational data. We also tested the 
model of \citet{TT22} against our numerical results, which led us to construct a new model that better compares with these numerical results.\\
In the following, we first present 
the stellar models and the oscillation code we used and discuss the mode identification in Sect.~\ref{Sec:method}. We then present the results of our numerical computations for three series of modes at different rotation rates and evolutionary stages in Sect.~\ref{Sec:Results}.
In Sect.~\ref{Sec:Analysis}, we provide an empirical model that relates the shape and position of dips to the inner structure of the stars.
We discuss the model of \citet{TT22} and present our new model in Sect.~\ref{Sec:Theory}. We then move to a discussion and the conclusion of paper in Sect.~\ref{Sec:Discussion}.

\section{Method}
\label{Sec:method}
In this section we describe the oscillation code and the stellar models we used to study properties of the dip. We then explain how oscillation modes are identified and selected, and finally we treat the question of numerical resolution.
\subsection{Stellar models}
\label{stellar_models}
In this subsection, we present the three main-sequence star models we used for our study. They were computed using the CESAM code \citep{Morel1997, Morel2008}. Their main characteristics are shown in Table \ref{table:model}. They are named after the CLES models used in \citet{Ouazzani2020}, as they share 
similar characteristics. 
For these models, we adopted a solar metal mixture \citep{Asplund2009}, an initial helium fraction $Y=0.27$, and a initial metallicity $Z=0.02$. We used opacity tables from OPAL \citep{OPAL_ref96} completed at low temperature with tables from \citet{Ferguson2005}. We used the OPAL2005 equation of state \citep{Rogers2002} and the nuclear reaction rates from the NACRE collaboration \citep{Angulo1999_NACRE} except for the $\rm {}^{14}N(p,\gamma){}^{15}O$ reaction, for which we used the LUNA reaction rate given in \citet{Imbriani2004}. Convection was treated using the mixing-length theory \citep{BV1958} with a mixing-length parameter  $\alpha_\mathrm{MLT} = 1.70$, close to a solar calibration. We included turbulent diffusion with a constant diffusion coefficient of $D_\mathrm{t}=700\:\mathrm{cm^2\, s^{-1}}$. The physical prescriptions adopted for these models are thus very similar to the ones used in \citet{Ouazzani2020}.

The three main-sequence models seek to cover the instability strip of $\gamma$ Dor stars \citep{Bouabid2013}.
The model labelled `1z' describes a young star of $1.40\:\mathrm{M_\odot}$ close to the zero age main sequence (ZAMS; age of $180$~My from the ZAMS, $X_{\rm c} = 0.68$, with $X_{\rm c}$ being the mass fraction of hydrogen in the core). The model labelled `2m' describes an evolved mid-main-sequence star of $1.60\:\mathrm{M_\odot}$ (age of $1800$~My from the ZAMS, $X_{\rm c} = 0.38$). The model labelled `3t' describes a star of $1.86\:\mathrm{M_\odot}$ at the end of the main sequence (age of $1480$~My from the ZAMS, $X_{\rm c} = 0.05$).

\begin{table}[!htp] 
\caption{Characteristics of the 1z, 2m, and 3t star models computed by CESAM.}\label{table:model}
\centering
\begin{tabular}{lcccc}
\hline \hline
Model name      & 1z    & 2m & 3t \\ \hline
$M/\rm M_{\odot}$      & 1.40  & 1.60 & 1.86 \\ 
$T_{\rm eff}$ (K)        & 6935 & 7289 & 6694 \\
$\log L/\rm L_{\odot}$ & 0.692 & 1.052 & 1.379\\ 
$\log g$           & 4.31  &  3.99 & 3.59\\ 
$R/\rm R_{\odot}$      & 1.38  & 2.10 & 3.64 \\
Age (Myr)          & 180   & 1800 & 1480 \\
$X_{\rm c}$              & 0.68  &  0.38 & 0.05 \\ [0.5ex]
\hline \end{tabular}
\tablefoot{Mass, effective temperature, luminosity, surface gravity, radius, age and central hydrogen mass fraction \emph{(from top to bottom)}}
\end{table}

\begin{figure}[!htp]
    \centering
    \includegraphics[width=1.03\linewidth]{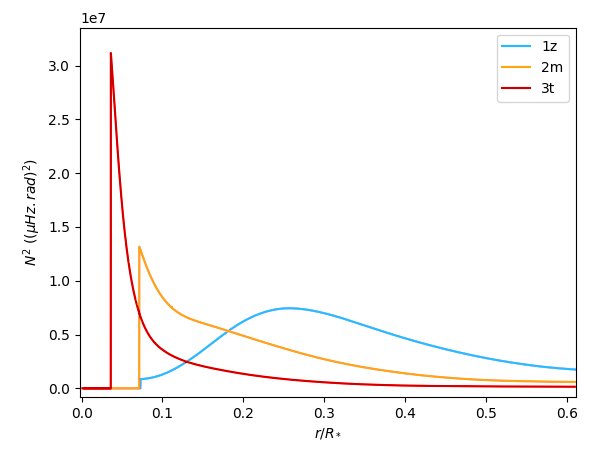}
    \caption{Squared Brunt-Väisälä frequency $N^2$ of the models presented in Table \ref{table:model} as a function of the relative radius. Model 1z is drawn in blue, 2m in orange, and 3t in red.}
    \label{fig:NN_1z_2m_3t}

\end{figure}

The Brunt-Väisälä frequency $N$ is an important parameter in our study since gravity modes depend directly on it. Moreover, the resonance with inertial modes in the convective core is greatly affected by the behaviour of $N$ at the bottom of the radiative zone, as we demonstrate later. 
As shown in Fig. \ref{fig:NN_1z_2m_3t}, the three models have different radial profiles of the Brunt-Väisälä frequency. They all present a discontinuity at the convective core-radiative zone interface, even the model that has just left the ZAMS (model 1z). This is due to the chemical composition gradient that develops on the outer edge of the convective core as the star evolves on the main sequence. This gradient, and thus the jump of the Brunt-Väisälä frequency, is sensitive to mixing processes such as rotational mixing, overshooting and microscopic diffusion. These mixing processes are modelled here through the turbulent diffusivity $D_{\rm t}$ mentioned above.




\subsection{Oscillation code}

To calculate the oscillation modes and frequencies, we used the code TOP \citep[Two-dimensional Oscillation Program,][]{2006PhDReese, Reese2009}, which we describe in this subsection.\\
We considered time-harmonic adiabatic small perturbations of a uniformly rotating spherical star. Thanks to the axial symmetry, the solutions are proportional to $ e^{i(m\phi +\omega t)}$, where $m$ is the azimuthal number and $\omega$ is the mode frequency. The governing equations are
\begin{gather}
   \rho + \Vec{\xi}\cdot\Vec{\nabla}\rho_0 + \rho_0 \Vec{\nabla}\cdot \Vec{\xi} = 0, \\
    2i\rho_0\Vec{\Omega}\times [\omega +m\Omega]\Vec{\xi} - [\omega +m\Omega]^2 \rho_0 \Vec{\xi} =  - \Vec{\nabla}p + \frac{\Vec{\nabla}P_0}{\rho_0}\rho - \rho_0 \Vec{\nabla}\psi, \\
   p - c_0^2 \rho + \Vec{\xi}\cdot\left ( \Vec{\nabla} P_0 - c_0^2 \Vec{\nabla} \rho_0 \right ) = 0, \\
   \Vec{\Delta}\psi - 4\pi G \rho =0,
\end{gather}
where $\Vec{\xi}$, $p$, $\rho$, and $\psi$ are the amplitude of the Lagrangian displacement, the Eulerian perturbations of pressure, density, and gravitational potential, respectively, and where $\rho_0$, $P_0$, and $c_0$ are the density, the pressure, and the sound velocity of the star model, respectively, with $c_0=\sqrt{\Gamma_1 p_0/\rho_0}$, $\Gamma_1$ being the first adiabatic index. The star rotation rate is $\Omega$, and $G$ is the gravitational constant. The mode frequency in the inertial frame $\omega$ is related to the mode frequency in the co-rotating frame $\omega_{\mathrm{co}}$ through
\begin{equation}
\omega = \omega_{\mathrm{co}} - m \Omega.
\end{equation}

The governing equations were discretised using a fourth-order finite difference method in the radial direction \citep[for details on the scheme, see][]{Reese2013} and spectral decomposition on spherical harmonics in the horizontal directions.
After discretisation, an algebraic eigenvalue problem was obtained and solved using the Arnoldi-Chebyshev algorithm \citep{chatelin1988valeurs, Reese2009}. Due the axial and equatorial symmetries of the problem, independent eigenvalue problems were solved for a given azimuthal order $m$ and a given parity with respect to the equator.

\subsection{Mode identification}

In practice, TOP finds a certain number of modes (typically two to eight) around a given frequency for a prescribed resolution. These modes have the same equatorial parity and azimuthal order, but modes associated with different degrees and radial orders are computed simultaneously. Unlike with 1D calculations, identification is thus non-trivial. 
Additionally, modes unresolved at the chosen resolution can be among the computed ones.\\
The series of gravito-inertial modes we computed for this work are the modes with the same degree $\ell$ and azimuthal number $m$ but different radial orders $n$. The degree $\ell$ corresponds to the associated Legendre polynomial $P_\ell^m$ that characterises the mode at zero rotation but remains relevant to label these same modes at higher rotation rates. 
For identification, we used the fact that the modes of the series are recognisable by their radial and latitudinal structure. 
We applied three criteria on the perturbed pressure $p$ 
to select the desired modes among the computed ones; two are related to the radial and latitudinal profiles, while the third criterion takes care of miscalculated modes. Using the asymptotic Wentzel-Kramers-Brillouin (WKB) formulation of the TAR \citep[see][]{Unno1989}, the number of radial nodes $n$ of a ($\ell, m$) gravito-inertial mode can be estimated by
\begin{equation}
    n + \epsilon_g \approx \frac{P_\mathrm{co} \sqrt{\Lambda_\ell^m}}{\Pi_0},
    \label{eq:radial_nodes}
\end{equation}
with $P_\mathrm{co}$ as the period of the mode in the co-rotating frame, $\Pi_0$ as the buoyancy radius, $\Lambda_\ell^m$ as the eigenvalue associated with the Hough function $H_\ell^m$ describing the latitudinal part of the mode in the TAR (see Appendix \ref{rad}), and $\epsilon_g$ as a small offset. We used this estimate of $n$ to remove modes with radial orders in the radiative zone noticeably higher than expected.
Additionally, we computed the correlation of the mode latitudinal profiles in the radiative zone with the corresponding Hough function and rejected modes with weak correlation coefficients.
Finally, we excluded badly resolved and spurious modes. This included modes with an extremely high amplitude in the core or at the surface as well as modes showing strong variations between consecutive radial grid points.

\subsection{Numerical resolution}
\begin{figure}[!htp]
    \centering
    \includegraphics[width=\linewidth]{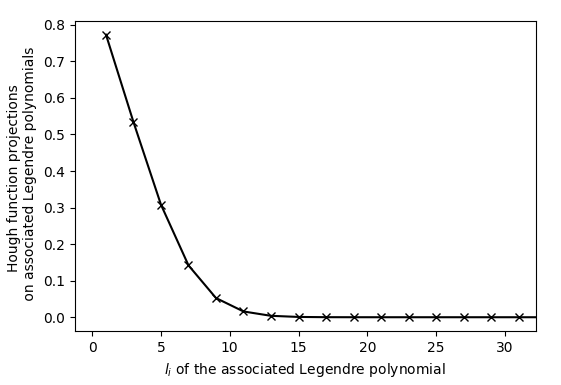}
    \caption{Projection of the Hough function describing the latitudinal profile of the mode $n=66$ ($\ell=1$, $m=-1$) in the radiative zone (TAR) on the associated Legendre polynomials $P_{\ell_i}^{m=-1}$.}
    \label{fig:hough_on_legendre}
\end{figure}

In this subsection, we estimate which radial and latitudinal resolutions are needed for our analysis.
The radial resolution was determined by the number of radial points $n_r$ and needed to be adjusted depending on the number of radial nodes of the considered mode. \\
The latitudinal resolution $n_\theta$ corresponds to the number of associated Legendre polynomials $P_\ell^m$ on which the modes are decomposed. The degree $\ell$ spans from $\ell_\mathrm{min}=|m|+i_p$ to $\ell_\mathrm{max}=\ell_{\rm min}+2(n_\theta-1)$, where $i_p=0$ or 1 when $\ell+m$ is even or odd, respectively.
For example, a mode of the series ($\ell=1$, $m=-1$) calculated with $n_\theta=3$ is decomposed latitudinally on the three first associated Legendre polynomials with the same azimuthal number and equatorial parity $P_1^{-1}$, $P_3^{-1}$ and $P_5^{-1}$.
The minimal latitudinal resolution needed to compute a given mode can be estimated by looking for the number of associated Legendre polynomials needed to represent the associated Hough function.
Figure \ref{fig:hough_on_legendre} shows that for the Hough function  $H_{\ell=1}^{m=-1}$ calculated at the spin parameter $s = 2\Omega/\omega_\mathrm{co} \sim 16.4$, 
the projection on Legendre polynomials peaks at  $P_1^{-1}$, while the contributions of polynomials above $P_{11}^{-1}$ are very small.
This suggests that a latitudinal resolution of at least $n_\theta=6$ is needed for this mode.

\begin{figure}[!htp]
    \centering
    \includegraphics[width=\linewidth]{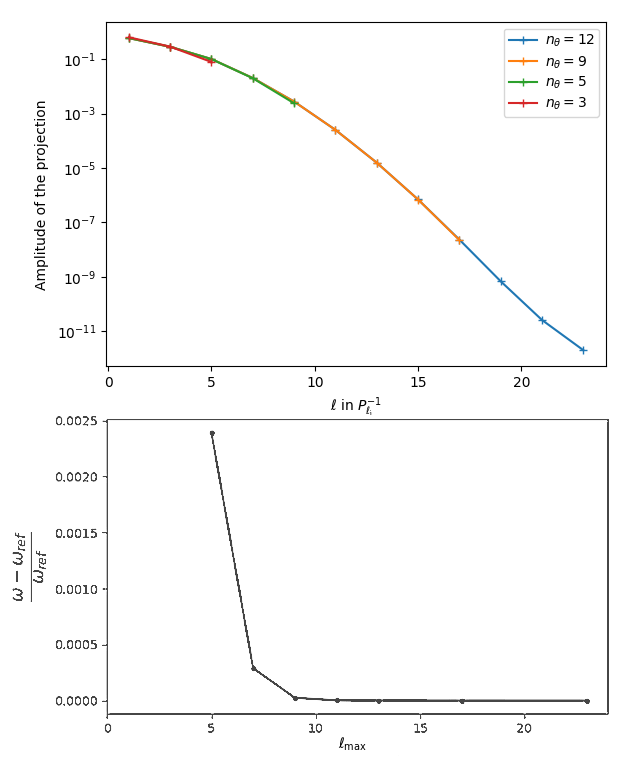}
    \caption{Effect of the latitudinal resolution on the mode $n=66$ ($s\sim16.4$) of the series ($\ell=1$, $m=-1$).
    Upper panel: Projections of the mean latitudinal profile in the radiative zone of this mode at different latitudinal resolution $n_\theta$ on the associated Legendre polynomials for a rotation rate $\Omega = 0.4 \Omega_\mathrm{K}$. Here, $\Omega_\mathrm{K}=\sqrt{G M / R^3}$ is the critical rotation rate of the star and $M$ and $R$ are its mass and radius, respectively. Lower panel:
    Evolution of the relative frequency of the same mode as a function of $\ell_\mathrm{max}$. The reference frequency $\omega_\mathrm{ref}$ corresponds to the frequency obtained using the highest latitudinal resolution ($n_\theta=12$). The model used is 1z.}
    \label{fig:example_repartitionlegendre}

\end{figure}

For a more accurate estimation, we applied the same process with 
modes calculated by TOP.
For example, in the upper panel of Fig. \ref{fig:example_repartitionlegendre}, one can see the decomposition of the mean latitudinal profile of a mode of the series ($\ell=1, m=-1$) onto the associated Legendre polynomials for different latitudinal resolutions. We first noticed that the amplitude decreases rapidly with the degree $\ell_i$. Nevertheless, when we used too few polynomials, we missed non-negligible contributions of polynomials of degree $\ell_i$ larger than $\ell_\mathrm{max}$, leading to inaccuracies. This translated into errors on the frequency determination. In the lower panel of the same figure, one can see how the computed frequency evolves with the latitudinal resolution and converges towards a stable value at a large $\ell_\mathrm{max}$. For this mode, the frequency varies by less than 0.01\% for an $\ell_\mathrm{max}$ higher than nine ($n_\theta=5$). 

\begin{figure}[!htp]
    \centering
    \includegraphics[width=\linewidth]{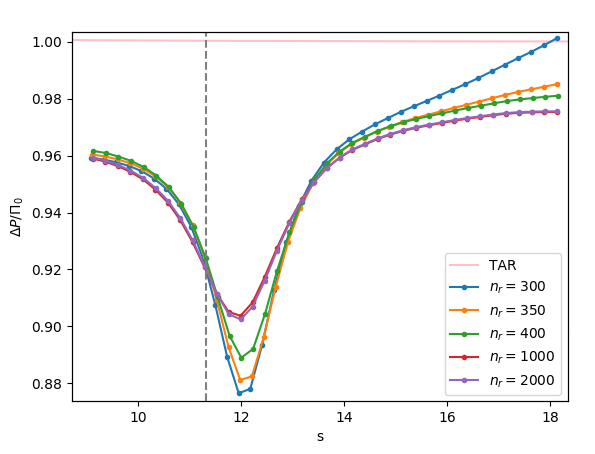}
    \caption{Period spacing $\Delta P_\mathrm{co}$, normalised by $\Pi_0$ (buoyancy radius), 
    plotted as a function of the spin parameter $s$ with different radial resolution $n_r$ for the series of modes ($\ell=1$, $m=-1$) and for a rotation rate $\Omega = 0.4 \Omega_\mathrm{K}$. Their radial order $n$ goes from $32$ to $72$. The model used is 1z. The pink line shows the traditional approximation of rotation. The vertical line shows the spin parameter of the resonant inertial mode calculated with an analytical uniform density model.}
    \label{fig:example_inadequate_nr}

\end{figure}

\begin{figure}[!htp]
    \centering
    \includegraphics[width=\linewidth]{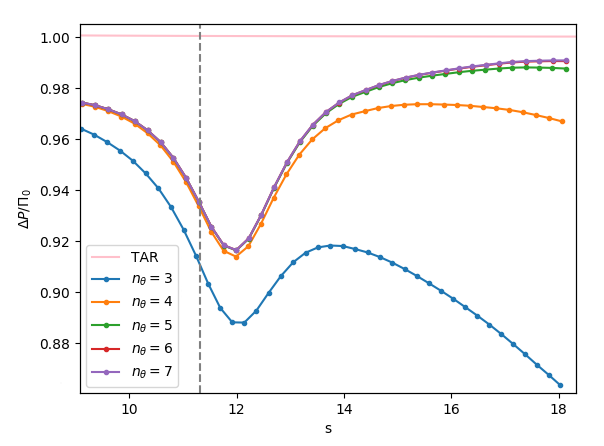}
    \caption{Period spacing $\Delta P_\mathrm{co}/\Pi_0$ plotted as a function of the spin parameter $s$ with different latitudinal resolutions $n_\theta$ for the series of modes ($\ell=1$, $m=-1$) and for a rotation rate $\Omega = 0.4 \Omega_\mathrm{K}$. Their radial order $n$ goes from 
    $32$ to $72$. The model used is 1z. The pink line shows the traditional approximation of rotation. The vertical line shows the spin parameter of the resonant inertial mode calculated with an analytical uniform density model.}
    \label{fig:example_inadequate_nt}

\end{figure}

Because we studied the dips, the resolution was chosen such that the dips are computed with enough accuracy.
Figures \ref{fig:example_inadequate_nr} and \ref{fig:example_inadequate_nt} show the impact of the radial and latitudinal  resolutions on a series of modes in diagrams depicting period spacing $\Delta P_\mathrm{co}$ (period differences of two modes with consecutive $n$ in the co-rotating frame) as a function of the spin parameter $s$ for the rotation rate $\Omega=0.4\Omega_\mathrm{K}$. The vertical line indicates the spin parameter of the expected resonant mode in the case of a core of uniform density (see Sect. \ref{Subsec:where_dips}). We observed that the dip deepens and that $\Delta P_\mathrm{co}$ increases for the higher $s$ as the radial resolution decreases, which increases the general slope of the curve. When the latitudinal resolution decreases, $\Delta P_\mathrm{co}$ decreases for the higher $s$, which leads to the opposite change in the general slope of the curve.
The curve barely evolves for $n_r>1000$ and $n_\theta>6$,
which confirms the primary guess we made using the projection of the Hough function on the associated Legendre polynomials. Based on Fig. \ref{fig:example_inadequate_nr} and knowing that the mode with the highest number of radial nodes is $n=72$, we deduced that about $15\times n$ points is enough to reach convergence.\\

To describe the dips accurately, we wanted the error on the period spacing $\delta \Delta P_\mathrm{co}$ not to exceed 1\% of the total depth of the dip. We determined the radial and latitudinal resolutions for our numerical calculations in order to meet this criterion on the frequency error $\delta \omega$ using
\begin{equation}
    \delta \omega = \omega_\mathrm{co}^2 \frac{\delta \Delta P_\mathrm{co}}{2\pi\sqrt{2}}.
\end{equation}

To ensure that $\delta\omega < 1\%$, we showed that it is sufficient to choose a radial resolution $n_r$ ranging from 5000 to 10000 for the ($\ell=1$, $m=-1$) series, from 3000 to 10000 for the ($\ell=2$, $m=-2$) series, and from 3000 to 10000 for the ($\ell=1$, $m=0$) series. These ranges cover the different resolutions that are needed since the radial order increases with the spin parameter and evolutionary status.
For the latitudinal resolution, we typically used $n_\theta = 6$ to $8$
for the series ($\ell=1$, $m=-1$) and ($\ell=2$, $m=-2$), and $n_\theta = 4$ to $6$ for the series ($\ell=1$, $m=0$).

 \begin{figure*}[!htp]
    \centering
    \includegraphics[width = \textwidth]{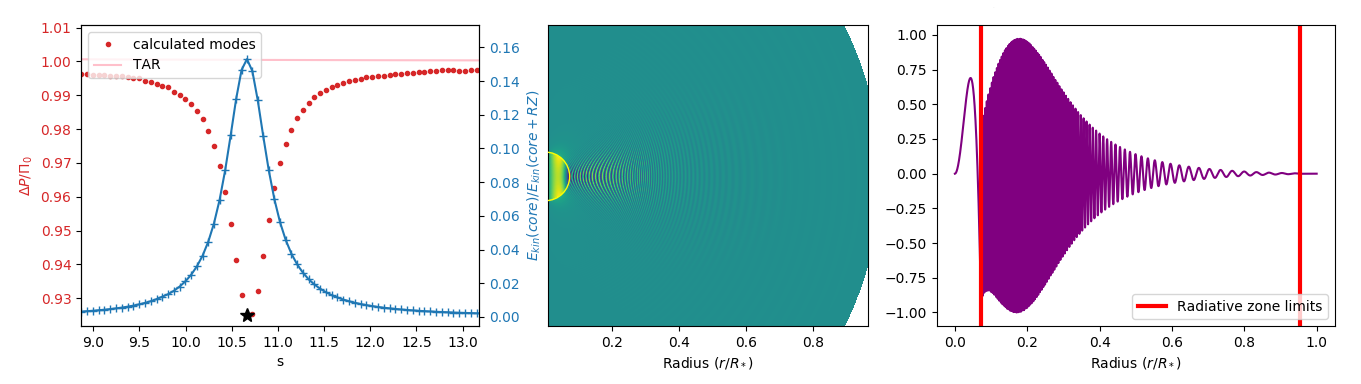}
    \caption{Example of a resonant mode.    
    Left panel: Period spacing $\Delta P_\mathrm{co}$ as a function of the spin parameter for the series ($\ell=1$, $m=-1$) and for the 1z model at a rotation rate of $\Omega = 0.1 \Omega_\mathrm{K}$.
    Red dots represent computed modes. The pink line shows the TAR. The blue line shows the proportion of kinetic energy in the core for each mode. Middle panel: 
    Quantity $p/\rho_0^{0.8}$ plotted in a meridional plane for the mode marked by a black star in the left panel. 
    Right panel: Radial profile of $p/\rho_0^{0.8}$ at the latitude $\pi$/2 (equator). The factor $\rho_0^{0.8}$ is a scaling factor used to make the visualisation easier. The red vertical lines show the radiative zone limits.}
    \label{fig:example_dip_Ekin}
\end{figure*}

\section{Results}
\label{Sec:Results}
In this section, we present the results of our numerical calculations. We studied the series of modes ($\ell = 1$, $m=-1$), ($\ell = 2$, $m=-2$), and ($\ell = 1$, $m=0$) at rotation rates ranging from $0.1$ to $0.5$ $\Omega/\Omega_\mathrm{K}$, with $\Omega_\mathrm{K}=\sqrt{G M / R^3}$, for the three stellar models described in Table \ref{table:model}. 
Among these three series, the ($\ell = 1$, $m=-1$) is by far the most frequently observed \citep{GangLiCatalogue}.
For each series, we computed the period spacing $\Delta P_{\mathrm{co}}$ in a frequency interval where the gravito-inertial modes strongly interact with an inertial mode of the convective core. We then described the properties of the dips obtained in the $\Delta P_\mathrm{co}-s$ diagram.
In the following, we first present the basic characteristics of a typical dip in Sect. \ref{Subsec:dips_properties}. The simple model  we used to determine the approximate spin parameter of the dips is described in Sect. \ref{Subsec:where_dips}. The results of our investigation of dips of the  ($\ell = 1$, $m=-1$), ($\ell = 2$, $m=-2$), and ($\ell = 1$, $m=0$) mode series are then reported in Sect. \ref{Subsec:Evo_sigma_sc}.

\subsection{Basic dip properties}
\label{Subsec:dips_properties}

We first describe the properties of a typical dip for the series ($\ell=1$, $m=-1$). The dip is located around $s=10.7$ and was calculated with the 1z model at rotation rate $\Omega=0.1\Omega_\mathrm{K}$ (shown in the left panel of Fig.~\ref{fig:example_dip_Ekin}). 
One can see the difference between our calculations and the TAR shown in pink. While there is a clear dip in the $\Delta P_\mathrm{co}-s$ curve around $s=10.7$ in our calculations, the TAR is nearly constant.
For prograde sectoral modes ($\ell=-m$), the TAR indeed leads to a constant $\Delta P_\mathrm{co}$ in the limit of high spin parameters. 
The middle and right panels of Fig. \ref{fig:example_dip_Ekin} show the Eulerian pressure perturbation $p$ for the mode located at the centre of the dip. It exhibits a very high amplitude in the convective core compared to the radiative zone.  
The presence of significant oscillations in both the convective core and the radiative zone indicates the mixed nature of the mode.
To confirm the mixed nature of the modes in the dip, we calculated the portion of kinetic energy in the core for each mode. It reads
\begin{equation}
    \frac{E_\mathrm{kin}(\mathrm{core})}{E_\mathrm{kin}(\mathrm{core + RZ})} =  \frac{\int_0^{r_c} \int_0^{2\pi} \int_0^\pi  \rho_0 \left |\Vec{v} \right |^2 r^2 \sin{\theta} \,\mathrm{d} \theta \,\mathrm{d} \phi \,\mathrm{d}r}{\int_0^{R} \int_0^{2\pi} \int_0^\pi  \rho_0 \left |\Vec{v} \right |^2 r^2 \sin{\theta} \,\mathrm{d} \theta \,\mathrm{d} \phi \,\mathrm{d}r},
\end{equation}
with $E_\mathrm{kin}(\mathrm{core})$ being the 
mode kinetic energy in the core and $E_\mathrm{kin}(\mathrm{core + RZ})$  as the mode kinetic energy in the core and the radiative zone. The mode perturbation velocity is denoted as $\Vec{v}$, the radius of the core as $r_c$, and the radius of the star as $R$. 
As observed in the left panel of Fig. \ref{fig:example_dip_Ekin}, the portion of kinetic energy in the core increases for the modes in the dip, reaching its maximum when $\Delta P_\mathrm{co}$ is minimum. 
Dips are formed by mixed (inertial and gravito-inertial) modes around the spin parameter of an inertial mode of the core that couples with the gravito-inertial modes of the $(\ell,m)$ series. In the dip, a frequency is added to the series. \\
To study the dips and quantify their evolution, we modelled them as inverse Lorentzian profiles by following \citet{TT22}:
\begin{equation}
        \frac{1}{\Delta P_\mathrm{co}} = x_1 + x_2 s + \frac{\Omega \sigma  / \pi^2 }{(s-s_c)^2 + \sigma^2},
        \label{eq:fit_lorentzienne_basique}
\end{equation}
where $s$ is the spin parameter of the gravito-inertial modes in the co-rotating frame and $x_1$, $x_2$, $s_c$, and $\sigma$ are free fitting parameters. With this parametrisation, $x_1$ and $x_2$ account for the general slope of the curve, $s_c$ is the spin parameter on which the dip is centred, and $\sigma$ is the width of the dip. 
\\
Adding one frequency locally modifies the period spacing but not the whole period gap between unaffected modes. This constraint imposes that the integral of the Lorenztian term  over $s$ is constant and thus that the width and the depth are anti-correlated \citep{TT22}.
We fit this model to our numerically computed dip using a least-squares method. 
The errors associated with the fit tend to increase with the rotation rate, as the dip implies less modes when the rotation increases, decreasing the quality of the fit. The relative error of $s_c$ is negligible; the one of $\sigma$ is always less than 1\% for the series ($\ell=1$, $m=-1$) and less than 2\% for the series ($\ell=2$, $m=-2$), but it goes up to 10\% with the rotation rate for the series ($\ell=1$, $m=0$). The latter case could be explained by the fact that $x_1$ and $x_2$ account for a linear slope of $1/\Delta P_\mathrm{co}$, which is only a good local approximation of the TAR. As the rotation rate increases, the dip widens, and this approximation becomes less valid, explaining the growing relative error.


\subsection{Model of \cite{Ouazzani2020} as an initial guess}
\label{Subsec:where_dips}

Assuming a convective core of uniform density, \citet{Ouazzani2020} proposed a simple analytical model to identify the inertial modes that produce significant dips in the period spacings of a gravito-inertial mode series.  \citet{Ouazzani2020} showed that this model provides a useful approximation of the dip spin parameters observed in full numerical computations.  

We thus used this model to choose the dips we wanted to study and to get an initial guess of their spin parameter. The model details are given in Appendix \ref{appendix:homogeneous_model}. Once a dip was found numerically, we could easily check that it is due to the expected inertial mode because the spatial structure of the inertial mode in our complete calculations (see for example the core region in the middle panel of Fig. \ref{fig:example_dip_Ekin}) is very similar to that of the inertial mode in the sphere of uniform density (see comparisons in Appendix \ref{appendix:homogeneous_model}).

For each series of gravito-inertial modes, we limited our study to resonances that could occur in spin-parameter ranges actually observed in $\gamma$ Dor stars. This condition was verified for one resonance in each series, which happens to be the one occurring at the smallest spin parameter.

\subsection{Evolution of the width and central spin parameter of the dips}
\label{Subsec:Evo_sigma_sc}

 \begin{figure*}[!htp]
    \centering
    \includegraphics[width = \textwidth]{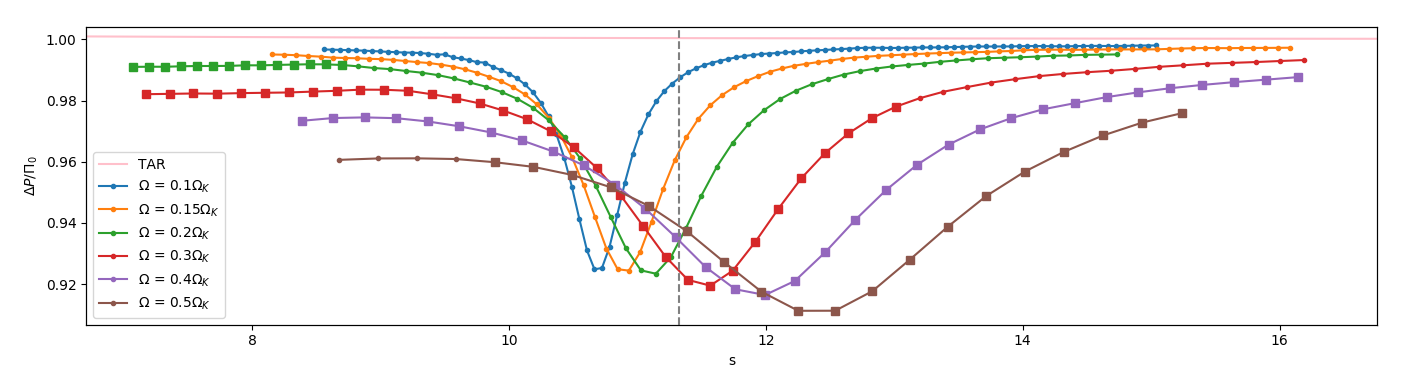}
    \caption{Period spacing in the co-rotating frame as a function of the spin parameter for the modes of the series ($\ell=1$, $m=-1$) at different rotation rates using the 1z model. The period spacing is normalised by the buoyancy radius $\Pi_0$. The pink line shows the traditional approximation of rotation. The square dots indicate the modes that are within the range of typically observed radial orders $30<n<70$. The dashed line corresponds to the frequency of the resonant inertial mode $\ell_i=3$, $m=-1$ calculated analytically with the uniform density model.}
    \label{fig:grosse_synthese}

\end{figure*}



 \begin{figure*}[!htp]
    \centering
    \includegraphics[width = \textwidth]{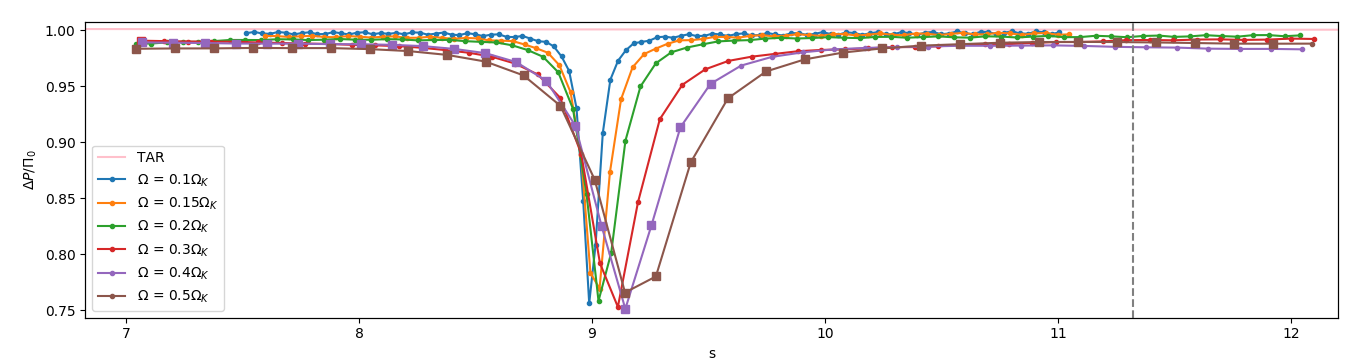}
    \caption{Same as Fig.~\ref{fig:grosse_synthese} but for the model 2m. The low-amplitude oscillations in $\Delta P$ are due to a glitch generated by a small numerical discontinuity in its Brunt-Väisälä frequency profile \citep[see][]{Miglio2008}.  }
    \label{fig:grosse_synthese_2m}

\end{figure*}

 \begin{figure*}[!htp]
    \centering
    \includegraphics[width = \textwidth]{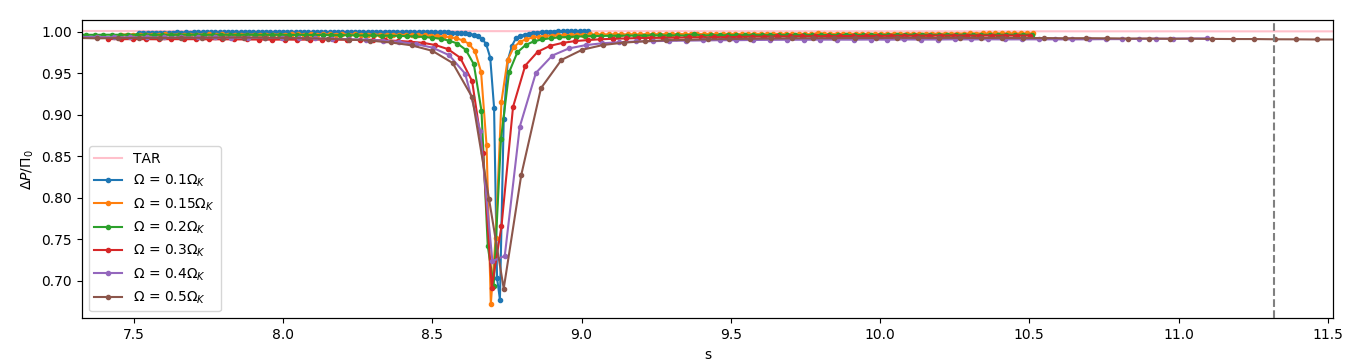}
    \caption{Same as Fig.~\ref{fig:grosse_synthese} but for the model 3t.
    }
    \label{fig:grosse_synthese_3t}

\end{figure*}

 \begin{figure}[!htp]
    \centering
    \includegraphics[width=\linewidth]{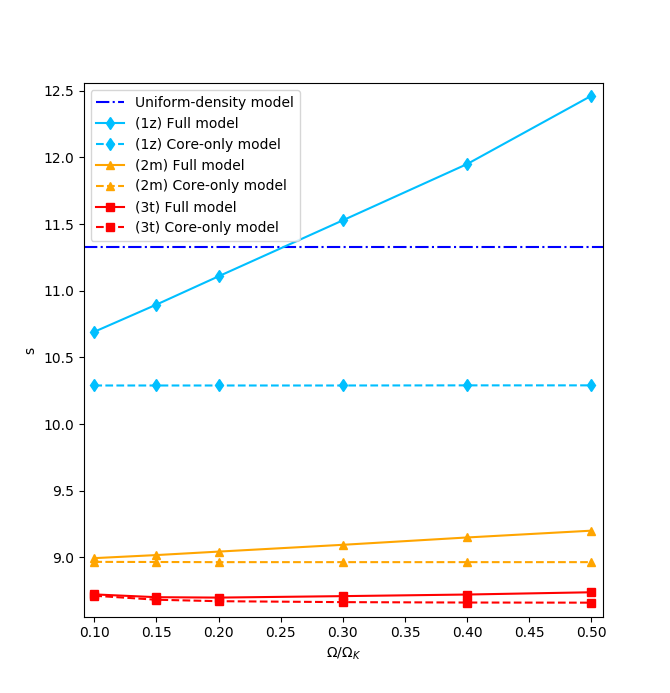}
    \caption{Evolution of the centre of the dip $s_\mathrm{c}$ obtained using a full model (full line) and evolution of the spin parameter of the inertial mode obtained using a core-only model (dotted line). The dash-and-dot blue line shows the spin parameter of the inertial mode in the case of a uniform density 
    core (analytical model). The light blue curves refer to the model 1z, orange curves 
    to the model 2m, and red curves 
    to the model 3t.}
    \label{fig:comparaison_s}
 \end{figure}

 \begin{figure}[!htp]
    \centering
    \includegraphics[scale = 0.59]{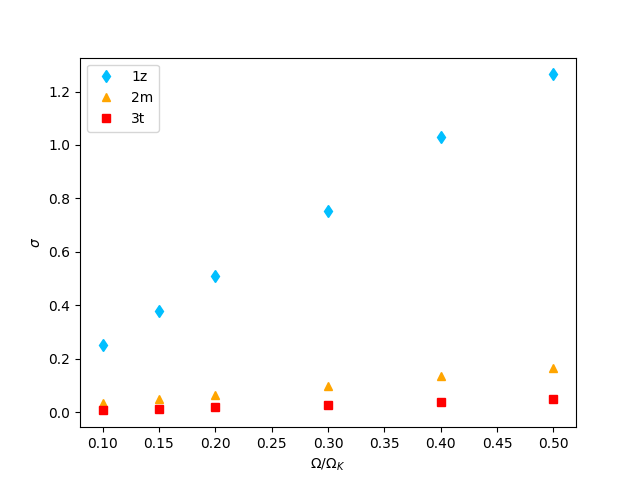}
    \caption{Width of the dips $\sigma$ as a function of the rotation rate for the series ($\ell=1$, $m=-1$) for the three stellar models described in Table \ref{table:model}. The error bars are too small to be visible with the scale used.}
    \label{fig:evo_sigma_rota}

\end{figure}

We show in Figs. \ref{fig:grosse_synthese},  \ref{fig:grosse_synthese_2m}, and \ref{fig:grosse_synthese_3t} the evolution of the first dip of the series ($\ell=1$, $m=-1$) with the star rotation rate for the 1z, 2m, and 3t model, respectively. The square dots indicate modes within an observable range of radial orders. 
Indeed, according to \citet{GangLiCatalogue}, the observed modes of the ($\ell=1$, $m=-1$) series are typically of a radial order $30<n<70$.
Analogous figures for the other two series, ($\ell=2$ $m=-2$) and ($\ell=1$, $m=0$), are shown in Appendix \ref{appendix:more_results}.

We fit the dips using Eq.~(\ref{eq:fit_lorentzienne_basique}) in order to determine their width $\sigma$ and central spin parameter $s_c$. 
Figures~\ref{fig:comparaison_s} and \ref{fig:evo_sigma_rota} show the evolution of $s_c$ and $\sigma$ with the star rotation for the three stellar models. The label `core-only model' refers to a model that we discuss later in Sect. \ref{subsec:empirical_sc} (some oscillation modes calculated with this model are shown in Appendix \ref{appendix:s_iner}).
One can see in Fig.~\ref{fig:comparaison_s} that $s_c$ varies with the rotation rate but becomes less sensitive to it as the star evolves. It generally increases with the rotation rate, with the exception of the 3t model between $\Omega=0.1 \Omega_\mathrm{K}$ and $\Omega=0.15 \Omega_\mathrm{K}$.
We also observed that $s_c$ decreases as the star evolves. The width $\sigma$ follows the same trend: It increases with the rotation rate but becomes smaller and less sensitive to the rotation rate as the star evolves (see Fig. \ref{fig:evo_sigma_rota}).
In summary, the rotation rate of the star tends to make the dips occur at a higher $s$ and widens them, whereas the dips become narrower and occur at a lower $s$ as the star ages.\\
We note that outside of the dip and for the three models we considered, the curve remains further and further from the TAR as the rotation rate grows. This behaviour has already been observed in previous studies \citep[e.g.][]{Ouazzani2017}.

\section{Analysis}
\label{Sec:Analysis}
In this section we analyse the evolution on $\sigma$ and $s_c$ and derive approximate empirical relations usable on observational data.

\subsection{Empirical expression of $\sigma$}

\begin{figure}[!ht]
    \centering
    \includegraphics[width=\linewidth]{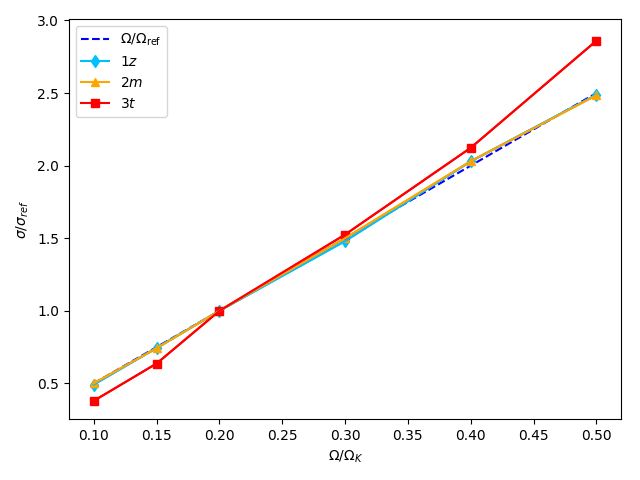}
    \caption{Evolution of $\sigma/\sigma_\mathrm{ref}$ with the rotation rate for the dip of the series ($\ell=1$, $m=-1$) for three different evolutionary stages (1z, 2m, 3t). The associated error bars are too small to be visible. The reference value $\sigma_\mathrm{ref}$ is the width for $\Omega = \Omega_\mathrm{ref} = 0.2\Omega_\mathrm{K}$. If, as predicted by \cite{TT22}, $\sigma \propto \Omega$, then $\sigma/\sigma_\mathrm{ref} = \Omega/\Omega_\mathrm{ref} =  5 \Omega$ (the dashed line).}
 \label{fig:rapport_sigma_T&T}
\end{figure}

According to the model of \citet{TT22}, $\sigma$ evolves linearly with the parameter $\epsilon = \Omega/N_0$, where 
$N_0$ is the jump of the Brunt-Väisälä frequency at the interface between the convective core and the radiative region. We tested whether $\sigma$ is proportional to $\Omega$ by plotting its evolution with $\Omega$ alongside the theoretical linear evolution predicted in \citet{TT22}.
Figure \ref{fig:rapport_sigma_T&T} shows the evolution of $\sigma/\sigma_{\mathrm{ref}}$ with the rotation rate, where $\sigma_\mathrm{ref}$ is the width computed at $\Omega=\Omega_\mathrm{ref}= 0.2\Omega_\mathrm{K}$. This figure shows that $\sigma$ is indeed proportional to the rotation rate of the star, with the exception of the 3t model for which one can see a deviation that is significant compared to the errors of $\sigma$. The origin of this particular behaviour remains unclear at the moment, and numerical effects cannot be excluded. Analogous figures for the series ($\ell=2$, $m=-2$) and ($\ell=1$, $m=0$) are shown in Appendix \ref{appendix:more_results}.

\begin{figure}[!htp]
    \centering
    \includegraphics[width=\linewidth]{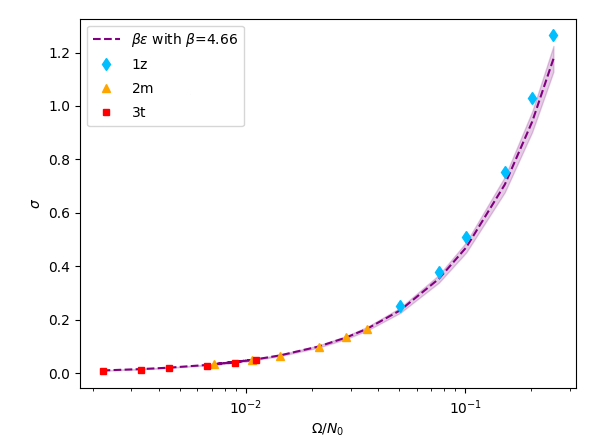}
    \caption{Width $\sigma$ as a function of $\epsilon$ for the series ($\ell=1$, $m=-1$) for three different evolutionary stages (1z, 2m, 3t). The purple dashed line shows the fit presented in Table \ref{table:beta_sigma} for the three models together, and the purple area shows the related error. The x-axis is logarithmic for visualisation purposes.}
    \label{fig:evo_sigma_log}
\end{figure}

Then, assuming a linear relation $\sigma = \beta\epsilon$, we performed a linear regression to determine the slope $\beta$. 
We used the least-squares method. Table \ref{table:beta_sigma} gathers the values of $\beta$ obtained for each dip and stellar model. While $\beta$ is expected to depend on the core structure \citep{TT22}, we find that it varies by less than $15$\% for the three stellar models considered.
For the dip of the ($\ell=1$, $m=-1$) series, we show in Fig.~\ref{fig:evo_sigma_log} the result of the fit obtained by considering the three stellar models together. When directly fitting with relation $\sigma = \beta \epsilon$, the 1z-model $\sigma$ values are weighted more in the determination of the free parameter $\beta$ because, due to the smaller $N_0$ of the 1z model, they extend over a larger $\epsilon$ range. To counter that, 
we actually fitted the relation $\sigma N_0 =\beta \Omega$.
Analogous figures for the dips of the ($\ell=2$, $m=-2$) and ($\ell=1$, $m=0$) series are shown in Appendix \ref{appendix:more_results}. 
Using our empirical relation, 
the mean value of $\beta$ could thus be used to estimate $N_0$ from $\sigma$ 
via
\begin{equation}
    N_0^\mathrm{calc} = \frac{\Omega \beta}{\sigma}.   \label{eq:calc_N0}
\end{equation}

\begin{table}[!htp] 
\centering
\caption{Values of $\beta$.}
\small
\begin{tabular}{crcccc}
\hline \hline
        $\ell$   &   $m$         &        1z       & 2m              &     3t       &  All \\[0.5ex]\hline
1 & -1    &  $5.03 \pm 0.06$  &  $4.63 \pm 0.05$  &   $4.33  \pm 0.37$  &  $4.66 \pm 0.19$ \\
1 & 0        &  $0.59 \pm 0.01$  &  $0.64 \pm 0.03$  &  $0.60 \pm 0.08$      &      $0.61 \pm 0.03$       \\
2 & -2       & $2.89 \pm 0.04$   &  $2.83 \pm 0.02$  &  $2.55 \pm 0.30$      &    $2.76 \pm 0.11$       \\
[0.5ex]
\hline
\end{tabular}
\tablefoot{Values of $\beta$ are obtained by fitting the relation $\sigma N_0 = \beta\Omega$
for the series of modes ($\ell=1$, $m=-1$), ($\ell=1$, $m=0$), and ($\ell=2$, $m=-2$) for the models 1z, 2m, and 3t separately and together.}
\label{table:beta_sigma}
\end{table}

\subsection{Empirical expression of $s_c$}
\label{subsec:empirical_sc}

In order to understand the variations of $s_c$, we first computed the spin parameters $s_*$ of inertial modes of truncated versions of our models limited to their convective cores, as done in \citet{Ouazzani2020}.
It allowed us to see the impact of varying the density as well as the effect of the rotation rate on the frequency of the resonant inertial modes. The modes were calculated with the boundary condition $\xi_r=0$, with $\xi_r$ being the radial displacement, and identified as the resonant modes using their number of latitudinal and radial nodes. 
The inertial modes and their spin parameter obtained for this core-only model are reported in Appendix \ref{appendix:s_iner}.
As already observed in \citet{Ouazzani2020}, the spin parameter of these modes remains independent of $\Omega$ as long as an acoustic term can be neglected in the governing equations (see Eq.~(\ref{eq:Wu2005})).

We could then compare the central spin parameter of the dip $s_c$ obtained in the previous section with
the spin parameter $s_*$ of the resonant inertial mode of the core-only model. We show the comparison for the series ($\ell=1$, $m=-1$) for the three stellar models in Fig. \ref{fig:comparaison_s}. The spin parameter of the inertial modes obtained using the uniform density model presented in Appendix \ref{appendix:homogeneous_model} is also indicated. Analogous figures for the ($\ell=2$, $m=-2$) and ($\ell=1$, $m=0$) series are shown in Appendix \ref{appendix:more_results}. We note that for the same inertial mode, the spin parameter decreases with the star evolution. This behaviour depends on the density stratification of the core, which is related to the stellar mass and the stellar evolution.
We observed that while the spin parameters of the eigenmodes of the uniform-density model and the core-only model remain constant with the rotation (with the exception of $s_*$ for the 3t model between $\Omega=0.1 \Omega_\mathrm{K}$ and $\Omega=0.15 \Omega_\mathrm{K}$), the central spin parameter of the dip
shows a significant evolution with the rotation rate. 
For the three models, $s_c$ tends towards $s_*$ as the rotation rate decreases. Additionally, the differences between the full model and the core-only model significantly decrease as the star ages. This is consistent with the fact that the Brunt-Väisälä jump $N_0$ grows as the star evolves on the main sequence, which makes the radial displacement $\xi_r$ at the core interface increase 
and thus makes the core more isolated and closer to the core-only model. This also explains the decrease of the dip width with evolutionary stage, as a less deformable interface weakens the coupling, which thus impacts fewer and fewer modes and leads to narrower dips. \\
While looking for an empirical relationship between $s_c$ and $\epsilon$ usable on observation data, we noticed that though $s_{\rm c}$ does not vary linearly with the rotation for all models (especially 3t), the difference $s_c-s_*$ appears to be proportional to $\Omega$ (figure not shown). We thus fit $s_c - s_*$ as a linear function of $\epsilon$: $s_c - s_* = \alpha \epsilon$.
We estimated $\alpha$ for the series of modes ($\ell=1$, $m=-1$), ($\ell=1$, $m=0$), and ($\ell=2$, $m=-2$) and for the models 1z, 2m, and 3t separately. Table \ref{table:alpha_sc} provides the estimate of $\alpha$ for each case.

\begin{table}[!htp]  
\caption{Values of $\alpha$.}
\centering
\begin{tabular}{crccc}
\hline \hline

      $\ell$  &    $m$        &       1z          &       2m      & 3t    \\[0.5ex]\hline
1 & -1       &  $8.36 \pm 0.24$  &  $6.29 \pm 0.58$  & $6.75 \pm 0.49$\\ 
1 & 0        &  $0.90 \pm 0.01$  &  $1.04 \pm 0.03$  & $1.16 \pm 0.05$\\  
2 & -2       &  $5.16 \pm 0.06$  &  $3.97 \pm 0.40$  & $4.29 \pm 0.49$ \\ 
[0.5ex]\hline 
\end{tabular}
\tablefoot{Values of $\alpha$ are obtained by fitting the relation $s_c(\epsilon) - s_*(\epsilon) = \alpha \epsilon$ for the series of modes ($\ell=1$, $m=-1$), ($\ell=1$, $m=0$), and ($\ell=2$, $m=-2$) for the models 1z, 2m, and 3t.}
\label{table:alpha_sc}
\end{table}

\begin{figure}[!htp]
    \centering
    \includegraphics[width=\linewidth]{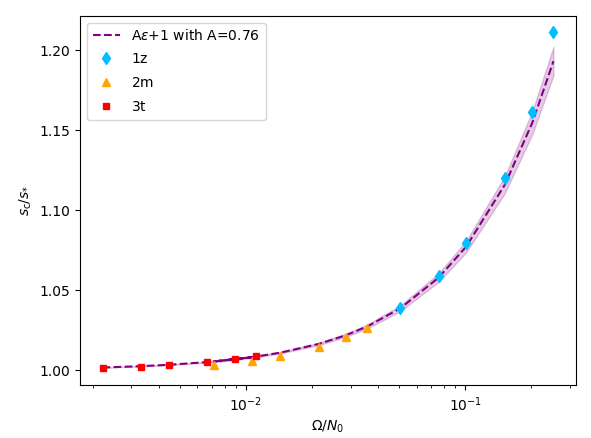}
    \caption{Ratio  
$s_c/s_*$ as a function of $\Omega/N_0$ ($=\epsilon$) for the series ($\ell=1$, $m=-1$) for the 1z, 2m, and 3t models. The purple dashed line shows the fit presented in Table \ref{table:A_sc/s*} for the three models together, and the purple area shows the related error. The x-axis is logarithmic for visualisation purposes.}
    \label{fig:fit_sc}
\end{figure}

One can see that $\alpha$ varies significantly (up to $\sim$30\%) with the star model for a same series of modes. 
However, we observed that the ratio $s_c/s_*$ plotted as a function of $\epsilon$ appears to almost gather on a same line for the three models (see Fig.~\ref{fig:fit_sc}). Since this quantity depends less on the model, we
fitted it as a linear function $s_c/s_* = A\epsilon + B$ for the three models, separately and together.
We obtained values of B compatible with 1, and this confirms that $s_c$ tends to $s_*$ when $\epsilon$ tends to 0. Therefore, we redid fits by fixing $B=1$, letting only $A$ be a free parameter. As we did for the estimation of $\beta$, we fitted $(s_c/s_* -1)N_0 = A \Omega$ so that all the models were weighted equally. The result of the fit for $A$ is shown in Table \ref{table:A_sc/s*}.

\begin{table}[!htp] 
\caption{Values of $A$.} 
\small
\centering
\begin{tabular}{crcccc}
\hline \hline

      $\ell$     & $m$ &       1z          &   2m       & 3t      &  All       \\[0.5ex]\hline
1 & -1       &  $0.81 \pm 0.02$  &  $0.70 \pm 0.06$  &  $0.78 \pm 0.06 $     & $0.76 \pm 0.04$   \\ 
1 & 0       &  $0.41 \pm 0.01$  &  $0.48 \pm 0.01$  &  $0.54 \pm 0.02$       &   $0.48 \pm 0.03$   \\
2 & -2      &  $0.65 \pm 0.01$  &  $0.57 \pm 0.05$  & $0.63 \pm 0.07$       &    $0.62 \pm 0.03$     \\
[0.5ex]
\hline 
\end{tabular}
\tablefoot{Values of $A$ are obtained by fitting the relation $(s_c/s_*-1)N_0=A\Omega$ for the series of modes ($\ell=1$, $m=-1$), ($\ell=1$, $m=0$), and ($\ell=2$, $m=-2$) for the models 1z, 2m, and 3t, separately and together.}

\label{table:A_sc/s*}
\end{table}
The fitting parameter $A$ varies by less than 15\% 
for the ($\ell=1$, $m=-1$) and the ($\ell=2$, $m=-2$) series between the three models, while it goes up to about 30\%
for the series ($\ell=1$, $m=0$). 

The global fit of all models together is plotted in Fig.~\ref{fig:fit_sc} for the series ($\ell=1$, $m=-1$).
Analogous figures for the ($\ell=2$, $m=-2$) and ($\ell=1$, $m=0$) series are shown in Appendix \ref{appendix:more_results}. 
As $A$ does not vary much with the stellar structure, we assumed it to be constant in order
to retrieve the frequency of the resonant inertial mode in the core, $s_*$, from the location of an observed dip through the following relation:
\begin{equation}
    s_*^{\mathrm{calc}} = \frac{N_0^\mathrm{calc} s_c}{A\Omega+N_0^\mathrm{calc}}. 
\end{equation}
Using Eq.~(\ref{eq:calc_N0}), we rewrote this relation as
\begin{equation}
    s_*^{\mathrm{calc}}= \frac{\beta s_c}{A\sigma +\beta}. \label{eq:calc_s_iner}
\end{equation}

Since $s_c$ and $\sigma$ are obtained from the $\Delta P_\mathrm{co}$-$s$ relation in the co-rotating frame, the rotation rate $\Omega$ is needed. It can be determined by fitting the TAR outside the dip \citep[e.g.][]{GangLiCatalogue}.

\subsection{Relative errors made on the estimations of $N_0$ and $s_*$}

\begin{figure}[!htp]
    \centering
    \includegraphics[width=1.07\linewidth]{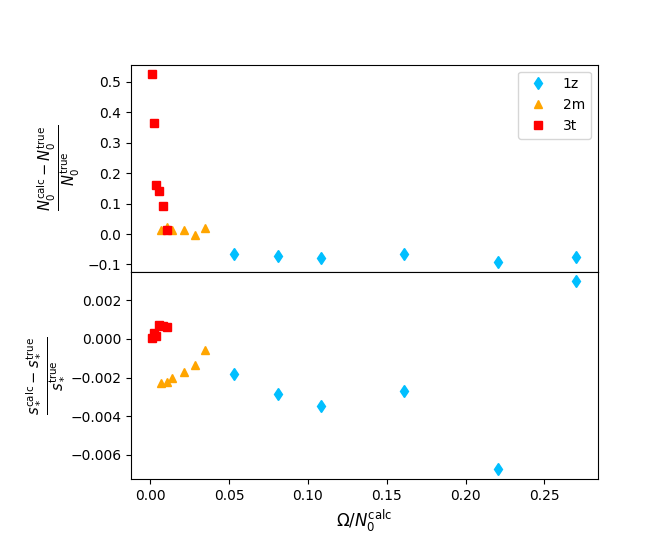}
    \caption{Relative errors for the series ($\ell=1$, $m=-1$) on the estimated $N_0^\mathrm{calc}$ (upper panel) and $s_*^\mathrm{calc}$ (lower panel) using the three models 1z, 2m, and 3t.}
    \label{fig:relative_error}
\end{figure}

\begin{figure}[!htp]
    \centering
    \includegraphics[width = 1.07\linewidth]{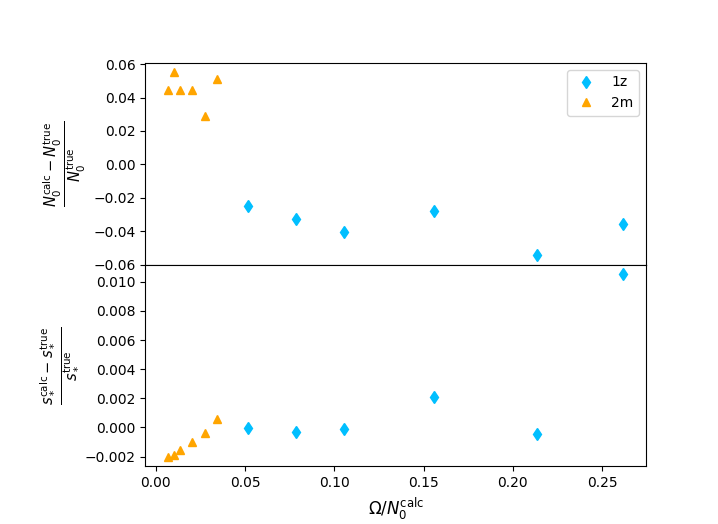}
    \caption{Relative errors for the series ($\ell=1$, $m=-1$) on the estimated $N_0^\mathrm{calc}$ (upper panel) and $s_*^\mathrm{calc}$ (lower panel) using the two models 1z and 2m.}
    \label{fig:relative_error_two_models}
\end{figure}

To estimate the errors made by our empirical method, we applied it to dips calculated numerically, and we obtained empirical measurements of
$N_0$ and $s_*$ using Eqs. (\ref{eq:calc_N0}) and (\ref{eq:calc_s_iner}), denoted as $N_0^\mathrm{calc}$ and $s_*^\mathrm{calc}$. We then calculated the errors relative to the real known values $N_0^\mathrm{true}$ and $s_*^\mathrm{true}$. 
We estimated $s_c$ and $\sigma$ by fitting the dips with the Eq. (\ref{eq:fit_lorentzienne_basique}), and the rotation rates 
were assumed to be perfectly known.
Figure~\ref{fig:relative_error} shows the relative error made on $N_0$ and $s_*$ as a function of $\epsilon^\mathrm{calc}=\Omega/N_0^\mathrm{calc}$ for the series ($\ell=1$, $m=-1$).
Analogous figures for the series ($\ell=2$, $m=-2$) and ($\ell=1$, $m=0$) are shown in Appendix \ref{appendix:more_results}. One can see that for the ($\ell=1$, $m=-1$) series, the relative error on $N_0$ is always inferior or equal to 10\% for the model 1z and 2m, while it increases up to 50\% for slowly rotating 3t models. 
The relative error for the 3t model is below 10\% for the rotation rates $\Omega = 0.4 , 0.5\Omega_\mathrm{K}$, between 10 and 20\% for the rotation rates $\Omega = 0.2 , 0.3\Omega_\mathrm{K}$, and between 30 and 50\% for the rotation rates $\Omega = 0.1 , 0.15\Omega_\mathrm{K}$. 
Such behaviour is due to the direct dependency between $N_0^\mathrm{calc}$ and $\sigma$. Indeed, as $\epsilon$ gets smaller, $\sigma$ goes to zero, making relative errors explode when we evaluate the difference between $\sigma$ and $\beta \epsilon$, despite decreasing absolute errors.
Nevertheless, it has to be noted that as typical radial orders for observable modes of the ($\ell=1$, $m=-1$) series range between 30 and 70 \citep[see][]{GangLiCatalogue}, none of the dips obtained with the 3t model are observable. Indeed, 
they exhibit radial orders lying out of the [30,70] range, going from $n\approx 90$ at $\Omega=0.5\Omega_\mathrm{K}$ to $n\approx 650$ at $\Omega = 0.1\Omega_\mathrm{K}$.
The same applies to the ($\ell=2$, $m=-2$) and ($\ell=1$, $m=0$) series, as the observed modes are typically of about the same radial order as the ($\ell=1$, $m=-1$) series, but the radial order of the modes calculated with the 3t model never goes lower than $n\approx170$ for the ($\ell=2$, $m=-2$) series and $n\approx~90$ for the ($\ell=1$, $m=0$) series.\\
The relative error of $s_*$ is 
far smaller and does not become greater than $\sim 0.7\%$. It tends to decrease as the stars evolves. As dips are not currently observed in stars as evolved as the 3t model, we also estimated the 
parameters $A$ and $\beta$ 
using only 
the 1z and 2m models (see the results in Table \ref{table:A_beta_two_models}). The relative error made on $N_0$ and $s_*$ for the series ($\ell=1$, $m=-1$) using these new estimates of $A$ and $\beta$ are shown in Fig. \ref{fig:relative_error_two_models}. The relative error of $N_0$ is always inferior to 6\%, and the relative error on $s_*$ is inferior to 0.2\%, except for the case of 1z and $\Omega/\Omega_\mathrm{K}=0.5$, where it reaches $\sim$ 1\%. The same approach was used for the series ($\ell=1$, $m=0$) and ($\ell=2$, $m=-2$), and the results are reported
in Appendix \ref{appendix:more_results}. 

\begin{table}[!htp]  
\caption{Values of $A$ and $\beta$ when only taking into account the models 1z and 2m together.
}
\centering
\begin{tabular}{lrcc}
\hline \hline
$\ell$  & $m$&       $A$         &       $\beta$   \\[0.5ex]\hline
1 & -1       &  $0.76 \pm 0.05$  &  $4.83 \pm 0.14$  \\ 
1 & 0        &  $0.44 \pm 0.03$  &  $0.61 \pm 0.02$  \\  
2 & -2       &  $0.61 \pm 0.04$  &  $2.86 \pm 0.03$   \\ 
[0.5ex]\hline 
\end{tabular}

\tablefoot{Values of $A$ and $\beta$ are obtained by 
fitting the relations $(s_c/s_*-1)N_0=A\Omega$ and $\sigma N_0 =\beta \Omega$ for the series of modes ($\ell=1$, $m=-1$), ($\ell=1$, $m=0$), and ($\ell=2$, $m=-2$).}

\label{table:A_beta_two_models}
\end{table}

We showed that for observable dips, our empirical method can accurately predict the frequency of the resonant inertial mode in the core, and it provides a reasonable estimate of the jump in the Brunt-Väisälä frequency at the interface between the core and the envelope. 
Both $N_0$ and $s_*$ are correlated to the evolutionary stage of the star. 
In particular, the frequency of the resonant inertial mode is only sensitive to the structure of the core. It is thus independent of the mixing and diffusion processes occurring in the overlying radiative layers, which, in contrast, may deeply affect the shape of the Brunt-Väisälä frequency profile.
\\
Further investigations are needed in order to characterise how the frequencies of inertial modes depend on the characteristics of the convective core. 
Inertial modes
could be a powerful tool in obtaining constraints on convective cores and thus precise determinations of stellar ages.



\section{Approximate analytical models}
\label{Sec:Theory}
We present the dip model of \cite{TT22} as well as a newly improved version of the model and compare them with the properties of the dip obtained by numerical calculations in the previous sections.

\subsection{Model of \citet{TT22}}

\cite{TT22} proposed an analytical model of the dips in $\Delta P = f(s)$ provoked by the coupling between gravito-inertial and inertial oscillations.
It is based on three main assumptions: the density is uniform in the convective core, the gravito-inertial oscillations are described by the TAR, and the ratio between the (uniform) rotation rate and the Brunt-Väisälä frequency at the bottom of the radiative zone is small. The assumption of uniform density is obviously unrealistic, but it enables one to get analytical solutions of the inertial oscillations in the core. In the absence of coupling with the radiative envelope, inertial modes in the uniform density core are identified by their spin parameter{, $s_*$, and the degree and azimuthal order}, $(\ell_i,m)$, of the Legendre polynomial involved in the solution (see Appendix~\ref{appendix:homogeneous_model}). Under the TAR, the latitudinal part of gravito-inertial $(\ell,m)$ modes is characterised by the Hough function $H_k^m (\theta,s)$ associated with the eigenvalue $\Lambda_k^m(s)$, where $k=\ell-|m|$ (see Eq.~\ref{Laplace}). The model results depend on the continuity of the  density and the Brunt-Väisälä frequency at the interface between the convective core and the radiative zone. For all stellar models of $\gamma$ Dor stars presented in Sect.~\ref{stellar_models}, the density is continuous at the convective-radiative interface, while the Brunt-Väisälä is discontinuous. Indeed, the jump from zero on the convective side to $N_0$ on the radiative side increases with age. According to \citet{TT22}, the coupling between the $(\ell,m)$ series of the gravito-inertial modes and a $(\ell_i,m)$ inertial mode produces a Lorentzian-shaped dip in the $\Delta P = f(s)$ that depends on two parameters, namely its width,\footnote{The expressions of $\sigma$ and $s_c$ in the case considered here are not explicitly written in \citet{TT22} but can easily be derived from their equations (89), (90), and (91) with $\Delta\rho=0$ and replacing $\tilde{F}$ by $F$ given by their equation (58).}
\begin{equation}
   \sigma = 
\frac{2 P_{\ell_i}^m(1/s_*)}{{\gamma'}_{\ell_i}^m(s_*)} \sqrt{\Lambda_{k}^m(s_*)}  \frac{\Omega}{N_0},
\end{equation}
and its central spin parameter,
\begin{equation}
   s_c = s_*,
\end{equation}
where ${\gamma'}_{\ell_i}^m$ is the derivative with respect to $s$ of the function
\begin{equation}
\gamma_{\ell_i}^m(s) =  {P'}_{\ell_i}^m (1/s) - \frac{m}{1 - 1/s^2} P_{\ell_i}^m(1/s),
\end{equation}
\noindent where ${P'}_{\ell_i}^m$  is the derivative with respect to $x$ of the Legendre polynomial ${P}_{\ell_i}^m (x)$.

A first success of this model is that the Lorentzian form of the dips is fully consistent with our numerical results (see Sect.~\ref{Subsec:dips_properties}).
The expressions of $\sigma$ and $s_c$ indicate that the dip parameters depend on the ratio $\Omega/N_0$, on the gravito-inertial oscillations through $\Lambda_k^m$,  and on the inertial oscillation through $s_*$ and $R(s_*) = 2 P_{\ell_i}^m (1/s_*)/{\gamma'}_{\ell_i}^m (s_*)$.
\cite{TT22} argued that these expressions remain relevant in the case of a realistic core with a non-uniform density, although the quantities related to the inertial oscillation, that is $s_*$ and the function $R(s)$, are no longer known analytically. 
We can nevertheless determine $s_*$ numerically by computing the inertial modes of stellar models truncated at the boundary of the convective core where the radial displacements vanish (see Appendix~\ref{appendix:s_iner}).
This allowed us to test the $s_c=s_*$ prediction against the numerical determination of $s_c$ for the dips studied in this paper. As shown in Fig. \ref{fig:comparaison_s}, our numerical results are not consistent with $s_c=s_*$. \\
The theoretical expression of $\sigma$ cannot be fully tested with our numerical results because the function $R(s)$, which is derived from the analytical form of free inertial oscillations at the convective-radiative interface, is unknown for cores of variable density. Nevertheless, the proportionality $\sigma \propto \Omega$ can be tested by considering the evolution of dip widths with rotation for a given star model and $(\ell,m)$ series. Figure~\ref{fig:rapport_sigma_T&T} shows that $\sigma \propto \Omega$ is indeed in agreement with our numerical results despite the small deviations observed in the case of the most evolved star model. \citet{TT22} also treated the case of a continuous Brunt-Väisälä frequency profile at the core interface and found a different scaling of the dip width with the rotation. We note that even the small jump of the Brunt-Väisälä in the 1z model produces an evolution of $\sigma$ corresponding to the discontinuous case.
A questionable property of the theoretical $\sigma$ is that each inertial mode with the same $m$ and equatorial parity as the gravito-inertial modes of a $(\ell,m)$ series produces a dip of non-negligible width. 
As already discussed in \cite{TT22}, this property is at odds with the \citet{Ouazzani2020} phenomenological model, where an efficient resonance requires a non-negligible geometrical matching between the gravito-inertial and inertial modes at the convective-radiative interface. This matching selects the significantly resonant inertial modes and thus the dips in the $\Delta P = f(s)$, a prediction that was successfully tested with full numerical calculations by \citet{Ouazzani2020}.

\subsection{The improved model}

\begin{figure}
\centering
\includegraphics[width=\hsize]{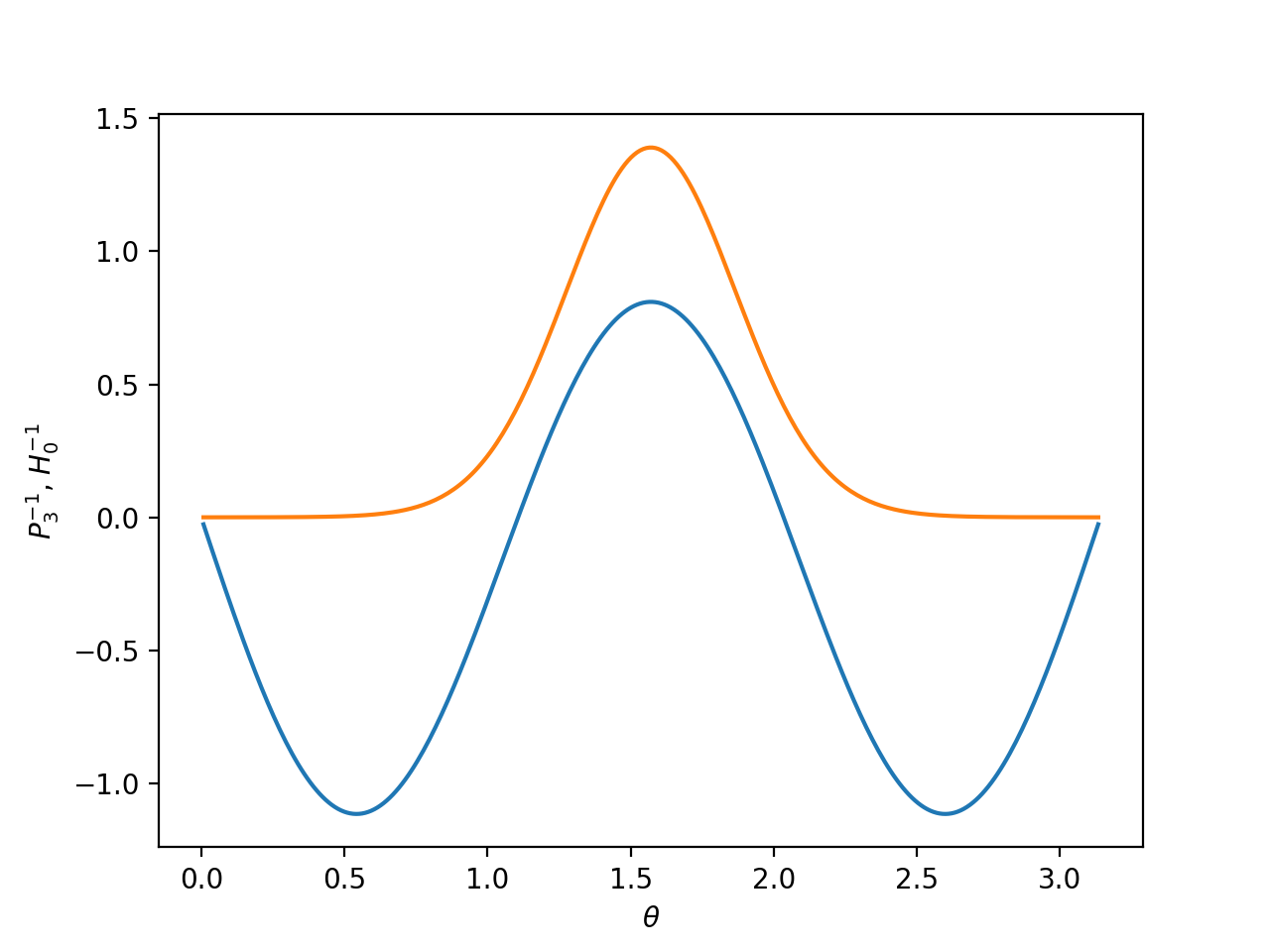}
\caption{Latitudinal profiles of the $\ell_i=3,m=-1, s_*=11.32$ inertial mode ($P_3^{-1}(\cos \theta)$, blue) and 
$\ell=1,m=-1$ gravito-inertial modes near $s_*$ ($H_0^{-1}(\theta,s_*)$, orange) at the convective-radiative interface. Mixed modes that are continuous at the interface can hardly be constructed with these two modes only. }
         \label{profile}
\end{figure}

In this subsection, we present a newly improved analytical model of the dips studied in this paper.
This model is similar to the \citet{TT22} model in that it uses the same three basic assumptions: uniform density core, TAR gravito-inertial oscillations, and $\Omega/N_0 \ll 1$. However, it goes a step further by improving on the geometrical matching of the inertial and gravito-inertial oscillations at the convective-radiative interface. This matching is indeed required by the continuity condition on the radial displacement and the pressure perturbations at this interface. 
Figure~\ref{profile} shows the latitudinal profiles of the gravito-inertial mode and the inertial mode involved in the $(\ell=1,m=-1)$ versus $(\ell_i=3, m=-1)$ resonance. It is clear that the continuity condition cannot be realised by only considering these two modes, as is done in \citet{TT22}. We improved the matching by allowing more complex eigenfunctions on both sides of the interface.
The detail of the derivation is given in Appendix \ref{model}.

We find that the coupling between the $(\ell=1,m=-1), (\ell=2,m=-2), (\ell=1,m=0)$ 
gravito-inertial modes and respectively the $(\ell_i=3,m=-1),(\ell_i=4,m=-2),(\ell_i=3,m=0)$ inertial modes produces Lorentzian-shaped dips in the $\Delta P = f(s)$ relation whose widths and central spin parameters are described by
\begin{align}
\label{sig_th}
\sigma = & \lp \frac{f_{k,\ell_i}^m (s_*)}{1+f_{k,\ell_i}^m (s_*)} \frac{2 P_{\ell_i}^m (1/s_*)}{{\gamma'}_{\ell_i}^m (s_*)} \sqrt{\Lambda_{k}^m (s_*)}\rp  \frac{\Omega}{N_0}, \\
\label{s_th} s_c = & s_* + \lp \frac{1}{1+f_{k,\ell_i}^m (s_*)} \frac{2 P_{\ell_i}^m (1/s_*)}{{\gamma'}_{\ell_i}^m (s_*)}  \sqrt{\Lambda_{k-2}^m (s_*)} \rp \frac{\Omega}{N_0},
\end{align}

\noindent where $f_{k,\ell_i}^m$ is defined by
\begin{equation}
  f_{k,\ell_i}^m (s_*)= -\frac{c_{k,\ell_i}^m (s_*) c_{k-2,\ell_i-2}^m (s_*)}{c_{k,\ell_i-2}^m (s_*) c_{k-2,\ell_i}^m (s_*)}. 
\end{equation}
Here, $c^m_{k,\ell_i}$ characterises the geometrical coupling between inertial and gravito-inertial oscillations along the convective-radiative interface:
\begin{equation}
    c^m_{k,\ell_i}(s) = \int_0^{\pi} \tilde{P}_{\ell_i}^m (\cos \theta) H_{k}^{m}(\theta,s) \sin \theta \; \mathrm{d}\theta,
\end{equation}
\noindent with $\tilde{P}_{\ell_i}^m (\cos \theta)$ and $H_{k}^{m}(\theta,s)$ being the normalised Legendre polynomial and Hough functions. In addition to $P_{\ell_i}^m$ and $H_k^m$, these mixed modes involve $P_{\ell_i-2}^m$ and $H_{k-2}^m$. In Appendix \ref{model}, we argue this is a good approximation for the first dip of the $(\ell=1,m=-1)$, $(\ell=2,m=-2)$, and $(\ell=1,m=0)$ series studied in this paper. This model cannot be generalised to all dips, as we expect some resonances to produce more complex mixed modes.\\
The dip properties predicted by this model have important differences with \citet{TT22}, and these differences appear to improve the model. First, the width $\sigma$ is multiplied by the factor $f_{k,\ell_i}^m/(1+f_{k,\ell_i}^m)$.  As $f_{k,\ell_i}^m$ is  proportional to the coupling coefficient $c_{k,\ell_i}(s_*)$ introduced in \citet{Ouazzani2020}, this new expression of $\sigma$ can account for the fact that significant dips only occur when the resonance involves modes with a non-negligible geometrical matching at the convective-radiative interface.
Second, the new form of the central spin parameter is now consistent with the numerical results as $s_c - s_* \propto \Omega$. 

When extrapolated to a realistic core with a non-uniform density, our model suggests that the first dip of the $(\ell=1,m=-1)$, $(\ell=2,m=-2)$, and $(\ell=1,m=0)$ series can be expressed as
\begin{align}
\sigma = & \beta \frac{\Omega}{N_0}, \\
s_c = & s_* + \alpha \frac{\Omega}{N_0},
\end{align}
where $\beta$, $s_*$, and $\alpha$ depend on both the gravito-inertial modes and on the resonant inertial mode. For a fixed $(\ell,m)$, these parameters only depend on the properties of the inertial mode, that is, on the radial gradient of the density profile in the convective core \citep[see Eq.~(1) in][]{Ouazzani2020}. In the absence of analytical expressions,
they can be derived by fitting the dips computed for different star models. In Sect.~\ref{Sec:Analysis}, using  three star models, it was found that $\beta$ remains roughly constant, while $\alpha$ is approximately proportional to $1/s_*$.   
An open question for future and more detailed empirical relations is whether $\beta$ and $\alpha$ can indeed be described by functions of $s_*$ only. In that is the case, investigating the link between the seismic observable $s_*$ and the star properties (age, mass, overshooting, etc.) would be much facilitated.
Computing $s_*$ for a given stellar model is indeed much easier than computing the frequency pattern and the dip parameters of a $(\ell, m)$ mode series.

Because of the uniform density assumption, a quantitative agreement between the theoretical and numerical values of $\sigma$ and $s_c$ is not expected. We nevertheless compared them by calculating the theoretical values of $\beta_{\rm th}(s_*)$,  $\alpha_{\rm th}(s_*)$, and $A_{\rm th}=\alpha_{\rm th}(s_*)s_*$.
Table~\ref{modelu} presents these parameters for the three dips we studied. It shows that despite the uniform density assumption, the theoretical values are in the range of those obtained by numerical calculations (see Tables~\ref{table:beta_sigma}, \ref{table:alpha_sc}, \ref{table:A_sc/s*}, and \ref{table:A_beta_two_models}). This is consistent with the fact that the spin parameter and the surface distribution of the inertial modes are not strongly modified by the core density stratification (see Appendix~\ref{appendix:s_iner}). We also note that as expected for $\alpha$, the deviations from the uniform density values increase for more evolved star models with higher density contrasts.

\begin{table}[!htp]  
    \caption{Parameters of the empirical relations for $\sigma$ and $s_c$ proposed in Sect.~\ref{Sec:Analysis} using a uniform density theoretical model.}
    \label{modelu}
\centering
\begin{tabular}{crccc}
\hline \hline
        $\ell$   &   $m$         &        $\alpha_{\rm th}$        &     $A_{\rm th}$   &        
        $\beta_{\rm th}$ \\\hline
1 & -1   &  $8.99$  &   $0.794$  &  $4.78$\\
1 & 0    &  $0.769$  &  $0.344$   &  $0.599$     \\  
2 & -2     &  $5.77$  &  $0.67$     & $2.74$  \\ 
\hline
\end{tabular}
\end{table}

\section{Discussion and conclusion}
\label{Sec:Discussion}
We have used the code TOP to study the resonances between inertial modes in the convective core and gravito-inertial modes in the radiative zone in $\gamma$ Dor stars. As in \cite{Ouazzani2020}, we obtained dips in the $\Delta P_\mathrm{co}-s$ relation around spin parameters of gravito-inertial modes approximately predicted by a simple analytical model. We computed those dips for different series of modes and different rotation rates and evolutionary stages across the main sequence. We quantified the dips using a Lorentzian function as suggested in \cite{TT22}, which allowed us to determine the evolution of their basic properties with the rotation and the evolutionary stage of the star. We then proposed an empirical 
model for the evolution of the dips.

We also compared our numerical results with the model of \cite{TT22}, showing that it correctly predicts that the dip width is proportional to the rotation rate, but it does not account for the variation of the dip central spin parameter with rotation. We have presented a new model that accounts for this variation and selects the significant dips by including the geometrical matching between the inertial mode and the gravito-inertial modes. 
The model is not fully predictive, but it is useful to support and discuss the proposed empirical relations.

Our empirical model gives access to the Brunt-Väisälä frequency jump at the core interface as well as the resonant inertial mode frequency for stars presenting a clear dip in their period spacing patterns. Such dips have been observed, as shown in \cite{Saio2021}, and further research could be led in order to find dips in a larger number of $\gamma$ Dor stars in the TESS and \textit{Kepler} data. This could be challenging, as modes in the dips may be harder to observe due to their higher inertia, as their energy is more concentrated in their core.
As our empirical model relies only on two different stellar models, it needs to be tested on more stellar models in order to estimate its accuracy. Such a study is not trivial, as computing and identifying series of gravito-inertial modes with 2D oscillation codes can be a tedious process.
The resonance phenomenon between the core and the envelope may not be exclusive to $\gamma$ Dor stars, and similar dip structures could arise from the $\Delta P$ - $P$ diagram of other types of stars. Slow pulsating B (SPB) stars are good candidates even though no dips have been observed yet \citep{Aerts2023}. A thorough investigation taking into account the expected spin parameters of the resonant inertial modes would be needed to establish whether resonances occur in the frequency ranges observed in SPB stars.

The main limitation of our present numerical study is related to the modelling of the rotation: We neglected the distortion induced by the centrifugal force and assumed a uniform rotation profile. For rotation rates observed in $\gamma$ Dor stars, we expect a weak impact of the centrifugal distortion on our results since it mainly affects the external layers of the stars. Indeed, the structure of convective cores and the surrounding regions where g modes have large amplitudes remains quasi-spherical \citep[see, e.g.,][]{Ballot2010,Ballot2012}. However, the influence of the differential rotation should be carefully investigated in a further study, especially when the core spins at a rotation that is different from the radiative zone. 
That kind of differential rotation has already been investigated and identified through the study of dips observed in 16 $\gamma$ Dor stars in \citet{Saio2021}. Although their differential rotation is shown to be small to non-significant in most cases, the core of one of the 16 stars was found to rotate 20\% faster than the surrounding radiative zone. Such differential rotation would cause the frequency of the resonant inertial mode to be misestimated, as its estimate relies on the use of the inner radiative zone rotation rate obtained through the TAR. Nevertheless, how such differential rotation would affect the width of the dips and the estimate of the Brunt-Väisälä jump at the core interface remains unclear. If $\beta$ remains constant, the width is only related to the radiative zone rotation rate rather than the core rotation rate. It implies that the estimate of the Brunt-Väisälä frequency jump through the dip width and the rotation rate of the inner radiative zone should remain correct despite the presence of such differential rotation.\\
We neglected non-adiabatic effects in our analysis. This approximation is supported by non-adiabatic calculations of high-order g modes, which have been performed to investigate the excitation mechanism of $\gamma$ Dor pulsations \citep{Guzik2000, Dupret2005, Bouabid2013}. These calculations showed that the adiabatic analysis provides precise enough oscillation frequencies \citep{Dupret2005}. The spatial distribution of the g modes is affected by non-adiabatic effects, but those are only significant in the upper layers of the star, where the thermal relaxation time is smaller or of the same order as the oscillation period \citep{Dupret2002,Dupret2005}. We thus expect that deep in the star, at the interface of the convective core and the radiative zone, the mode geometry is not affected by non-adiabatic effects. Thus, the adiabatic approximation we used to model the coupling between modes in the convective core and in the radiative envelope should be precise enough.\\
Our stellar structure models include turbulent diffusion to reproduce the different mixing processes that occur in stars. The value of the diffusion coefficient we used is the one proposed by \citet{Ouazzani2020}, which is based on a previous calibration to Genova models \citep{Miglio2008}. The mixing is thus very efficient and prevents the formation of sharp features in the Brunt-Väisälä frequency profile in the g-mode cavity. Such features are known to generate glitches, visible in $\Delta P - s$ diagrams as oscillations or periodic dips \citep[see for example][]{Miglio2008}. While absent in our computations, glitches may be present in real data and perturb the characterisation or even the detection of dips.
Conversely, dips caused by the resonance phenomenon can potentially be mistaken for a glitch signature and could lead to wrong estimations of stellar parameters \citep[see][]{Mombarg2021}.

The frequency and the width of the dips depends on the structure of the convective core through the spin parameter and the spatial distribution of the resonant core inertial modes. As shown by Eq.~(\ref{eq:Wu2005}), it is the density stratification of the convective core that controls the inertial mode characteristics. Thus, the evolutionary stage or the stellar mass affects the dips as far as they affect the convective core density stratification. It is worth noting that the core size itself has no direct impact on the inertial modes.
Further investigations are needed 
to understand the relation between the frequencies of the inertial modes in the core and the stellar properties. Such studies would thus allow one to use the said frequencies extracted by our empirical model to constrain stellar parameters. In particular, coupled with the jump in the Brunt-Väisälä frequency at the core interface we extracted from our model and the buoyancy radius deduced from the TAR, the inertial mode frequencies could be a powerful tool to determine stellar ages. Determining precise and accurate ages will be crucial for the PLATO mission \citep{Rauer2014_PLATO}, which will use $\gamma$~Dor stars as scientific calibrators.

\begin{acknowledgements}
We thank the anonymous referee for carefully reading our manuscript and providing helpful and constructive comments.
We thank R.M. Ouazzani, M. Takata and D.R. Reese for very useful discussions and for performing some tests and comparisons, which were crucial to validate our computations with TOP. We acknowledge support from the Centre National d’Etudes Spatiales (CNES). This work was supported by the "Programme National de Physique Stellaire" (PNPS) of CNRS/INSU co-funded by CEA and CNES.
\end{acknowledgements}

\bibliographystyle{aa} 
\bibliography{biblio} 

\begin{thebibliography}{50}
\expandafter\ifx\csname natexlab\endcsname\relax\def\natexlab#1{#1}\fi

\bibitem[{{Aerts}(2021)}]{Aerts2021}
{Aerts}, C. 2021, Reviews of Modern Physics, 93, 015001

\bibitem[{{Aerts} \& {Mathis}(2023)}]{Aerts2023}
{Aerts}, C. \& {Mathis}, S. 2023, \aap, 677, A68

\bibitem[{{Aerts} {et~al.}(2019){Aerts}, {Mathis}, \& {Rogers}}]{Aerts2019}
{Aerts}, C., {Mathis}, S., \& {Rogers}, T.~M. 2019, \araa, 57, 35

\bibitem[{{Angulo} {et~al.}(1999){Angulo}, {Arnould}, {Rayet}, {Descouvemont}, {Baye}, {Leclercq-Willain}, {Coc}, {Barhoumi}, {Aguer}, {Rolfs}, {Kunz}, {Hammer}, {Mayer}, {Paradellis}, {Kossionides}, {Chronidou}, {Spyrou}, {degl'Innocenti}, {Fiorentini}, {Ricci}, {Zavatarelli}, {Providencia}, {Wolters}, {Soares}, {Grama}, {Rahighi}, {Shotter}, \& {Lamehi Rachti}}]{Angulo1999_NACRE}
{Angulo}, C., {Arnould}, M., {Rayet}, M., {et~al.} 1999, \nphysa, 656, 3

\bibitem[{{Asplund} {et~al.}(2009){Asplund}, {Grevesse}, {Sauval}, \& {Scott}}]{Asplund2009}
{Asplund}, M., {Grevesse}, N., {Sauval}, A.~J., \& {Scott}, P. 2009, \araa, 47, 481

\bibitem[{{Ballot} {et~al.}(2012){Ballot}, {Ligni{\`e}res}, {Prat}, {Reese}, \& {Rieutord}}]{Ballot2012}
{Ballot}, J., {Ligni{\`e}res}, F., {Prat}, V., {Reese}, D.~R., \& {Rieutord}, M. 2012, in Astronomical Society of the Pacific Conference Series, Vol. 462, Progress in Solar/Stellar Physics with Helio- and Asteroseismology, ed. H.~{Shibahashi}, M.~{Takata}, \& A.~E. {Lynas-Gray}, 389

\bibitem[{{Ballot} {et~al.}(2010){Ballot}, {Ligni{\`e}res}, {Reese}, \& {Rieutord}}]{Ballot2010}
{Ballot}, J., {Ligni{\`e}res}, F., {Reese}, D.~R., \& {Rieutord}, M. 2010, \aap, 518, A30

\bibitem[{{B{\"o}hm-Vitense}(1958)}]{BV1958}
{B{\"o}hm-Vitense}, E. 1958, \zap, 46, 108

\bibitem[{{Borucki} {et~al.}(2010){Borucki}, {Koch}, {Basri}, {Batalha}, {Brown}, {Caldwell}, {Caldwell}, {Christensen-Dalsgaard}, {Cochran}, {DeVore}, {Dunham}, {Dupree}, {Gautier}, {Geary}, {Gilliland}, {Gould}, {Howell}, {Jenkins}, {Kondo}, {Latham}, {Marcy}, {Meibom}, {Kjeldsen}, {Lissauer}, {Monet}, {Morrison}, {Sasselov}, {Tarter}, {Boss}, {Brownlee}, {Owen}, {Buzasi}, {Charbonneau}, {Doyle}, {Fortney}, {Ford}, {Holman}, {Seager}, {Steffen}, {Welsh}, {Rowe}, {Anderson}, {Buchhave}, {Ciardi}, {Walkowicz}, {Sherry}, {Horch}, {Isaacson}, {Everett}, {Fischer}, {Torres}, {Johnson}, {Endl}, {MacQueen}, {Bryson}, {Dotson}, {Haas}, {Kolodziejczak}, {Van Cleve}, {Chandrasekaran}, {Twicken}, {Quintana}, {Clarke}, {Allen}, {Li}, {Wu}, {Tenenbaum}, {Verner}, {Bruhweiler}, {Barnes}, \& {Prsa}}]{Borucki2010}
{Borucki}, W.~J., {Koch}, D., {Basri}, G., {et~al.} 2010, Science, 327, 977

\bibitem[{{Bouabid} {et~al.}(2013){Bouabid}, {Dupret}, Salmon, Montalbán, Miglio, \& Noels}]{Bouabid2013}
{Bouabid}, M.-P., {Dupret}, M.-A., Salmon, S., {et~al.} 2013, Monthly Notices of the Royal Astronomical Society, 429, 2500

\bibitem[{{Chatelin}(1988)}]{chatelin1988valeurs}
{Chatelin}, F. 1988, Valeurs propres de matrices, Collection Math{\'e}matiques appliqu{\'e}es pour la ma{\^\i}trise (Masson)

\bibitem[{{Christophe} {et~al.}(2018){Christophe}, {Ballot}, {Ouazzani}, {Antoci}, \& {Salmon}}]{Christophe2018}
{Christophe}, S., {Ballot}, J., {Ouazzani}, R.~M., {Antoci}, V., \& {Salmon}, S.~J.~A.~J. 2018, \aap, 618, A47

\bibitem[{{Dupret} {et~al.}(2002){Dupret}, {De Ridder}, {Neuforge}, {Aerts}, \& {Scuflaire}}]{Dupret2002}
{Dupret}, M.~A., {De Ridder}, J., {Neuforge}, C., {Aerts}, C., \& {Scuflaire}, R. 2002, \aap, 385, 563

\bibitem[{{Dupret} {et~al.}(2005){Dupret}, {Grigahc{\`e}ne}, {Garrido}, {Gabriel}, \& {Scuflaire}}]{Dupret2005}
{Dupret}, M.~A., {Grigahc{\`e}ne}, A., {Garrido}, R., {Gabriel}, M., \& {Scuflaire}, R. 2005, \aap, 435, 927

\bibitem[{{Ferguson} {et~al.}(2005){Ferguson}, {Alexander}, {Allard}, {Barman}, {Bodnarik}, {Hauschildt}, {Heffner-Wong}, \& {Tamanai}}]{Ferguson2005}
{Ferguson}, J.~W., {Alexander}, D.~R., {Allard}, F., {et~al.} 2005, \apj, 623, 585

\bibitem[{{Garcia} {et~al.}(2022){Garcia}, {Van Reeth}, {De Ridder}, \& {Aerts}}]{Garcia2022}
{Garcia}, S., {Van Reeth}, T., {De Ridder}, J., \& {Aerts}, C. 2022, \aap, 668, A137

\bibitem[{{Guzik} {et~al.}(2000){Guzik}, {Kaye}, {Bradley}, {Cox}, \& {Neuforge}}]{Guzik2000}
{Guzik}, J.~A., {Kaye}, A.~B., {Bradley}, P.~A., {Cox}, A.~N., \& {Neuforge}, C. 2000, \apjl, 542, L57

\bibitem[{{Iglesias} \& {Rogers}(1996)}]{OPAL_ref96}
{Iglesias}, C.~A. \& {Rogers}, F.~J. 1996, \apj, 464, 943

\bibitem[{{Imbriani} {et~al.}(2004){Imbriani}, {Costantini}, {Formicola}, {Bemmerer}, {Bonetti}, {Broggini}, {Corvisiero}, {Cruz}, {F{\"u}l{\"o}p}, {Gervino}, {Guglielmetti}, {Gustavino}, {Gy{\"u}rky}, {Jesus}, {Junker}, {Lemut}, {Menegazzo}, {Prati}, {Roca}, {Rolfs}, {Romano}, {Rossi Alvarez}, {Sch{\"u}mann}, {Somorjai}, {Straniero}, {Strieder}, {Terrasi}, {Trautvetter}, {Vomiero}, \& {Zavatarelli}}]{Imbriani2004}
{Imbriani}, G., {Costantini}, H., {Formicola}, A., {et~al.} 2004, \aap, 420, 625

\bibitem[{{Kaye} {et~al.}(1999){Kaye}, {Handler}, {Krisciunas}, {Poretti}, \& {Zerbi}}]{1999Kaye}
{Kaye}, A.~B., {Handler}, G., {Krisciunas}, K., {Poretti}, E., \& {Zerbi}, F.~M. 1999, \pasp, 111, 840

\bibitem[{{Lebreton} {et~al.}(2014){Lebreton}, {Goupil}, \& {Montalb{\'a}n}}]{Lebreton2014b}
{Lebreton}, Y., {Goupil}, M.~J., \& {Montalb{\'a}n}, J. 2014, in EAS Publications Series, Vol.~65, EAS Publications Series, ed. Y.~{Lebreton}, D.~{Valls-Gabaud}, \& C.~{Charbonnel}, 177--223

\bibitem[{{Lee} \& {Saio}(1989)}]{Lee1989}
{Lee}, U. \& {Saio}, H. 1989, \mnras, 237, 875

\bibitem[{{Lee} \& {Saio}(1997)}]{Lee1997}
{Lee}, U. \& {Saio}, H. 1997, \apj, 491, 839

\bibitem[{{Li} {et~al.}(2019){Li}, {Van Reeth}, {Bedding}, {Murphy}, \& {Antoci}}]{Li2019b}
{Li}, G., {Van Reeth}, T., {Bedding}, T.~R., {Murphy}, S.~J., \& {Antoci}, V. 2019, \mnras, 487, 782

\bibitem[{{Li} {et~al.}(2020){Li}, {Van Reeth}, {Bedding}, {Murphy}, {Antoci}, {Ouazzani}, \& {Barbara}}]{GangLiCatalogue}
{Li}, G., {Van Reeth}, T., {Bedding}, T.~R., {et~al.} 2020, Monthly Notices of the Royal Astronomical Society, 491, 3586

\bibitem[{{Miglio} {et~al.}(2008){Miglio}, {Montalb{\'a}n}, {Noels}, \& {Eggenberger}}]{Miglio2008}
{Miglio}, A., {Montalb{\'a}n}, J., {Noels}, A., \& {Eggenberger}, P. 2008, \mnras, 386, 1487

\bibitem[{{Mombarg} {et~al.}(2021){Mombarg}, {Van Reeth}, \& {Aerts}}]{Mombarg2021}
{Mombarg}, J.~S.~G., {Van Reeth}, T., \& {Aerts}, C. 2021, \aap, 650, A58

\bibitem[{{Mombarg} {et~al.}(2019){Mombarg}, {Van Reeth}, {Pedersen}, {Molenberghs}, {Bowman}, {Johnston}, {Tkachenko}, \& {Aerts}}]{Mombarg2019}
{Mombarg}, J.~S.~G., {Van Reeth}, T., {Pedersen}, M.~G., {et~al.} 2019, \mnras, 485, 3248

\bibitem[{{Morel}(1997)}]{Morel1997}
{Morel}, P. 1997, \aaps, 124, 597

\bibitem[{{Morel} \& {Lebreton}(2008)}]{Morel2008}
{Morel}, P. \& {Lebreton}, Y. 2008, \apss, 316, 61

\bibitem[{{Ouazzani} {et~al.}(2020){Ouazzani}, {Ligni\`eres}, {Dupret}, {Salmon}, {Ballot}, {Christophe}, \& {Takata}}]{Ouazzani2020}
{Ouazzani}, R.-M., {Ligni\`eres}, F., {Dupret}, M.-A., {et~al.} 2020, A\&A, 640, A49

\bibitem[{{Ouazzani} {et~al.}(2019){Ouazzani}, {Marques}, {Goupil}, {Christophe}, {Antoci}, {Salmon}, \& {Ballot}}]{Ouazzani2019}
{Ouazzani}, R.~M., {Marques}, J.~P., {Goupil}, M.~J., {et~al.} 2019, \aap, 626, A121

\bibitem[{{Ouazzani} {et~al.}(2017){Ouazzani}, {Salmon}, {Antoci}, {Bedding}, {Murphy}, \& {Roxburgh}}]{Ouazzani2017}
{Ouazzani}, R.-M., {Salmon}, S.~J.~A.~J., {Antoci}, V., {et~al.} 2017, \mnras, 465, 2294

\bibitem[{{Rauer} {et~al.}(2014){Rauer}, {Catala}, {Aerts}, {Appourchaux}, {Benz}, {Brandeker}, {Christensen-Dalsgaard}, {Deleuil}, {Gizon}, {Goupil}, {G{\"u}del}, {Janot-Pacheco}, {Mas-Hesse}, {Pagano}, {Piotto}, {Pollacco}, {Santos}, {Smith}, {Su{\'a}rez}, {Szab{\'o}}, {Udry}, {Adibekyan}, {Alibert}, {Almenara}, {Amaro-Seoane}, {Eiff}, {Asplund}, {Antonello}, {Barnes}, {Baudin}, {Belkacem}, {Bergemann}, {Bihain}, {Birch}, {Bonfils}, {Boisse}, {Bonomo}, {Borsa}, {Brand{\~a}o}, {Brocato}, {Brun}, {Burleigh}, {Burston}, {Cabrera}, {Cassisi}, {Chaplin}, {Charpinet}, {Chiappini}, {Church}, {Csizmadia}, {Cunha}, {Damasso}, {Davies}, {Deeg}, {D{\'\i}az}, {Dreizler}, {Dreyer}, {Eggenberger}, {Ehrenreich}, {Eigm{\"u}ller}, {Erikson}, {Farmer}, {Feltzing}, {de Oliveira Fialho}, {Figueira}, {Forveille}, {Fridlund}, {Garc{\'\i}a}, {Giommi}, {Giuffrida}, {Godolt}, {Gomes da Silva}, {Granzer}, {Grenfell}, {Grotsch-Noels}, {G{\"u}nther}, {Haswell}, {Hatzes}, {H{\'e}brard}, {Hekker}, {Helled}, {Heng}, {Jenkins},
  {Johansen}, {Khodachenko}, {Kislyakova}, {Kley}, {Kolb}, {Krivova}, {Kupka}, {Lammer}, {Lanza}, {Lebreton}, {Magrin}, {Marcos-Arenal}, {Marrese}, {Marques}, {Martins}, {Mathis}, {Mathur}, {Messina}, {Miglio}, {Montalban}, {Montalto}, {Monteiro}, {Moradi}, {Moravveji}, {Mordasini}, {Morel}, {Mortier}, {Nascimbeni}, {Nelson}, {Nielsen}, {Noack}, {Norton}, {Ofir}, {Oshagh}, {Ouazzani}, {P{\'a}pics}, {Parro}, {Petit}, {Plez}, {Poretti}, {Quirrenbach}, {Ragazzoni}, {Raimondo}, {Rainer}, {Reese}, {Redmer}, {Reffert}, {Rojas-Ayala}, {Roxburgh}, {Salmon}, {Santerne}, {Schneider}, {Schou}, {Schuh}, {Schunker}, {Silva-Valio}, {Silvotti}, {Skillen}, {Snellen}, {Sohl}, {Sousa}, {Sozzetti}, {Stello}, {Strassmeier}, {{\v{S}}vanda}, {Szab{\'o}}, {Tkachenko}, {Valencia}, {Van Grootel}, {Vauclair}, {Ventura}, {Wagner}, {Walton}, {Weingrill}, {Werner}, {Wheatley}, \& {Zwintz}}]{Rauer2014_PLATO}
{Rauer}, H., {Catala}, C., {Aerts}, C., {et~al.} 2014, Experimental Astronomy, 38, 249

\bibitem[{{Reese}(2006)}]{2006PhDReese}
{Reese}, D. 2006, PhD thesis, Universite de Toulouse Paul Sabatier, France

\bibitem[{{Reese}(2013)}]{Reese2013}
{Reese}, D.~R. 2013, \aap, 555, A148

\bibitem[{{Reese} {et~al.}(2009){Reese}, {MacGregor}, {Jackson}, {Skumanich}, \& {Metcalfe}}]{Reese2009}
{Reese}, D.~R., {MacGregor}, K.~B., {Jackson}, S., {Skumanich}, A., \& {Metcalfe}, T.~S. 2009, \aap, 506, 189

\bibitem[{{Ricker} {et~al.}(2014){Ricker}, {Winn}, {Vanderspek}, {Latham}, {Bakos}, {Bean}, {Berta-Thompson}, {Brown}, {Buchhave}, {Butler}, {Butler}, {Chaplin}, {Charbonneau}, {Christensen-Dalsgaard}, {Clampin}, {Deming}, {Doty}, {De Lee}, {Dressing}, {Dunham}, {Endl}, {Fressin}, {Ge}, {Henning}, {Holman}, {Howard}, {Ida}, {Jenkins}, {Jernigan}, {Johnson}, {Kaltenegger}, {Kawai}, {Kjeldsen}, {Laughlin}, {Levine}, {Lin}, {Lissauer}, {MacQueen}, {Marcy}, {McCullough}, {Morton}, {Narita}, {Paegert}, {Palle}, {Pepe}, {Pepper}, {Quirrenbach}, {Rinehart}, {Sasselov}, {Sato}, {Seager}, {Sozzetti}, {Stassun}, {Sullivan}, {Szentgyorgyi}, {Torres}, {Udry}, \& {Villasenor}}]{Ricker2014}
{Ricker}, G.~R., {Winn}, J.~N., {Vanderspek}, R., {et~al.} 2014, in Society of Photo-Optical Instrumentation Engineers (SPIE) Conference Series, Vol. 9143, Space Telescopes and Instrumentation 2014: Optical, Infrared, and Millimeter Wave, ed. J.~{Oschmann}, Jacobus~M., M.~{Clampin}, G.~G. {Fazio}, \& H.~A. {MacEwen}, 914320

\bibitem[{{Rogers} \& {Nayfonov}(2002)}]{Rogers2002}
{Rogers}, F.~J. \& {Nayfonov}, A. 2002, \apj, 576, 1064

\bibitem[{{Saio} {et~al.}(2018){Saio}, {Kurtz}, {Murphy}, {Antoci}, \& {Lee}}]{Saio2018}
{Saio}, H., {Kurtz}, D.~W., {Murphy}, S.~J., {Antoci}, V.~L., \& {Lee}, U. 2018, \mnras, 474, 2774

\bibitem[{{Saio} {et~al.}(2021){Saio}, {Takata}, {Lee}, Li, \& Van Reeth}]{Saio2021}
{Saio}, H., {Takata}, M., {Lee}, U., Li, G., \& Van Reeth, T. 2021, Monthly Notices of the Royal Astronomical Society, 502, 5856

\bibitem[{{Takata} {et~al.}(2020){Takata}, {Ouazzani}, {Saio}, {Christophe}, {Ballot}, {Antoci}, {Salmon}, \& {Hijikawa}}]{Takata2020vrai}
{Takata}, M., {Ouazzani}, R.~M., {Saio}, H., {et~al.} 2020, \aap, 635, A106

\bibitem[{{Tokuno} \& {Takata}(2022)}]{TT22}
{Tokuno}, T. \& {Takata}, M. 2022, \mnras, 514, 4140

\bibitem[{{Townsend}(2003)}]{Townsend2003}
{Townsend}, R. 2003, Monthly Notices of the Royal Astronomical Society, 340, 1020

\bibitem[{{Unno} {et~al.}(1989){Unno}, {Osaki}, {Ando}, {Saio}, \& {Shibahashi}}]{Unno1989}
{Unno}, W., {Osaki}, Y., {Ando}, H., {Saio}, H., \& {Shibahashi}, H. 1989, {Nonradial oscillations of stars}

\bibitem[{{Uytterhoeven} {et~al.}(2011){Uytterhoeven}, {Moya}, {Grigahc{\`e}ne}, {Guzik}, {Guti{\'e}rrez-Soto}, {Smalley}, {Handler}, {Balona}, {Niemczura}, {Fox Machado}, {Benatti}, {Chapellier}, {Tkachenko}, {Szab{\'o}}, {Su{\'a}rez}, {Ripepi}, {Pascual}, {Mathias}, {Mart{\'\i}n-Ru{\'\i}z}, {Lehmann}, {Jackiewicz}, {Hekker}, {Gruberbauer}, {Garc{\'\i}a}, {Dumusque}, {D{\'\i}az-Fraile}, {Bradley}, {Antoci}, {Roth}, {Leroy}, {Murphy}, {De Cat}, {Cuypers}, {Kjeldsen}, {Christensen-Dalsgaard}, {Breger}, {Pigulski}, {Kiss}, {Still}, {Thompson}, \& {van Cleve}}]{Uytterhoeven2011}
{Uytterhoeven}, K., {Moya}, A., {Grigahc{\`e}ne}, A., {et~al.} 2011, \aap, 534, A125

\bibitem[{{Van Reeth} {et~al.}(2018){Van Reeth}, {Mombarg}, {Mathis}, {Tkachenko}, {Fuller}, {Bowman}, {Buysschaert}, {Johnston}, {Garc{\'\i}a Hern{\'a}ndez}, {Goldstein}, {Townsend}, \& {Aerts}}]{VanReeth2018}
{Van Reeth}, T., {Mombarg}, J.~S.~G., {Mathis}, S., {et~al.} 2018, \aap, 618, A24

\bibitem[{{Van Reeth} {et~al.}(2016){Van Reeth}, {Tkachenko}, \& {Aerts}}]{VanReeth2016}
{Van Reeth}, T., {Tkachenko}, A., \& {Aerts}, C. 2016, \aap, 593, A120

\bibitem[{{Van Reeth} {et~al.}(2015){Van Reeth}, {Tkachenko}, {Aerts}, {P{\'a}pics}, {Triana}, {Zwintz}, {Degroote}, {Debosscher}, {Bloemen}, {Schmid}, {De Smedt}, {Fremat}, {Fuentes}, {Homan}, {Hrudkova}, {Karjalainen}, {Lombaert}, {Nemeth}, {{\O}stensen}, {Van De Steene}, {Vos}, {Raskin}, \& {Van Winckel}}]{VanReeth2015b}
{Van Reeth}, T., {Tkachenko}, A., {Aerts}, C., {et~al.} 2015, \apjs, 218, 27

\bibitem[{{Wu}(2005)}]{W05}
{Wu}, Y. 2005, \apj, 635, 674

\end{thebibliography}

%
%

\begin{appendix}

\section{Uniform density model}
\label{appendix:homogeneous_model}

Here we briefly present the uniform-density resonance model proposed in \citet{Ouazzani2020} and apply it to the gravito-inertial mode series studied in this paper.
When there is no density stratification, the wave equation reduces $\nabla ^2 \Psi - s^2 \frac{\partial^2 \Psi}{\partial s^2}$=0, the Poincaré equation. It is separable using the ellipsoidal coordinates $(x_1, x_2, \phi)$ (see Eqs.~\ref{D1}, \ref{D2}, \ref{D3}). 

Using regularity conditions and the boundary condition $\xi_r=0$ applied at the sphere surface, it becomes possible to calculate the eigenfrequencies of the inertial modes as they are the non-trivial positive roots of
\begin{equation}
\label{eigenspin}
    \frac{dP_{\ell_i}^{m}(\mu)}{d\mu} = \frac{m}{1-\mu^2}P_{\ell_i}^{m}(\mu),
\end{equation}
where $\mu = 1/s=\omega_{\mathrm{co}}/2\Omega$ and $\ell_i$ is the degree of the inertial mode.
The spin parameters of inertial modes that could couple with gravito-inertial series studied here are presented in Table~\ref{table:s_homogene_correl}. For an inertial mode in the core to be significantly coupled with a gravito-inertial mode in the radiative zone, there needs to be spatial and temporal correspondences. The two coupling modes must share a similar latitudinal profile at the interface between the core and the radiative zone and must oscillate with a similar frequency. The condition on the frequency is easily fulfilled as the frequency spectrum of gravito-inertial modes is dense. To quantify the spatial correspondence between an inertial and a gravito-inertial mode, \citet{Ouazzani2020} uses a correlation coefficient between the associated Legendre polynomial (which describes the inertial mode in the core in the analytical model) and the Hough function (which describes the gravito-inertial mode in the envelope in the TAR approximation). Those coefficient are presented in parentheses next to the spin parameter of the modes in Table~\ref{table:s_homogene_correl}. The modes $(\ell_i=3$ $m=-1$), $(\ell_i=3$ $m=0$), and $(\ell_i=4$ $m=-2$) with the highest correlation coefficients are shown in Table~\ref{table:comparaison_inertiel}.

\begin{table}[!htp]
\caption{Spin parameters of inertial modes obtained using an analytical uniform density core model. The correlation coefficient between the inertial mode and the Hough function of the same symmetry class and spin parameter is shown in parentheses.} 
\centering 
\begin{tabular}{ccccc}
\hline \hline

$\ell_i$& $m = 0$            &                          &                            & \\ \hline
    1   & $\emptyset$ &                          &                            &  \\
    3   & $2.236$ $(0.62)$    &                        &                            &  \\ 
    5   & $3.506$ $(0.51)$    & $1.307$ $(0.014)$  &            &  \\ 
    7   & $4.778$ $(0.44)$    & $1.690$ $(3.3 \hspace{3pt} 10^{-3})$  & $1.147$ $(1.6 \hspace{3pt} 10^{-4})$   &  \\[0.5ex]

\hline \hline


$\ell_i$& $m = -1$           &                          &                            &  \\ \hline
    1   & $\emptyset$   &                          &                            &  \\
    3   & $11.325$ $(0.50)$    &                 &                            &  \\ 
    5   & $29.330$ $(0.39)$    & $1.690$ $(4.3 \hspace{3pt} 10^{-3})$        &                            &  \\
    7   & $55.332$ $(0.34)$    & $2.318$ $(3.8 \hspace{3pt} 10^{-5})$        & $1.293$ $(1.89 \hspace{3pt} 10^{-5})$ &  \\[0.5ex]
    \hline \hline


$\ell_i$& $m = -2$           &                          &          \\ \hline
    2   & $\emptyset$        &                          &            \\
    4   & $8.624$ $(0.48)$    &                          &            \\ 
    6   & $19.648$ $(0.38)$   & $1.831$ $(1.6 \hspace{3pt} 10^{-2})$ &            \\ 
    8   & $34.662$ $(0.32)$   & $2.482$ $(3.3 \hspace{3pt} 10^{-3})$ & $1.374$ $(2.4 \hspace{3pt} 10^{-4})$   \\ [0.5ex]



\hline

\end{tabular}
\label{table:s_homogene_correl}
\end{table}

\section{Core-only model}
\label{appendix:s_iner}
\begin{figure}[htp!]
    \centering
    \includegraphics[width=\linewidth]{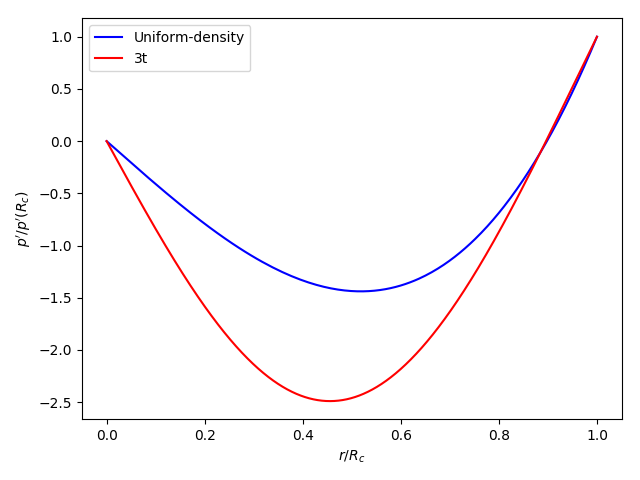}
    \caption{Eulerian pressure perturbations normalised by their value at the core radius $R_\mathrm{c}$ of the ($\ell_i=3$, $m=-1$) mode, as a function of the radius at the colatitude $\theta=\pi/2$ (equator), for the uniform density model (blue) and for the 3t model (red).}
    \label{fig:uniform_vs_3t_equator}
\end{figure}

We run TOP with a truncated version of our models to obtain the resonant frequencies.
Table \ref{table:s_iner} shows the spin parameters of the inertial modes coupling with the series of gravito-inertial modes  ($\ell=1$ $m=-1$), ($\ell=1$ $m=0$), and ($\ell=2$ $m=-2$) for the 3 models. They are labelled as ($\ell_i=3$ $m=-1$), ($\ell_i=3$ $m=0$), and ($\ell_i=4$ $m=-2$) respectively as they remain similar to the modes calculated with the uniform density model presented in Appendix \ref{appendix:homogeneous_model} (see Table \ref{table:comparaison_inertiel}). Despite their apparent resemblance, it has to be noted that their spin parameters vary significantly with the density stratification as the relative differences between the 2m and 3t models’ spin parameters and the uniform-density model exceed  20\%. To stress their differences, we show the mode profiles along the equator for both the uniform-density and 3t models 
in Fig. \ref{fig:uniform_vs_3t_equator}.

The rotation rate has little to no impact on the spin parameter and spatial distribution of inertial modes, as expected from the equation governing adiabatic waves in an isentropic convective core under the Cowling approximation \citep{W05}:
\begin{equation}
    \nabla^2 \Psi -s^2 \frac{\partial^2 \Psi}{\partial z^2} = - \frac{1}{\rho_0}\frac{\mathrm{d}\rho_0}{\mathrm{d}r} \left ( \frac{\partial\Psi}{\partial r} - s^2 \cos{\theta} \frac{\partial\Psi}{\partial z} + \frac{ms}{r} \Psi \right ) - (1-s^2)\frac{\omega_\mathrm{co}^2}{c_0^2}\Psi
    \label{eq:Wu2005}
\end{equation}
where $\Psi=p'/(\rho_0\omega_\mathrm{co}^2)$ and $z=r\cos{\theta}$ is the coordinate alongside the rotation axis. Indeed, the acoustic term, the last on the right-hand side, which could introduce an explicit dependence on the rotation rate, is negligible with respect to the first term on the left-hand side in the convective core of $\gamma$ Dor stars.

\begin{table}[!htp]  
\caption{Spin parameters $s_*$ of the resonant inertial modes coupling with the series of modes  ($\ell=1$ $m=-1$), ($\ell=1$ $m=0$), and ($\ell=2$ $m=-2$), for the models 1z, 2m, and 3t using a core-only model at $\Omega = 0.1 \Omega_K$ and $\Omega = 0.5 \Omega_K$. \label{table:s_iner}}
\centering
\begin{tabular}{lrccc}
\hline \hline

 $\ell_i$          &     $m$        &     1z    &    2m    &  3t \\[0.5ex]\hline
3 & -1       &  10.289 - 10.290 &  8.966 - 8.964 &  8.712 - 8.660 \\ 
3 & 0        &  2.207 - 2.207  &  2.163 - 2.163 &  2.153 - 2.152 \\ 
4 & -2       &  7.920 - 7.921  &  6.980 - 6.978 &  6.805 - 6.755 \\ 
[0.5ex]\hline 
\end{tabular}
\end{table}

\begin{table}[!htp] 
\caption{Inertial modes in the core coupling with the ($\ell=1$ $m=-1$), ($\ell=1$ $m=0$), and ($\ell=2$ $m=-2$) series of gravito-inertial modes in the envelope for the uniform density and 3t models.}
\centering
\begin{tabular}{c r p{2.5cm} p{2.5cm}} 
\hline \hline
$\ell_i$   & $m$              & Uniform density &  \multicolumn{1}{c}{3t} \\
\hline
3 & -1      &  \includegraphics[width=2.3cm]{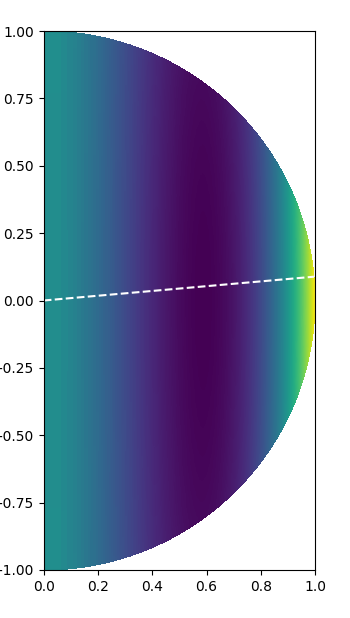} & \includegraphics[width=2.3cm]{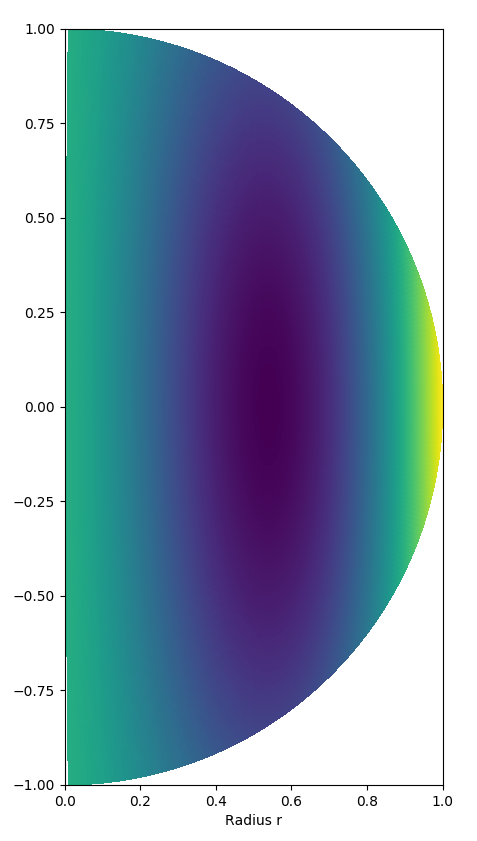}  \\ 
3 & 0        & \includegraphics[width=2.3cm]{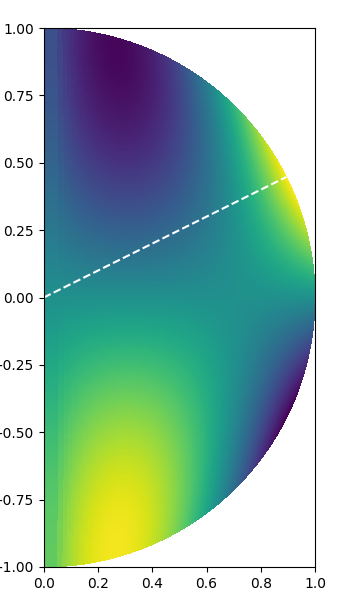} &   \includegraphics[width=2.3cm]{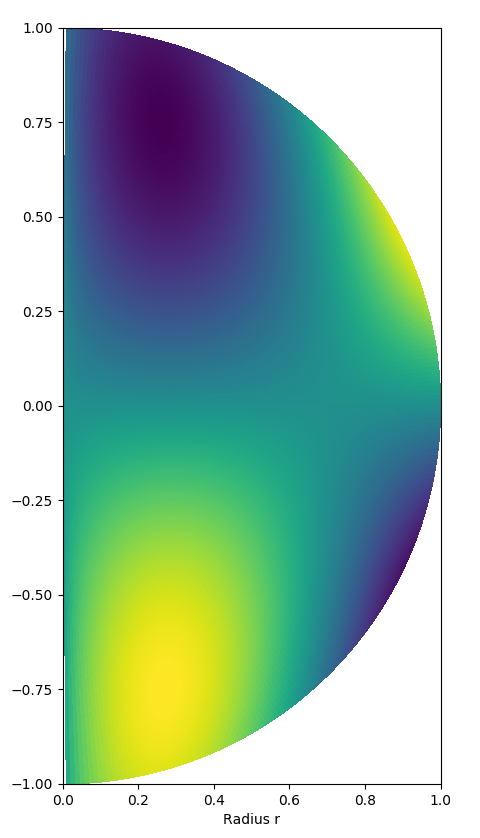}      \\ 
4 & -2       &  \includegraphics[width=2.3cm]{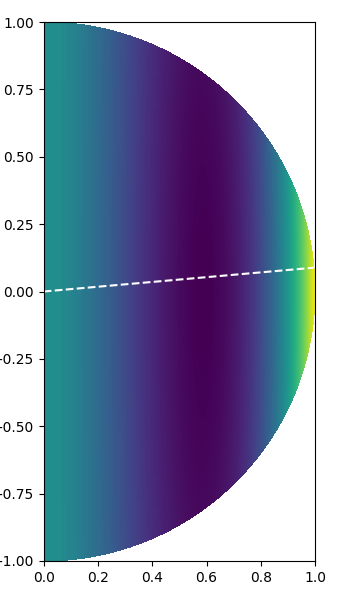}&  \includegraphics[width=2.39cm]{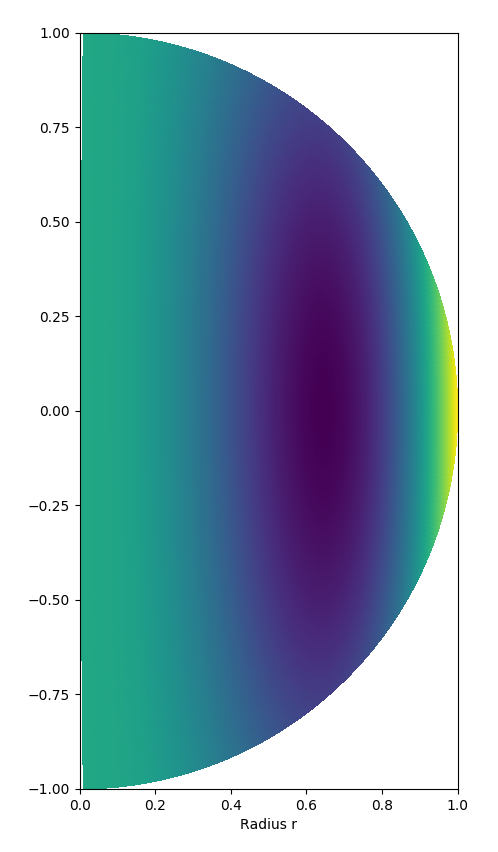} \\ 
[0.5ex]\hline 
\end{tabular}
\tablefoot{The white line shows the critical latitude $\theta_c = \arccos(1/s)$ in the case of the uniform density model. $\Omega = 0.3 \Omega_K$.}

\label{table:comparaison_inertiel}
\end{table}

\section{Results and analysis for the ($\ell=1$, $m=0$) and ($\ell=2$, $m=-2$) series}
\label{appendix:more_results}

In this appendix, we present our results for ($\ell=1$, $m=0$) and ($\ell=2$, $m=-2$) series.
We computed and analysed these series of modes in the same way as the ($\ell=1$, $m=-1$) series, which is extensively described in the main text.

\subsection{($\ell=1$, $m=0$) series}

\begin{figure}[!htp]
    \centering
    \includegraphics[width=\linewidth]{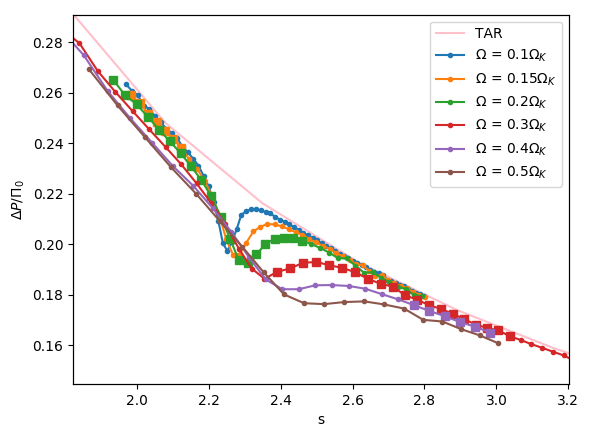}
    \caption{Period spacing in the co-rotating frame as a function of the spin parameter for the modes of the series ($\ell=1$, $m=0$) at different rotation rates using the 1z model. The period spacing is normalised by the buoyancy radius $\Pi_0$. The pink line shows the traditional approximation of rotation. Modes that are observable are marked with a square-shaped dot \citep{GangLiCatalogue}.}
    \label{fig:synthese_1z_l1_m0}
\end{figure}

\begin{figure}[!htp]
    \centering
    \includegraphics[width=\linewidth]{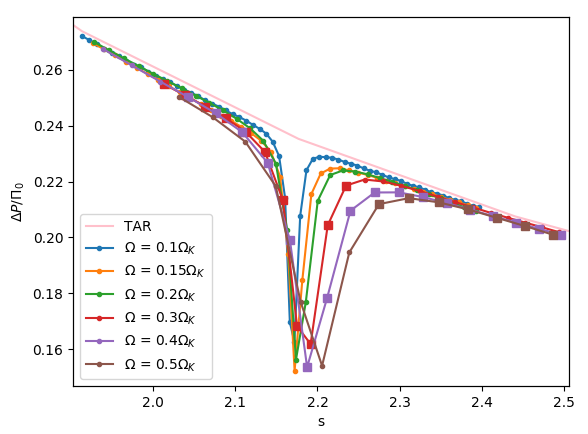}
    \caption{Same as Fig. \ref{fig:synthese_1z_l1_m0} for the model 2m.}
        \label{fig:synthese_2m_l1_m0}

\end{figure}

\begin{figure}[!htp]
    \centering
    \includegraphics[width=\linewidth]{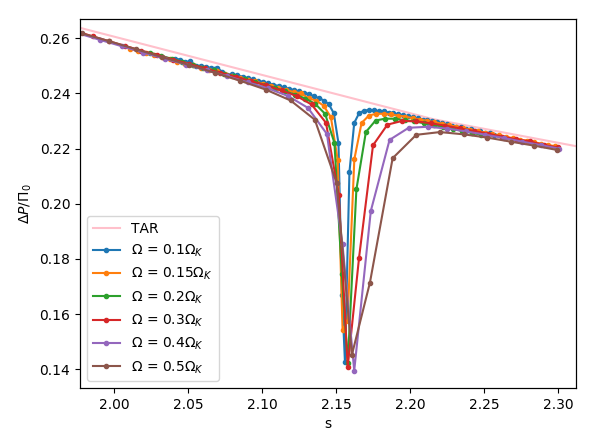}
    \caption{Same as Fig. \ref{fig:synthese_1z_l1_m0} for the model 3t.}
    \label{fig:synthese_3t_l1_m0}
\end{figure}

The dips in the series of modes ($\ell=1$, $m=0$) computed for the models 1z, 2m, and 3t at different rotations are presented in Figs.~\ref{fig:synthese_1z_l1_m0}, \ref{fig:synthese_2m_l1_m0}, and \ref{fig:synthese_3t_l1_m0}.

\begin{figure}[!htp]
    \centering
    \includegraphics[width=\linewidth]{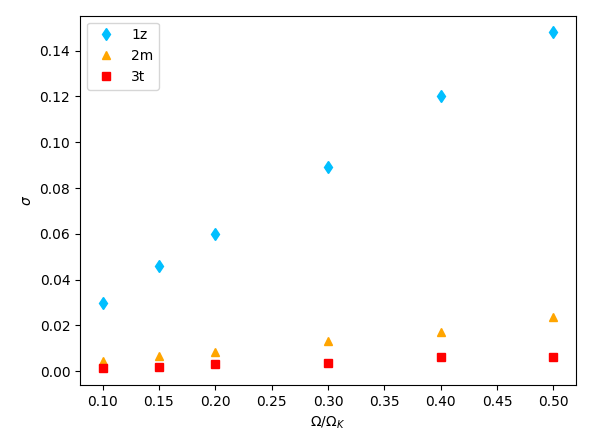}
    \caption{Width of the dips $\sigma$ as a function of the rotation rate for the series ($\ell=1$, $m=0$), for the three stellar models described in Table \ref{table:model}.}
    \label{fig:evo_sigma_rota_l1_m0}
\end{figure}

\begin{figure}[!htp]
    \centering
    \includegraphics[width=1.05\linewidth]{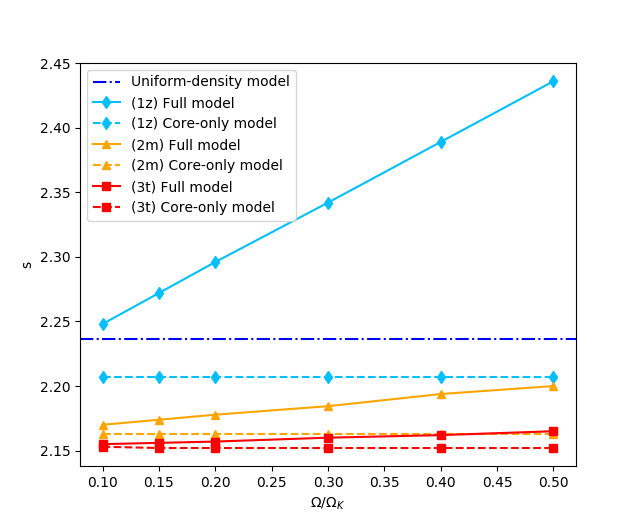}
    \caption{Evolution of the centre of the dip $s_\mathrm{c}$ obtained using a full model (full line) and evolution of the spin parameter of the inertial mode obtained using a core-only model (dotted line). The point-and-dot blue line shows the spin parameter of the inertial mode in the case of a uniform-density 
    core (analytical model). Light blue curves refer to the model 1z, orange curves 
    to the model 2m, and red curves 
    to the model 3t.}
    \label{fig:comparaison_s_l1_m0}
 \end{figure}

The evolution of the width $\sigma$ and location $s_c$ of dips as a function of the rotation are shown in Figs.~\ref{fig:evo_sigma_rota_l1_m0} and \ref{fig:comparaison_s_l1_m0}. In the latter, we also show the evolution of the spin parameter $s_*$ of the resonant inertial mode of core-only models and the spin parameter of the same mode in a uniform density sphere.

\begin{figure}[!htp]
    \centering
    \includegraphics[width=1.05\linewidth]{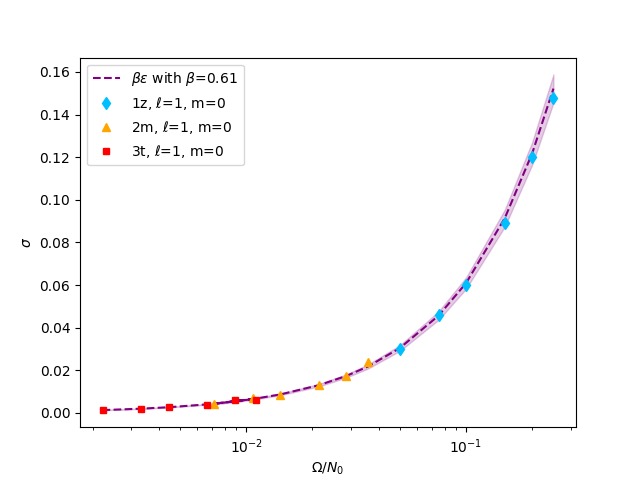}
    \caption{Width $\sigma$ as a function of $\epsilon$ for the series ($\ell=1$ $m=0$) for 3 different evolutionary stages (1z, 2m, 3t). The purple dashed line shows the fit presented in Table \ref{table:beta_sigma} for the three models together and the purple area shows the related error. The x-axis is logarithmic for visualisation purposes.}
    \label{fig:evo_sigma_log_l1_m0}. 
\end{figure}

\begin{figure}[!htp]
    \centering
    \includegraphics[width=1.05\linewidth]{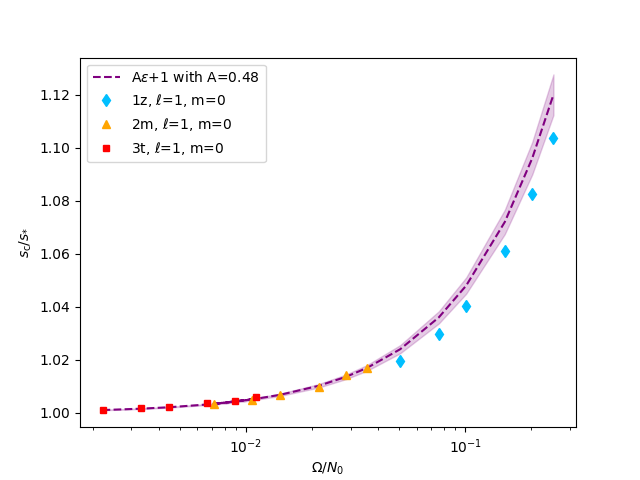}
    \caption{Ratio $s_c/s_*$ as a function of $\Omega/N_0$ ($\epsilon$) for the series ($\ell=1$ $m=0$), for the 1z, 2m, and 3t models. The purple dashed line shows the fit presented in Table \ref{table:A_sc/s*} for the three models together and the purple area shows the related error. The x-axis is logarithmic for visualisation purposes.}
    \label{fig:evo_sc_s_iner_log_l1_m0}
\end{figure}

We plot $\sigma$ and $s_c/s_*$ as a function of $\epsilon=\Omega/N_0$ for the three models in
Figs.~\ref{fig:evo_sigma_log_l1_m0} and \ref{fig:evo_sc_s_iner_log_l1_m0}, along with linear fits of the form $\sigma=\beta\epsilon$ and $s_c/s_* = A \epsilon +1$. 
We can see that $\sigma$ is a lot smaller for the ($\ell=1$, $m=0$) series than it is for the ($\ell=1$, $m=-1$), which is consistent with our analytical model (see the $\beta_\mathrm{th}$ column of Table \ref{modelu}). 
The ratio $s_c/s_*$ is also smaller, which is in agreement with our analytical model ($A_\mathrm{th}$ column of Table \ref{modelu}).

\begin{figure}[!htp]
    \centering
    \includegraphics[width=\linewidth]{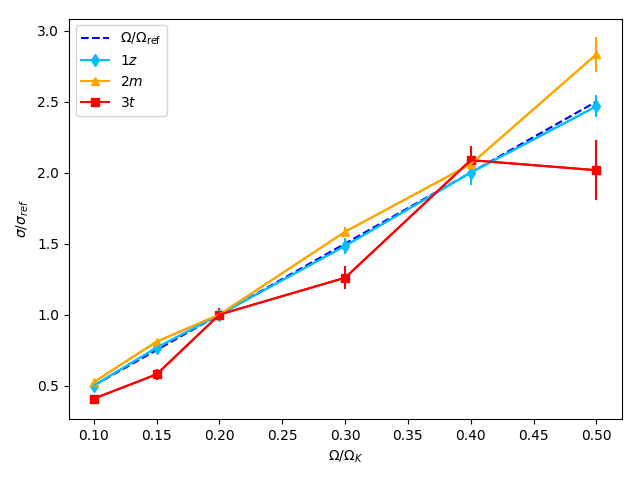}
        \caption{Evolution of $\sigma/\sigma_\mathrm{ref}$ with the rotation rate, for the dip of the series ($\ell=1$, $m=0$) for three different evolutionary stages (1z, 2m, 3t), with the associated error bars. The reference value $\sigma_\mathrm{ref}$ is the width for $\Omega = \Omega_\mathrm{ref} = 0.2\Omega_\mathrm{K}$. If, as predicted by \cite{TT22}, $\sigma \propto \Omega$, $\sigma/\sigma_\mathrm{ref} = \Omega/\Omega_\mathrm{ref} =  5 \Omega$ (the dashed line).}
    
    \label{fig:rapport_sigma_l1_m0}
\end{figure}

In Fig.~\ref{fig:rapport_sigma_l1_m0}, we compare the relative evolution of $\sigma$ with rotation in our numerical computations and in the theoretical model of \citet{TT22}.
Compared to the ($\ell=1$, $m=-1$) series, the non-linearity of $\sigma$ for the 3t model is less clear. As the dip were less well described by a Lorentzian function in this case, the fit errors are significant.

\begin{figure}[!htp]
    \centering
    \includegraphics[width=1.08\linewidth]{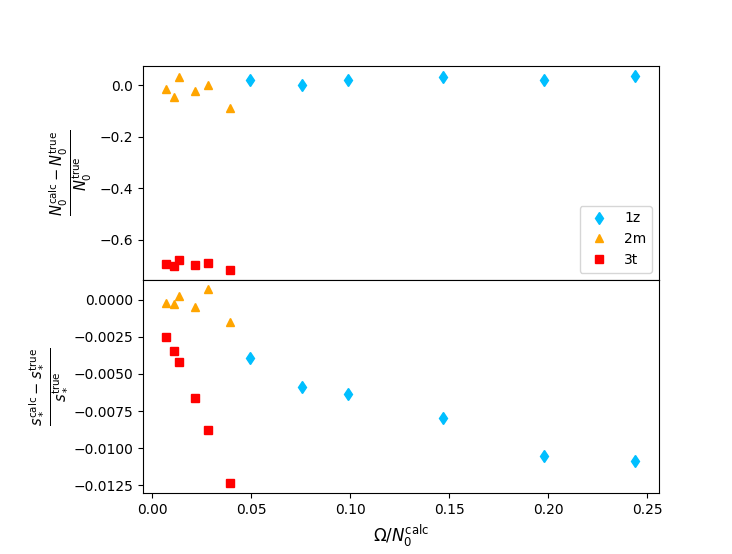}
    \caption{Relative errors for the series ($\ell=1$ $m=0$) on the estimated $N_0^\mathrm{calc}$ (upper panel) and $s_*^\mathrm{calc}$ (lower panel) using the three models 1z, 2m, and 3t.}
    \label{fig:relative_error_l1_m0}
\end{figure}

\begin{figure}[!htp]
    \centering
    \includegraphics[width=1.08\linewidth]{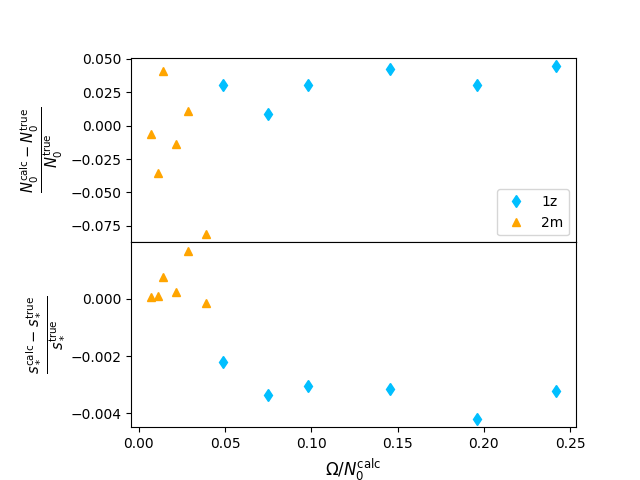}
    \caption{Relative errors for the series ($\ell=1$ $m=0$) on the estimated $N_0^\mathrm{calc}$ (upper panel) and $s_*^\mathrm{calc}$ (lower panel) using the two models 1z and 2m.}
    \label{fig:relative_error_l1_m0_2models}
\end{figure}

We used the empirical relations (\ref{eq:calc_N0}) and (\ref{eq:calc_s_iner}) on our numerical calculations to estimate $s_*$ and $N_0$ and compared them to their true values. The errors made are shown in Figs.~\ref{fig:relative_error_l1_m0} and \ref{fig:relative_error_l1_m0_2models}. In the first one, we used the three models and in the latter only the models 1z and 2m.
Relative errors behave quite differently when compared to the $\ell=|m|$ (Kelvin) modes. The estimate for $N_0$ is bad for the 3t model regardless of the rotation rate, making the estimate of $s_*$ also worse for the 3t model.

 \subsection{($\ell=2$, $m=-2$) series}

\begin{figure}[!htp]
    \centering
    \includegraphics[width=\linewidth]{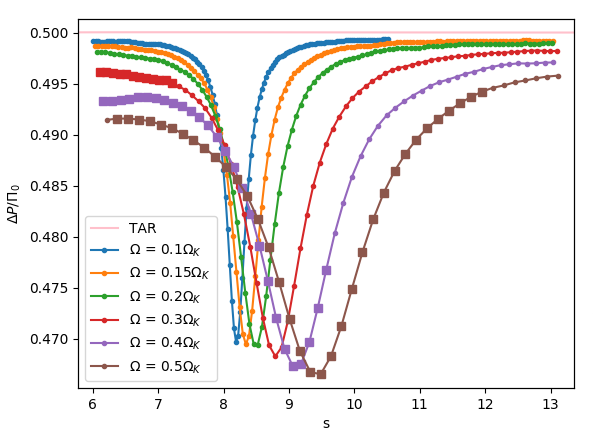}
    \caption{Same as Fig. \ref{fig:synthese_1z_l1_m0} for the series ($\ell=2$, $m=-2$).}
    \label{fig:synthese_1z_l2_m-2}
\end{figure}

\begin{figure}[!htp]
    \centering
    \includegraphics[width=\linewidth]{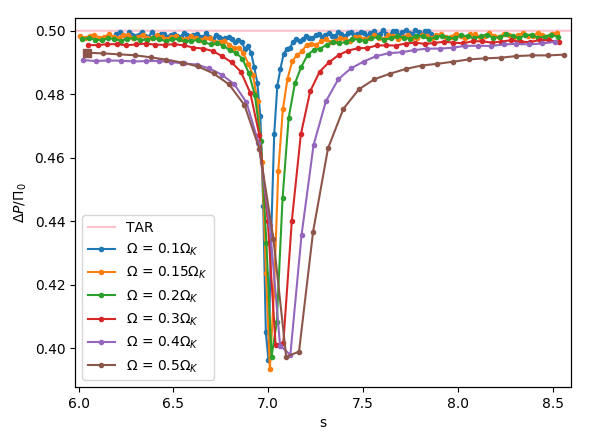}
    \caption{Same as Fig. \ref{fig:synthese_2m_l1_m0} for the series ($\ell=2$, $m=-2$).}
        \label{fig:synthese_2m_l2_m-2}

\end{figure}

\begin{figure}[!htp]
    \centering
    \includegraphics[width=\linewidth]{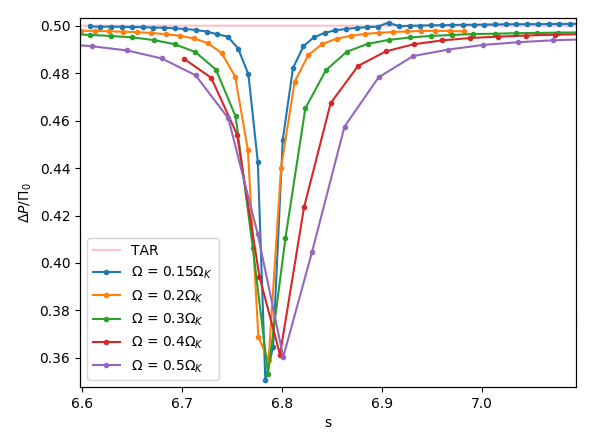}
    \caption{Same as Fig. \ref{fig:synthese_3t_l1_m0} for the series ($\ell=2$, $m=-2$).}
        \label{fig:synthese_3t_l2_m-2}

\end{figure}

The dips in the series of modes ($\ell=2$, $m=-2$) computed for the models 1z, 2m, and 3t at different rotations are presented in Figs.~\ref{fig:synthese_1z_l2_m-2}, \ref{fig:synthese_2m_l2_m-2} and \ref{fig:synthese_3t_l2_m-2}.

\begin{figure}[!htp]
    \centering
    \includegraphics[width=\linewidth]{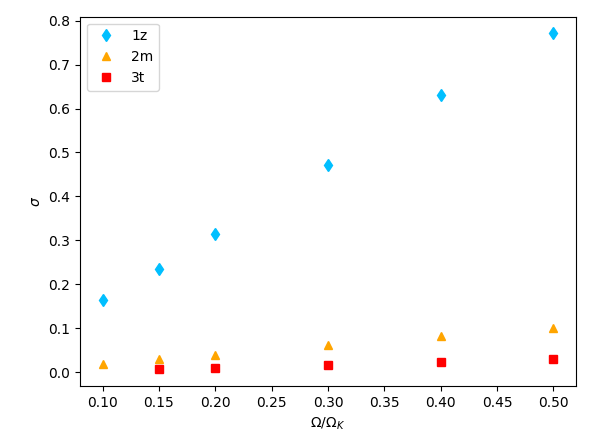}
    \caption{Same as Fig. \ref{fig:evo_sigma_rota_l1_m0} for the series ($\ell=2$, $m=-2$).}
    \label{fig:evo_sigma_rota_l2_m-2}
\end{figure}

\begin{figure}[!htp]
    \centering
    \includegraphics[width=1.05\linewidth]{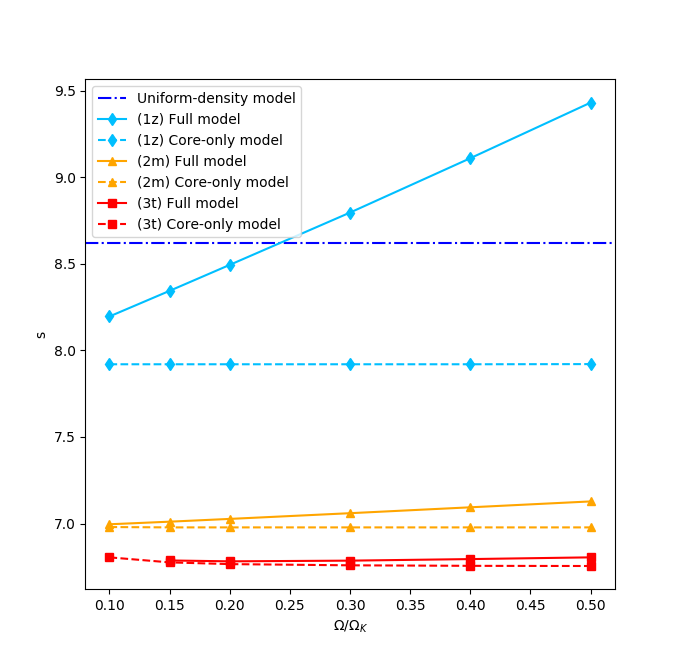}
    \caption{Same as Fig. \ref{fig:comparaison_s_l1_m0} for the series ($\ell=2$, $m=-2$).}
    \label{fig:comparaison_s_l2_m-2}
 \end{figure}

The evolution of the width $\sigma$ and location $s_c$ of dips as a function of the rotation are shown in Figs.~\ref{fig:evo_sigma_rota_l2_m-2} and \ref{fig:comparaison_s_l2_m-2}. In the latter, we also show the evolution of the spin parameter $s_*$ of the resonant inertial mode of core-only models and the spin parameter of the same mode in a uniform density sphere.

\begin{figure}[!htp]
    \centering
    \includegraphics[width=1.05\linewidth]{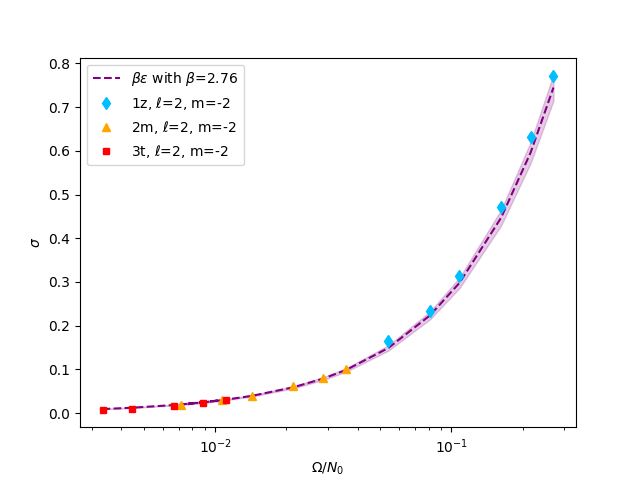}
    \caption{Same as Fig. \ref{fig:evo_sigma_log_l1_m0} for the series ($\ell=2$, $m=-2$).}
    \label{fig:evo_sigma_log_l2_m-2}
\end{figure}

\begin{figure}[!htp]
    \centering
    \includegraphics[width=1.05\linewidth]{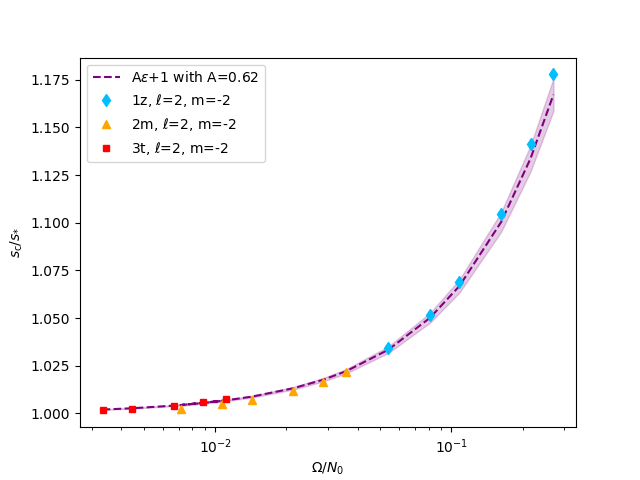}
    \caption{Same as Fig. \ref{fig:evo_sc_s_iner_log_l1_m0} for the series ($\ell=2$, $m=-2$).}
    \label{fig:evo_sc_s_iner_log_l2_m-2}
\end{figure}

We plot $\sigma$ and $s_c/s_*$ as a function of $\epsilon=\Omega/N_0$ for the three models in
Figs.~\ref{fig:evo_sigma_log_l2_m-2} and \ref{fig:evo_sc_s_iner_log_l2_m-2}, along with linear fits of the form $\sigma=\beta\epsilon$ and $s_c/s_* = A \epsilon +1$.
At a given $\epsilon$, both $\sigma$ and $s_c/s_*$ are slightly smaller for the ($\ell=2$, $m=-2$) series compared to the ($\ell=1$, $m=-1$) series. This behaviour is consistent with our analytical model (see the $\beta_\mathrm{th}$ and $A_\mathrm{th}$ column of Table \ref{modelu}).

\begin{figure}[!htp]
    \centering
    \includegraphics[width=\linewidth]{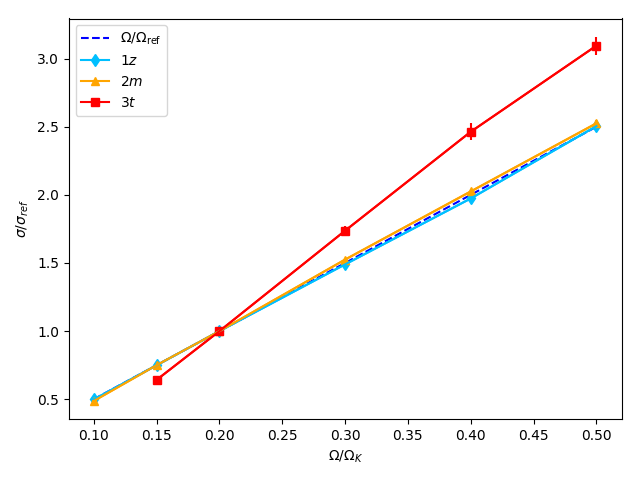}
    \caption{Same as Fig. \ref{fig:rapport_sigma_l1_m0} for the series ($\ell=2$, $m=-2$).}
    \label{fig:rapport_sigma_l2_m-2}
\end{figure}

In Fig.~\ref{fig:rapport_sigma_l2_m-2}, we compare the relative evolution of $\sigma$ with rotation in our numerical computations and in the theoretical model of \citet{TT22}.
As for the series ($\ell=1$, $m=-1$, while $\sigma$ is clearly linear for the 1z and 2m models, it is not the case for the 3t model. One can see that this tendency is more pronounced for the ($\ell=2$, $m=-2$) series than it is for the ($\ell=1$, $m=-1$) series.

\begin{figure}[!htp]
    \centering
    \includegraphics[width=1.08\linewidth]{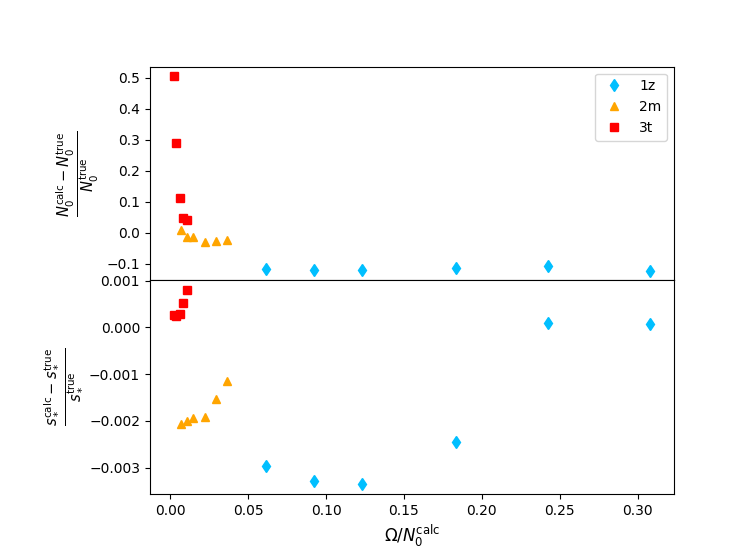}
    \caption{Same as Fig. \ref{fig:relative_error_l1_m0} for the series ($\ell=2$, $m=-2$).}
    \label{fig:relative_error_l2_m-2}
\end{figure}

\begin{figure}[!htp]
    \centering
    \includegraphics[width=1.08\linewidth]{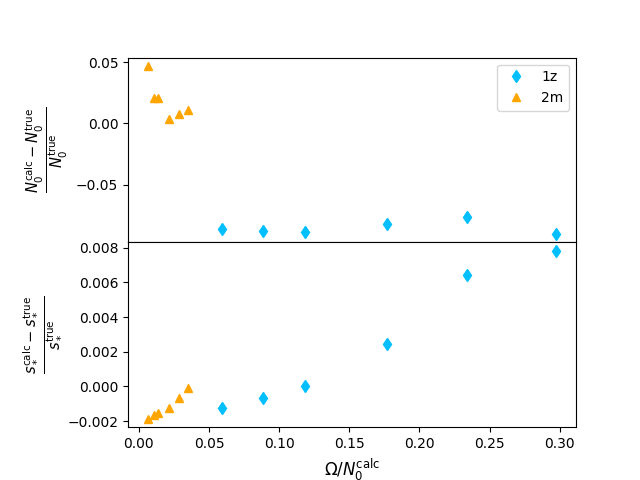}
    \caption{Same as Fig. \ref{fig:relative_error_l1_m0_2models} for the series ($\ell=2$, $m=-2$).}
    \label{fig:relative_error_l2_m-2_2models}
\end{figure}

We used the empirical relations (\ref{eq:calc_N0}) and (\ref{eq:calc_s_iner}) on our numerical calculations to estimate $s_*$ and $N_0$ and compared them to their true values. The errors made are shown in Figs.~\ref{fig:relative_error_l2_m-2} and \ref{fig:relative_error_l2_m-2_2models}. In the first one, we used the three models and in the latter only the models 1z and 2m.
One can see that the relative errors on $N_0$ and $s_*$ follow the same kind of trend as for the ($\ell=1$, $m=-1$) series.

\clearpage

\section{Analytical model of the resonant coupling between inertial and gravito-inertial oscillations.}
\label{model}

We present an analytical model of mixed (inertial and gravito-inertial modes) when the density is assumed to be uniform in the convective core. This model is similar to
the one proposed by \citet{TT22} in that its uses the same hypothesis to derive analytical solutions in the convective core and in the radiative zone respectively. However, it goes a step further 
by improving the
continuity
of the mixed modes 
at the convective-radiative interface. 

In the following, we determine analytical solutions of free oscillations in the convective core (Sect.~\ref{conv}) and then in the radiative envelope (Sect.~\ref{rad}). In Sect.~\ref{both}, the continuity conditions at the interface are written and approximated in the case of $(\ell,m) =(1,-1)$ gravito-inertial modes interacting with the $(\ell_i,m) =(3,-1), s_*=11.32$ inertial mode. The effect of this mode coupling on the $\Delta P =f(s)$ relation is then derived.

\subsection{Analytical solutions in the convective core}
\label{conv}

In this subsection, we describe small amplitude inertial oscillations in the convective core with no specified boundary conditions at  the interface with the radiative zone. In an axially symmetric star, we search for solutions proportional to $ e^{i(m\phi +\omega t)}$ where $m$ is the azimuthal number. The radial displacement and the Eulerian pressure perturbations are thus expressed as $\xi^{\rm conv}_r(r,\theta) e^{i(m\phi +\omega t)}$ and ${p'}^{\rm conv}(r,\theta) e^{i(m\phi +\omega t)}$ respectively. 
In an isentropic convective core of uniform density, perturbations of low frequency are governed by the Poincaré equation $\Delta \Psi = s^2 \frac{\partial \Psi}{\partial z}$ where $\Psi$ is proportional to ${p'}^{\rm conv}$ \citep{W05}. In a full sphere, separable solutions are obtained using the ellipsoidal coordinates ($x_1,x_2,\phi$) defined in terms of the Cartesian coordinates as

\begin{eqnarray}
    x/r_c &=& \left(\frac{(1-x_1^2)(1-x_2^2)}{1-\mu_s^2}\right)^{1/2}  \cos(\phi) \label{D1}\\
    y/r_c & =& \left(\frac{(1-x_1^2)(1-x_2^2)}{1-\mu_s^2}\right)^{1/2}\sin(\phi)
    \label{D2}\\
    z/r_c &=& \frac{x_1 x_2}{\mu_s}
    \label{D3}
\end{eqnarray}
\noindent where $\mu_s =1/s$, $x_1 \in [\mu_s, 1]$, and $x_2 \in [-\mu_s, \mu_s]$. With these coordinates, the surface of the sphere is given by  $\{x_1 \in [\mu_s, 1], x_2 = \mu_s\} \cup \{x_1 = \mu_s, x_2 \in [-\mu_s, \mu_s]\}$.

The separable solutions read:
 \begin{align}
\Psi = r_c^2 P_{\ell_i}^m(x_1) P_{\ell_i}^m(x_2)
\end{align}
\noindent where $P_{\ell_i}^m$ is the associated Legendre polynomial of degree $\ell_i$. The radial displacement $\xi^{\rm conv}_r$ and the Eulerian pressure perturbations $p'^{\rm conv}$ are then obtained from the relations \citep{W05}:
 \begin{align}
\xi^{\rm conv}_r = & \frac{1}{1-s^2} \lp \frac{\partial \Psi}{\partial r} + \frac{ms}{r} \Psi -s^2 \cos \theta \frac{\partial \Psi}{\partial z} \rp \\
p'^{\rm conv} = & \rho_0 \omega^2  \Psi.
\end{align}
At the outer radius of the convective core $r=r_c$, it reads \citep{TT22}:
\begin{align}
\xi^{\rm conv}_r(r=r_c) =& r_c \frac{1}{s}  \gamma_{\ell_i}^m(s) P_{\ell_i}^m (\cos \theta) \;\;\;{\rm with} \\
\gamma_{\ell_i}^m(s) = & {P'}_{\ell_i}^m (1/s) - \frac{m}{1 - 1/s^2} P_{\ell_i}^m(1/s) 
\end{align}
\noindent where ${P'}_{\ell_i}^m$ is the derivative of the polynomial ${P}_{\ell_i}^m$, and 
\begin{align}
p'^{\rm conv}(r=r_c) = & \rho_0 r_c^2 \omega^2 P_{\ell_i}^m(1/s) P_{\ell_i}^m (\cos \theta). 
\end{align}

These solutions oscillate freely at the convective-radiative interface because no boundary condition is imposed there. If the $\xi^{\rm conv}_r(r=r_c)=0$ boundary condition is enforced, one recovers the eigenfrequencies of pure inertial modes presented in Appendix~\ref{appendix:homogeneous_model}  which satisfy Eq.~\ref{eigenspin} or equivalently $\gamma_{\ell_i}^m(s) = 0$.

\subsection{Analytical solutions in the radiative zone}
\label{rad}

In the radiative zone, analytical gravito-inertial oscillations with no prescribed boundary conditions at the convective interface can also be found in the framework of the TAR and in the limit of short-radial wavelengths. These solutions are separable in the spherical coordinates and can thus be written  $\xi^{\rm rad}_r(r,\theta) = \zeta(r)H(\theta)$ and 
${p'}^{\rm rad}(r,\theta) = p'(r)H(\theta)$, where $\zeta$ and $p'$ denote the radial part of the radial displacement and Eulerian pressure perturbation respectively, and $H(\theta)$ their latitudinal part.

The equation governing the latitudinal part of the solution is the Laplace's tidal equation:
\begin{equation}
\label{Laplace}
{\cal L_{\rm T}} (H) + \Lambda H =0 
\end{equation}
\noindent with
\begin{align*}
\label{Laplace_b}
{\cal L_{\rm T}}(H) =& \frac{d}{d \mu} \left[\frac{1-\mu^2}{(1-s^2\mu^2)} \frac{d}{d \mu} H \right] \\
& - \frac{1}{1-s^2\mu^2} \left( \frac{m^2}{1-\mu^2} + ms\frac{1+s^2\mu^2}{1-s^2\mu^2} \right) H, 
\end{align*}

\noindent where $\mu=\cos \theta$ and $\Lambda$ is a constant. Applying regularity conditions at the polar axis amounts to solve an eigenvalue problem. For each doublet $(m,s)$, the eigenfunctions and their associated eigenvalues can be labelled with an integer $k$ \citep{Lee1997}. The eigenvalues and the eigenfunctions, known as the Hough functions, are thus denoted $\Lambda^{m}_{k}(s)$ and $H^{m}_{k}(\theta, s)$ respectively. They form an orthogonal set of functions as $\int_0^{\pi} H^{m}_{k} (\theta,s) H^{m}_{k'} (\theta,s)\sin \theta \; d\theta $=0 if $k \neq k'$.  In the following, we shall use normalised Hough functions such that $\int_0^{\pi} \lp H^{m}_{k} (\theta,s)\rp^2 \sin \theta \; d \theta =1$. Physically, a positive or null $k$ corresponds to a gravity mode modified by the Coriolis force. In this case, the index $k$ relates to the degree $\ell$ of the spherical harmonics of the gravity mode in a non-rotating star through $\ell = |m| + k$. Thus, for example, the prograde sectoral modes, also called Kelvin modes, can be equivalently labelled $(\ell=|m|, m<0)$ or $(k=0,m<0)$. Negative $k$ in turn are associated to solutions that only exist in rotating stars either the (retrograde) Rossby modes or the $\Lambda <0$ modes that do not propagate in the radial direction.
For the prograde or axisymmetric modes that we shall consider below, the $k<0$ modes correspond to $\Lambda <0$ non propagative modes. 

Within the Cowling approximation, the equations governing the radial part of the oscillation, $\zeta$ and $p'$, are:
\begin{align}
    \frac{d \zeta}{dr} = & - \lp \frac{2}{r} - \frac{1}{\Gamma_1 H_P} \rp \zeta + \frac{1}{\rho_0c_0^2} \lp \frac{\Lambda c_0^2}{r^2 \omega^2} - 1 \rp p' \\
    \frac{d p'}{dr} = & \rho_0 (\omega^2- N^2) \xi_r - \frac{1}{\Gamma_1 H_P} p'
\end{align}

\noindent where $\Lambda^{m}_{k}(s)$ is the eigenvalue of the Laplace's tidal equation.

In the limit of short radial wavelengths, the WKB solutions of these equations depends on the sign of $k_r^2  = \frac{\omega^2 - N^2}{c_0^2}\lp 1 - \frac{\Lambda_{k}^{m}(s) c_0^2}{r^2 \omega^2} \rp$ :
\begin{align}
\label{D13}
{\rm if}\; k_r^2 >0 & \;\; \zeta_{k}^{m}(r,s) = -c \frac{|1-\frac{\Lambda c_0^2}{r^2 \omega^2}|^{1/2}}{r c_0 (\pi \rho_0 k_r)^{1/2}} \sin \! \lp \int_{r}^{r_b} k_r \;dr -\frac{\pi}{4} \rp\\
&  \;\; \frac{{p'}_{k}^{m}(r,s)}{\rho_0} = c \frac{(N^2-\omega^2)^{1/2}}{r (\pi \rho_0 k_r)^{1/2}} \cos \! \lp \int_{r}^{r_b} k_r \;dr -\frac{\pi}{4} \rp \\
{\rm if}\; k_r^2 <0 & \;\;
\zeta_{k}^{m}(r,s) = c' \frac{\lp 1-\frac{\Lambda c_0^2}{r^2 \omega^2}\rp^{1/2}}{r c_0 (\pi \rho_0 |k_r|)^{1/2}} \exp \! \lp - \int_{r_c}^{r} |k_r| \;dr \rp  \\
& \;\; \frac{{p'}_{k}^{m}(r,s)}{\rho_0}= c' \frac{(N^2-\omega^2)^{1/2}}{r (\pi \rho_0 |k_r|)^{1/2}} \exp \! \lp - \int_{r_c}^{r} |k_r| \;dr \rp 
\end{align}
\noindent where $c$ and $c'$ are arbitrary constants and the labels ${k,m,s}$ of $\Lambda$ and $k_r$ have been omitted for clarity. In these expressions, $r_b$ is the outer radius of the oscillation cavity which satisfies $k_r(r_b) =0$ that is $\frac{\Lambda_{k}^{m}(s) c_0(r_b)^2}{r_b^2 \omega^2}\approx 1$ as $\omega \ll N$. The solution exponentially decreasing from $r_c$ corresponds to Hough modes with negative $\Lambda$.

These oscillations are free at the convective/radiative interface. If a boundary condition $\xi_r(r=r_c)=0$ were to be enforced, solutions $\zeta_{k}^{m}(r,s) H_{k}^{m}(\theta,s)$ with $\zeta_{k}^{m}(r=r_c, s)=0$ would be the eigenmodes. According to Eq.~\ref{D13}, this leads to the  quantisation condition $\int_{r_c}^{r_b} k_r \; dr - \pi/4 = n \pi$. As $\Lambda c_0^2/(r^2 \omega^2) \gg 1$ 
away from $r_b$ and $\omega \ll N$, the condition can be simplified into $\frac{\pi^2 s \sqrt{\Lambda}}{\Omega \Pi_0} = \pi (n+1/4)$ with $\Pi_0 = \frac{2 \pi^2}{\int_{r_i}^{r_o} \frac{N}{r} \, \mathrm{d}r }$.

\subsection{Modes of mixed inertial and gravito-inertial nature}
\label{both}

To produce
mixed modes, the inertial and gravito-inertial oscillations have to match at the convective-radiative interface in the sense that $\xi_r$ and $\delta p$ have to be continuous there. However, as illustrated in Fig.~\ref{profile} for the $(\ell=1,m=-1)$ vs $(\ell_i=3, m=-1)$ resonance, these matching conditions can not be fulfilled with only one separable solutions in each domain.
We thus need more general solutions on both sides of the interface. This can be done by considering linear combinations of the separable solutions derived in the two previous sections. Namely, in the radiative zone :
 \begin{align}
{\xi}^{\rm rad}_r =& \sum_{k=-\infty}^{k=+\infty} a_k \zeta_{k}^{m}(r,s) H_{k}^{m}(\theta,s) \\
p'^{\rm rad} =& \sum_{k=-\infty}^{k=+\infty} a_k {p'}_{k}^{m}(r,s) H_{k}^{m}(\theta,s)
\end{align}
and in the convective zone :
 \begin{align}
\Psi \propto & r_c^2 \sum_{\ell_i=1}^{+\infty} b_{\ell_i} P_{\ell_i}^m(x_1) P_{\ell_i}^m(x_1).
\end{align}
\noindent

At the interface between the convective core and the radiative zone, the density is continuous while its radial derivative is discontinuous. In such a case, the continuity condition on $\xi_r$ and $\delta p$ is equivalent to the continuity of $\xi_r$ and $p'$, thus to ${\xi}^{\rm conv}_r (r=r_c^-) = {\xi}^{\rm rad}_r (r=r_c^+)$
and $p'^{\rm conv}(r=r_c^-)=p'^{\rm rad}(r=r_c^+)$.

The expressions of ${\xi}^{\rm rad}_r(r_c^+)$ and $p'^{\rm rad}(r_c^+)$ are 
\begin{align}
{\xi}^{\rm rad}_r(r_c^+) =& \sum_{k=-\infty}^{k=+\infty} a_k \zeta_{k}^{m}(r_c^+,s) H_{k}^{m}(\theta,s) \;\;\;{\rm with} \\
\zeta_{k}^{m}(r_c^+,s) = & - \frac{1}{\rho_0^{1/2} \omega r_c^3} \lp \frac{\Lambda}{\pi k_r} \rp^{1/2} \sin(\psi) \;\;\;{\rm if}\;\;\; k_r^2 >0 \\
\zeta_{k}^{m}(r_c^+,s) = & \frac{1}{\rho_0^{1/2} \omega r_c^3} \lp \frac{|\Lambda|}{\pi |k_r|} \rp^{1/2} \;\;\;{\rm if}\;\;\; k_r^2 <0
\end{align}

and 
\begin{align}
p'^{\rm rad}_r(r_c^+) =& \sum_{k=-\infty}^{k=+\infty} a_k {p'}_{k}^{m}(r_c^+,s) H_{k}^{m}(\theta,s) \;\;\;{\rm with}\\
{p'}_{k}^{m}(r_c^+,s) = & \frac{\rho_0^{1/2}}{r_c} \lp \frac{N_0^2 - \omega^2}{\pi k_r} \rp^{1/2} \cos (\psi) &\;\;\;{\rm if}\;\;\; k_r^2 >0 \\
{p'}_{k}^{m}(r_c^+,s) = & \frac{\rho_0^{1/2}}{r_c} \lp \frac{N_0^2 - \omega^2}{\pi |k_r|} \rp^{1/2} &\;\;\;{\rm if}\;\;\; k_r^2 <0
\end{align}
\noindent where $N_0=N(r=r_c^+)$ and $\psi_{k}^{m}(s) = \int_{r_c}^{r_b} k_r \; dr - \pi/4 \approx \frac{\pi^2 s \sqrt{\Lambda}}{\Omega \Pi_0} - \pi/4$  because $\Lambda c_0^2/(r^2 \omega^2) \gg 1$ 
away from $r_b$ and $\omega \ll N$.

The expressions of ${\xi}^{\rm conv}_r(r_c^-)$ and $p'^{\rm conv}(r_c^-)$ are 
\begin{align}
{\xi}^{\rm conv}_r(r_c^-) =& \frac{r_c}{s} \sum_{\ell_i=1}^{+\infty} b_{\ell_i} \gamma_{\ell_i}^m(s) \tilde{P}_{\ell_i}^m (\cos \theta) \\
p'^{\rm conv}(r_c^-) = & \rho_0 r_c^2 \omega^2 \sum_{\ell_i=1}^{+\infty} b_{\ell_i} P_{\ell_i}^m(1/s) \tilde{P}_{\ell_i}^m (\cos \theta) 
\end{align}
\noindent where, for convenience, we used $\tilde{P}_{\ell_i}^m$, the normalised form of the associated Legendre polynomial, i.e. $\int_0^{\pi} \lp \tilde{P}_{\ell_i}^m (\cos \theta)\rp^2 \sin \theta \; d \theta =1$.

The continuity conditions then read:
\begin{align}
\frac{r_c}{s} \sum_{\ell_i=1}^{+\infty} b_{\ell_i} \gamma_{\ell_i}^m(s) \tilde{P}_{\ell_i}^m (\cos \theta) =& \sum_{k=-\infty}^{k=+\infty} a_k \zeta_{k}^{m}(r_c^+,s) H_{k}^{m}(\theta,s)\\
\rho_0 r_c^2 \omega^2 \sum_{\ell_i=1}^{+\infty} b_{\ell_i} P_{\ell_i}^m(1/s) \tilde{P}_{\ell_i}^m (\cos \theta) = & \sum_{k=-\infty}^{k=+\infty} a_k {p'}_{k}^{m}(r_c^+,s) H_{k}^{m}(\theta,s)
\end{align}

Multiplying both equations by $H_{k}^{m}(\theta,s)$ and taking the integral $\int_0^{\pi} ... \sin \theta \; d\theta$ leads to :
\begin{align}
\frac{r_c}{s} \sum_{\ell_i=1}^{+\infty} b_{\ell_i} \gamma_{\ell_i}^m(s) c^m_{k,\ell_i} =& a_k \zeta_{k}^{m}(r_c^+,s) \\
\rho_0 r_c^2 \omega^2 \sum_{\ell_i=1}^{+\infty} b_{\ell_i} P_{\ell_i}^m(1/s) c^m_{k,\ell_i} = & a_k  {p'}_{k}^{m}(r_c^+,s) 
\end{align}
\noindent for each integer $k$, where $c^m_{k,\ell_i}(s) = \int_0^{\pi} \tilde{P}_{\ell_i}^m (\cos \theta) H_{k}^{m}(\theta,s) \sin \theta \; d\theta$ is a coupling coefficient between $\tilde{P}_{\ell_i}^m$ and $H_{k}^{m}(\theta,s)$. Eliminating $a_k$ we obtain:
\begin{equation}
\label{full_cont}
\sum_{\ell_i=1}^{+\infty} \lp \frac{r_c}{s}\gamma_{\ell_i}^m(s)  {p'}_{k}^{m}(r_c^+,s) - \rho_0 r_c^2 \omega^2 P_{\ell_i}^m(1/s)\zeta_{k}^{m}(r_c^+,s)\rp  c^m_{k,\ell_i}(s) b_{\ell_i} = 0 
\end{equation} 
\noindent for each integer $k$. Introducing the expressions of ${p'}_{k}^{m}(r_c^+,s)$ and $\zeta_{k}^{m}(r_c^+,s)$, we obtain for $k \ge 0$:
\begin{align}
\label{sys1}
&\sum_{\ell_i=1}^{+\infty} \lp \gamma_{\ell_i}^m(s) \frac{N_0}{2 \Omega} \cos(\psi_{k}^{m}(s)) 
+P_{\ell_i}^m(1/s) \Lambda_{k}^{m}(s)^{1/2} \sin(\psi_{k}^{m}(s)) \rp c^m_{k,\ell_i}(s) b_{\ell_i}   \nonumber \\
&= 0 
\end{align}
\noindent and for $k < 0$:
\begin{align}
\label{sys2}
\sum_{\ell_i=1}^{+\infty} \lp \gamma_{\ell_i}^m(s) \frac{N_0}{2 \Omega} - P_{\ell_i}^m(1/s) |\Lambda_{k}^{m}(s)|^{1/2}\rp c^m_{k,\ell_i}(s) b_{\ell_i} = 0 
\end{align}
\noindent where we used the approximation $(1 -\omega^2/N_0^2)^{1/2} \approx 1$.

\begin{figure}
\centering
\includegraphics[width=\hsize]{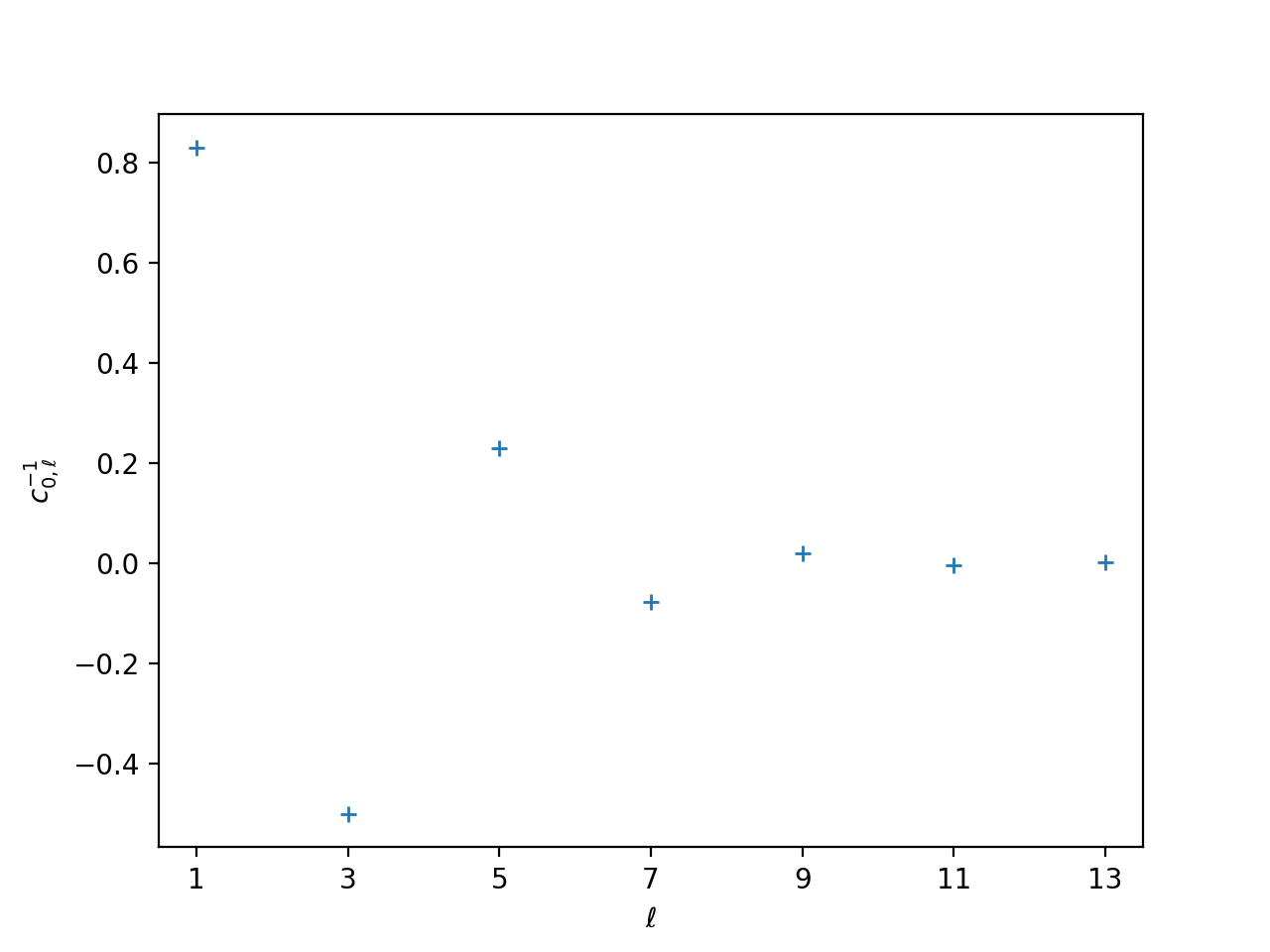}

\includegraphics[width=\hsize]{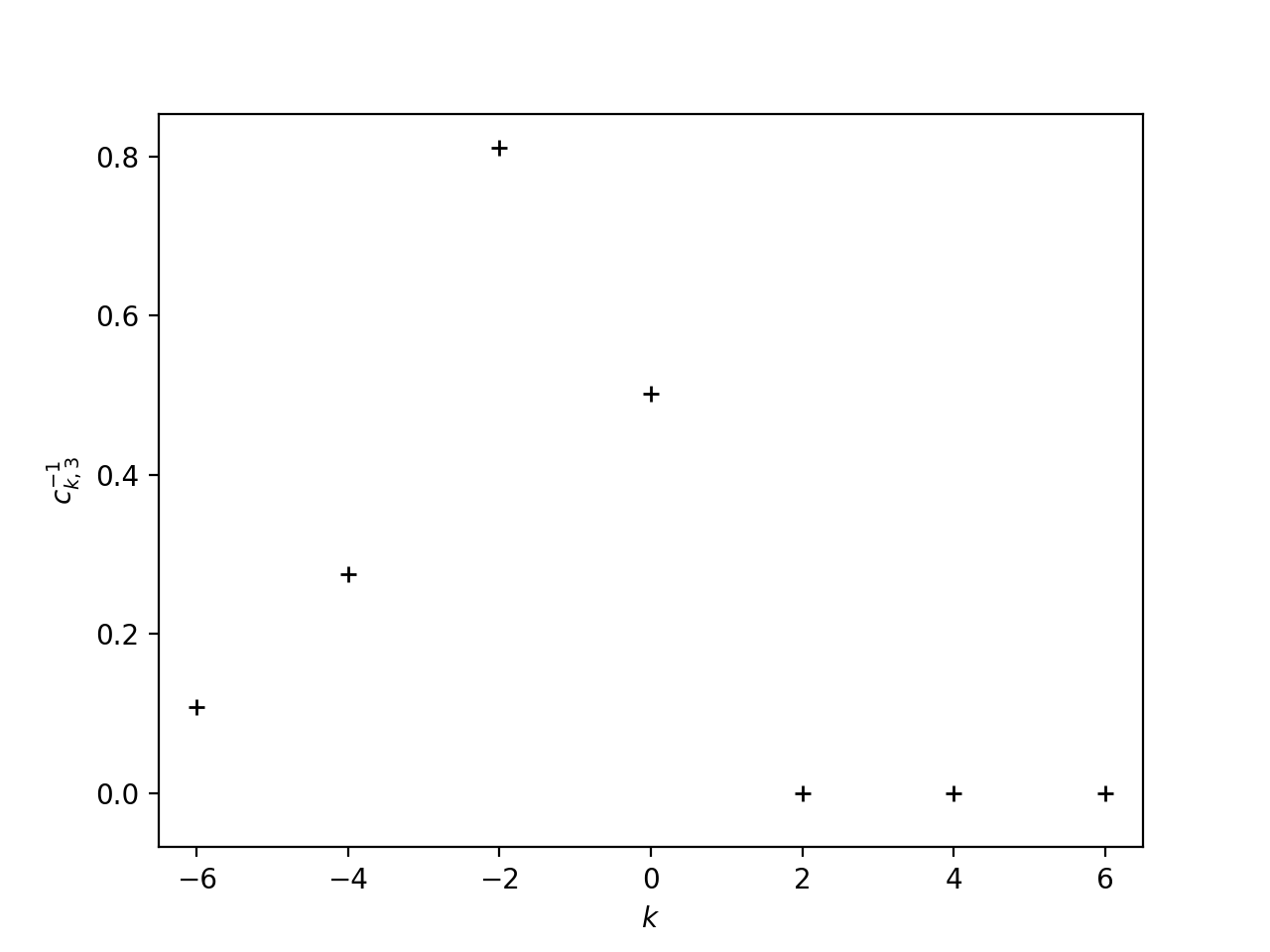}
\caption{Upper panel: Projection of the Hough function $H_{k=0}^{m=-1}(s_*=11.32)$ on the associated Legendre polynomial of the same $m$ and parity. Lower panel: Projection of the associated Legendre polynomial $P_{\ell_i=3}^{m=-1}$ on the Hough functions of the same $m$ and parity determined at $s_*= 11.32$.}
\label{H0_1}
\end{figure}

\begin{figure}
\centering

\includegraphics[width=\hsize]{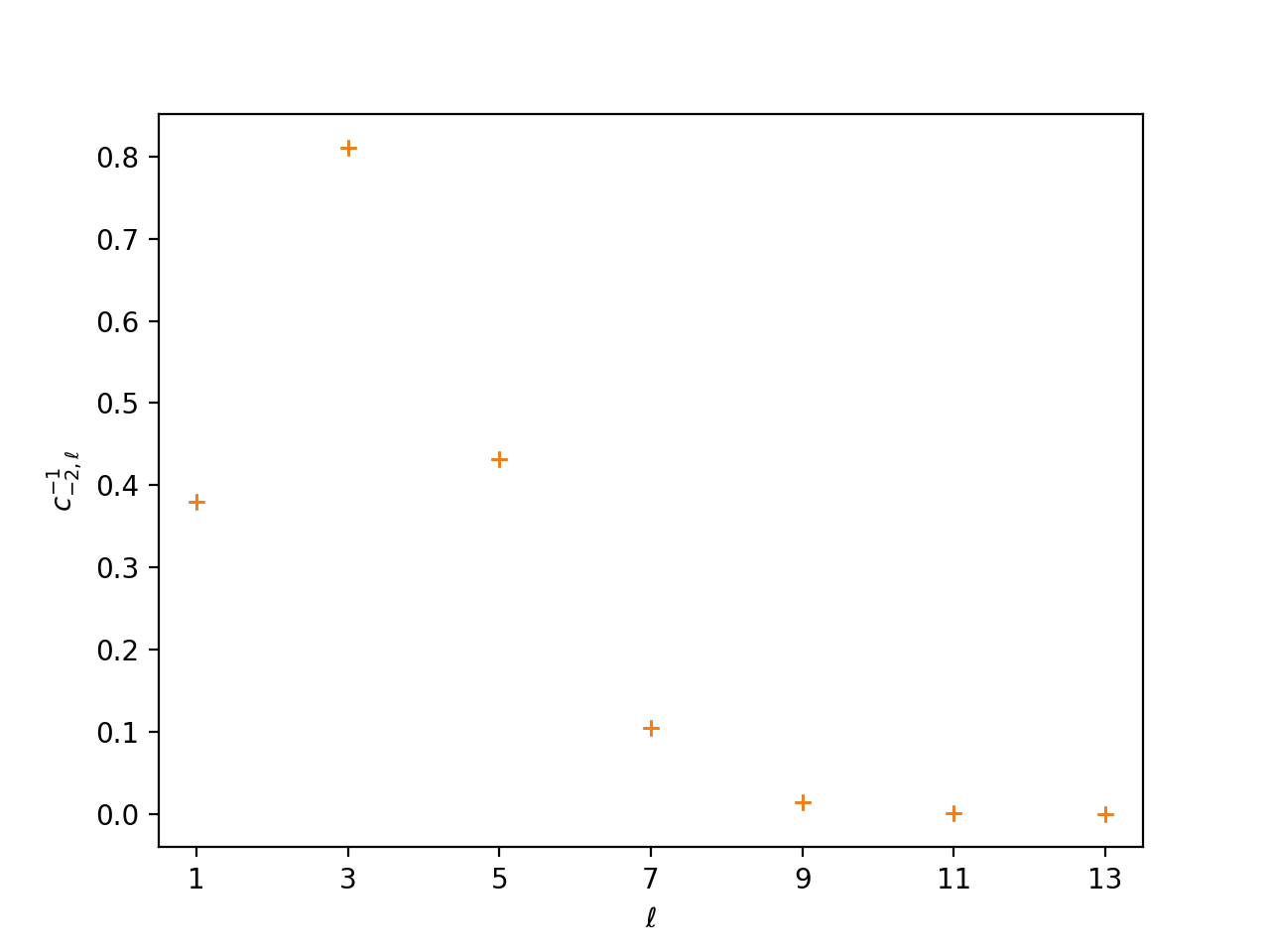}
\includegraphics[width=\hsize]{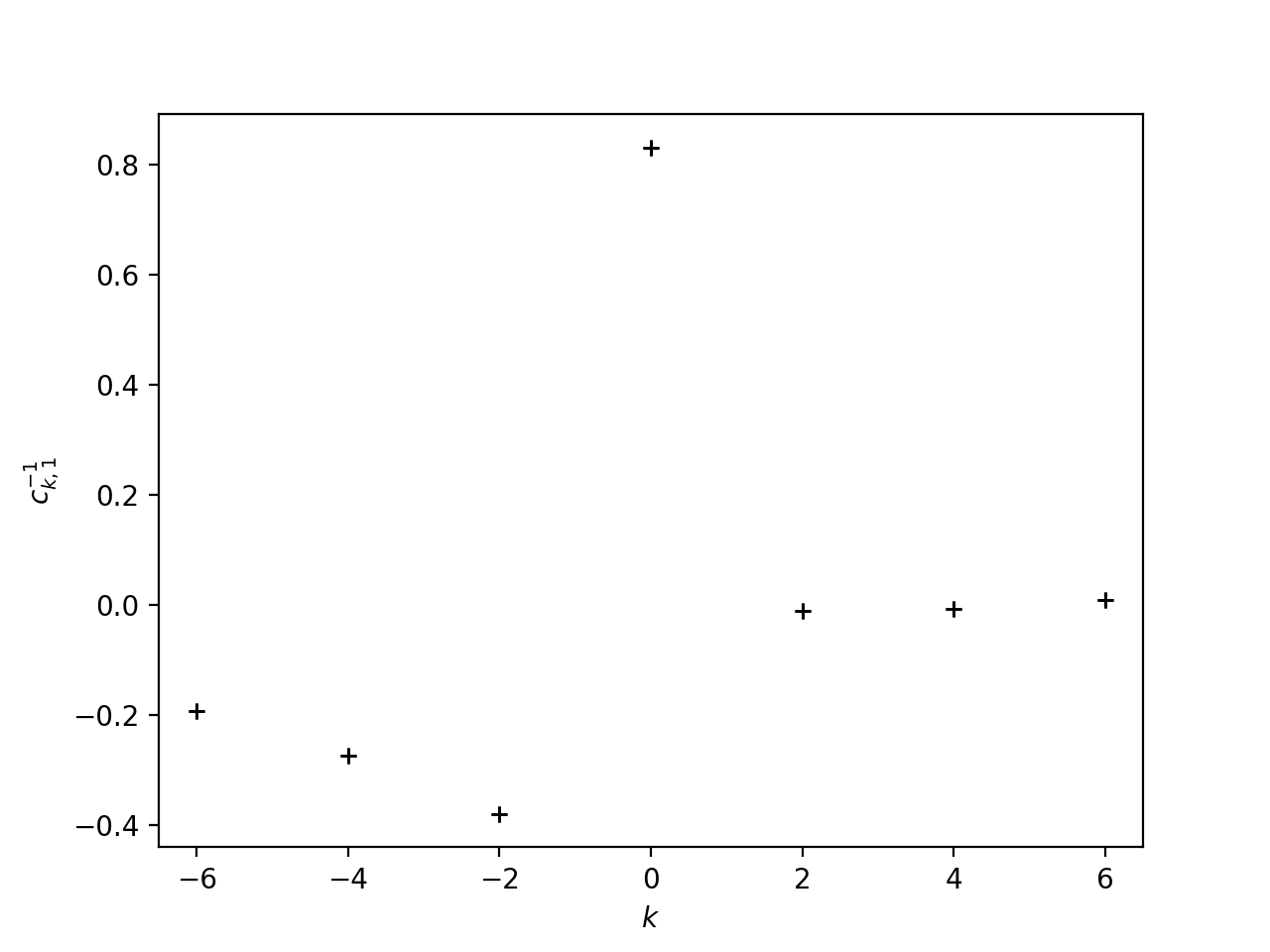}
\caption{Upper panel: Projection of the Hough function $H_{k=-2}^{m=-1}(s_*=11.32)$ on the associated Legendre polynomial of the same $m$ and parity. Lower panel: Projection of the associated Legendre polynomial $P_{\ell_i=1}^{m=-1}$ on the Hough functions of the same $m$ and parity determined at $s_*= 11.32$.}
         \label{Hm2_1}
\end{figure}

To get approximate analytical solutions, we may try to truncate this infinite system of equations to a finite system. We first consider the case of the resonance between the Kelvin modes $(\ell=1,m=-1)$ and the $(\ell_i=3,m=-1)$ inertial mode near $s_*=11.32$. To analyse the geometrical matching between these modes at $r=r_c$, the upper panel of Fig.~\ref{H0_1} shows how  $H_{0}^{-1}(\theta,s_*)$, projects onto Legendre polynomials of same $m$ and equatorial parity. We observe that the two main contributions come from $P_1^{-1}$ and $P_3^{-1}$. Conversely, the lower panel of  Fig.~\ref{H0_1} 
shows that the inertial mode $P_{3}^{-1}$ mostly projects onto $H_{0}^{-1}(s_*)$ and $H_{-2}^{-1}(s_*)$. This suggests to truncate the infinite system to linear combinations of $P_3^{-1}$ and $P_1^{-1}$ on the convective side and of $H_{0}^{-1}(s)$ and $H_{-2}^{-1}(s)$ on the radiative side. For consistency one should also consider the modes that are excited by $H_{-2}^{-1}(s_*)$ and $P_1^{-1}$ at the interface. The projections of $H_{-2}^{-1}(s)$ and $P_1^{-1}$ onto Legendre polynomials and Hough functions respectively are displayed in Fig.~\ref{Hm2_1}.
It shows that, except for the contribution of $P_5^{-1}$ to $H_{-2}^{-1}(s)$, the two functions $H_{-2}^{-1}(s)$ and $P_1^{-1}$ are dominantly projected onto respectively 
$P_3^{-1}$ and $P_1^{-1}$, and 
$H_{0}^{-1}(s_*)$ and $H_{-2}^{-1}(s)$.
We thus expect that the truncation of the infinite system to $P_1^{-1}$, $P_3^{-1}$, $H_{0}^{-1}(s)$, and $H_{-2}^{-1}(s)$ provides an approximate representation of the mixed modes near $s_*=11.32$. 

With this assumption, the system of Eqs.~\ref{sys1} and ~\ref{sys2} reduces to the two equations corresponding to $k=0$ and $k=-2$:
\begin{align}
\label{sys_l1m1_1}
\lp \gamma_{1} \frac{N_0}{2 \Omega} \cos(\psi_0) + P_1 \Lambda_{0}^{1/2} \sin(\psi_0) \rp c_{0,1} b_1 +& \nonumber \\
\lp \gamma_{3} \frac{N_0}{2 \Omega} \cos(\psi_0) + P_3 \Lambda_{0}^{1/2} \sin(\psi_0) \rp c_{0,3} b_3 &= 0\\
\label{sys_l1m1_2}
\lp \gamma_{1} \frac{N_0}{2 \Omega} -
P_1 |\Lambda_{-2}^{1/2}|\rp c_{-2,1} b_1 + & \nonumber\\
\lp \gamma_{3} \frac{N_0}{2 \Omega} -
P_3 |\Lambda_{-2}^{1/2}|\rp c_{-2,3} b_3& = 0 
\end{align}
\noindent where the $m=-1$ indices and the $s$ dependence have been omitted for clarity.

Having a non-trivial solution for this system requires that its determinant vanishes. This can be written as:
\begin{equation}
\label{det}
{\cal A} \cos(\psi_0) + {\cal B} \sin(\psi_0) =0 
\end{equation}
\noindent with
\begin{align}
 {\cal A} = &c_{0,1}c_{-2,3} \gamma_1 P_3 \Lambda_{-2}^{1/2} \frac{2 \Omega}{N_0}\nonumber \\
& +  \lc (c_{0,3}c_{-2,1} - c_{0,1}c_{-2,3}) \gamma_1 - c_{0,3}c_{-2,1} P_1 \Lambda_{-2}^{1/2} \frac{2 \Omega}{N_0} \rc \gamma_3 \\
{\cal B} = & \Biggl[ c_{0,3}c_{-2,1} \gamma_1 P_3 - (c_{0,3}c_{-2,1} - c_{0,1}c_{-2,3}) \Lambda_{-2}^{1/2} P_1 P_3 \frac{2 \Omega}{N_0} \Biggr. \nonumber\\
& - \Biggl.  c_{0,1}c_{-2,3} \gamma_3 P_1 \Biggr] \Lambda_{0}^{1/2} \frac{2 \Omega}{N_0} 
\end{align}

We can simplify this equation using the fact that $\Omega/N_0 \ll 1$ in $\gamma$ Dor stars. However, one should pay attention that $\gamma_3$ is also small as we consider spin parameters close to $s_*$ and $\gamma_3(s_*)=0$. We obtain:
\begin{align}
{\cal A}  \approx & c_{0,1}c_{-2,3} \gamma_1 P_3 \Lambda_{-2}^{1/2} \frac{2 \Omega}{N_0} + (c_{0,3}c_{-2,1} - c_{0,1}c_{-2,3}) \gamma_1 \gamma_3 \\
{\cal B} \approx & \lp c_{0,3}c_{-2,1} \gamma_1 P_3 - c_{0,1}c_{-2,3} \gamma_3 P_1 \rp \Lambda_{0}^{1/2} \frac{2 \Omega}{N_0}
\end{align}

Close to $s_*$ we can write $\gamma_3 \approx \gamma'_3 (s-s_*)$, while the other terms in ${\cal A}$ and ${\cal B}$ remain approximately equal to their value at $s_*$. Keeping the dominant terms, we get:
\begin{align}
{\cal A}  \approx & c_{0,1}c_{-2,3} \gamma_1 P_3 \Lambda_{-2}^{1/2} \frac{2 \Omega}{N_0} + (c_{0,3}c_{-2,1} - c_{0,1}c_{-2,3}) \gamma_1 \gamma'_3 (s-s_*)\\
{\cal B} \approx & c_{0,3}c_{-2,1} \gamma_1 P_3  \Lambda_{0}^{1/2} \frac{2 \Omega}{N_0}
\end{align}
\noindent so that Eq.~\ref{det} becomes:
\begin{equation}
\label{cot}
  \cot (\psi_0(s)) = - \frac{\epsilon/V}{s-s_c}  
\end{equation}
\noindent where
\begin{align}
\psi_0 = & \frac{\pi^2 s \sqrt{\Lambda_0}}{\Omega \Pi_0} -\frac{\pi}{4} \\
\epsilon = & \Omega/N_0 \\
s_c = & s_* + \lp \frac{1}{1+f(s_*)} \frac{2 P_3(1/s_*)}{\gamma'_3(s_*)} \sqrt{\Lambda_{-2}(s_*)} \rp \epsilon \\
\frac{1}{V} = & \frac{f(s_*)}{1+f(s_*)} \frac{2 P_3(1/s_*)}{\gamma'_3(s_*)} \sqrt{\Lambda_{0}(s_*)} \\
f(s_*)= &-\frac{c_{0,3}(s_*)c_{-2,1}(s_*)}{c_{0,1}(s_*)c_{-2,3}(s_*)} 
\end{align}

Following \citet{TT22}, the relation Eq.~\ref{cot} translates into a Lorentzian-shaped dip in the $\Delta P = f(s)$, the dip being centred on $s_c$ and of width $\sigma = \epsilon/V$. 

We obtain similar dip models for the $(\ell=1,m=0)$ vs $(\ell_i=3, m=0)$ resonance near $s_*= \sqrt{5}$ and the $(\ell=2,m=-2)$ vs $(\ell_i=4, m=-2)$ resonance near $s_*=8.624$. Indeed, as for the $(\ell=1,m=-1)$ vs $(\ell_i=3, m=-1)$ resonance, the projection of $H_{k}^m$ on the Legendre polynomials and the projection of $P_{\ell_i}^{m}$ on the Hough functions suggest it is sufficient to add two modes to get approximate continuity conditions. These are $H_{k-2}^m$, the Hough function associated with the smallest exponential decrease, and $P_{\ell_i -2}^{m}$.
When expressed for these three dips, the width and location of the dips are given by Eqs.~(\ref{sig_th}) and (\ref{s_th}). 

To conclude, we note that up to the (infinite) system of equations Eq.~\ref{full_cont} describing the continuity conditions, the present model is identical to that of \citet{TT22}. From there, we improved the continuity conditions by adding the contribution of the Hough function $H_{k-2}^m$ on the radiative side and of the Legendre polynomial $P_{\ell_i -2}^{m}$ mode on the convective side. While we argue this is enough to approximate the resonances considered in this paper, we stress that this model cannot be generalised to all dips. Indeed, for other resonances (e.g. the $(\ell=2,m=0)$ vs $(\ell_i=4,m=0)$ resonance), the projections of the $H_{k}^m$ Hough function on the Legendre polynomials and of the $P_{\ell_i}^{m}$ Legendre polynomial on the Hough functions strongly suggest that many more free oscillation modes will be involved leading to much more complex mixed modes than the ones considered here.

\end{appendix}

\end{document}